\documentclass[]{interact}
\usepackage{xcolor}
\usepackage[caption=false]{subfig}

\usepackage[natbibapa,nodoi]{apacite}
\theoremstyle{plain}
\newtheorem{theorem}{Theorem}[section]

\theoremstyle{definition}
\newtheorem{definition}[theorem]{Definition}

\theoremstyle{remark}

\begin{document}

\articletype{RESEARCH ARTICLE}

\title{Beyond the Bid--Ask: Strategic Insights into Spread Prediction and the Global Mid-Price Phenomenon}

\author{
\name{
Yifan He\textsuperscript{a}\thanks{CONTACT Yifan He. Email: Yifan.He@ttu.edu}, 
Abootaleb Shirvani\textsuperscript{b},
Barret Shao\textsuperscript{c},
Svetlozar Rachev\textsuperscript{a}, and
Frank Fabozzi\textsuperscript{d}}
\affil{
\textsuperscript{a}Department of Mathematics and Statistics, Texas Tech University, Lubbock, TX, USA\\
\textsuperscript{b}Department of Mathematical Sciences, Kean University, Union, NJ, USA\\
\textsuperscript{c}Tudor Investment Corporation, New York, NY, USA\\
\textsuperscript{d}Carey Business School, Johns Hopkins University, Baltimore, MD, USA}
}

\maketitle

\begin{abstract}
This research extends the conventional concepts of the bid--ask spread (BAS) and mid-price to include the total market order book bid--ask spread (TMOBBAS) and the global mid-price (GMP). Using high-frequency trading data, we investigate these new constructs, finding that they have heavy tails and significant deviations from normality in the distributions of their  log returns, which are confirmed by three different methods. We shift from a static to a dynamic analysis, employing the ARMA(1,1)-GARCH(1,1) model to capture the temporal dependencies in the return time-series, with the normal inverse Gaussian distribution used to capture the heavy tails of the returns.  We apply an option pricing model to address the risks associated with the low liquidity indicated by the TMOBBAS and GMP.  Additionally, we employ the Rachev ratio to evaluate the risk--return performance at various depths of  the  limit order book and examine tail risk interdependencies across spread levels. This study provides insights into the dynamics of financial markets, offering tools for trading strategies and systemic risk management.
\end{abstract}



\section{Introduction}
The  bid--ask spread (BAS)  has been an integral aspect of financial market transactions throughout history, from the early bazaars to the sophisticated electronic exchanges of today. This spread, which represents the difference between the highest price a buyer is willing to pay (bid price) and the lowest price a seller is willing to accept (ask price), has always influenced the cost of trading.

In ancient times, marketplaces functioned without the formalized BAS as we understand it today. Traders would haggle over the price of goods, with an inherent spread emerging from the difference between the highest price a buyer (bid price) was willing to pay and the lowest price a seller (ask price) was willing to accept. While not explicitly documented in the same way as modern financial markets, the principle of the spread was inherently present \citep{smith2002inquiry}.

The formalization of the BAS came much later, with the development of organized exchanges. The Amsterdam Stock Exchange of the 17th century is often cited as one of the first to list actual prices for stocks, which implicitly included the spread \citep{goetzmann2005origins}. The spread served as a compensation for market makers who took on the risk of holding inventory and providing liquidity.

The mid-price, defined as the arithmetic mean of the bid and ask prices, has long been used as an estimate of the fair value of a security. This simple calculation provides a reference point for traders and investors, although it does not account for the actual transaction costs associated with crossing the spread \citep{hansen2005forecast}.

 With the advent of computerized trading in the 20th trading, the BAS became more transparent and measurable. The spread narrowed as technology improved, reducing the cost of trading and increasing the market's efficiency \citep{hasbrouck2009trading}.  However, the mid-price remains a key indicator used in financial analysis and trading algorithms.

Today, the BAS and the mid-price continue to be critical in understanding market dynamics. High-frequency trading (HFT) and the use of complex algorithms have further reduced the spread in many markets, although it remains an important source of revenue for market makers and an inherent cost for traders \citep{cont2013price}.

In this research, we expand traditional concepts of the BAS and mid-price to encompass the \textit{total market order book bid–ask spread} (TMOBBAS) and the \textit{global mid-price} (GMP). Using HFT data, we explore how extreme events in individual spreads impact the global spread index, represented by the GMP.

The organization of this paper is as follows.
Section~\ref{sec: prelim} introduces TMOBBAS and GMP with a numerical example. Section~\ref{sec: data} describes the dataset.
In Section~\ref{sec: rtn}, we analyze the log returns for TMOBBAS and GMP, finding deviations from normality and evidence of heavy tails. Section~\ref{sec: dynamics} takes a dynamic approach, modeling the returns with an autoregressive moving average-generalized autoregressive conditional heteroskedasticity model incorporating the normal inverse Gaussian (NIG) distribution to account for heavy tails. In Section~\ref{sec: option pricing}, we apply an option pricing model to hedge against liquidity risks in the limit order book (LOB). Section~\ref{sec: rachev} uses the Rachev ratio to assess the risk–return performance of TMOBBAS and GMP, while Section~\ref{sec: CoETL} examines the interdependencies between tail risks at varying depths of the LOB.
Finally, Section~\ref{sec: conclusion} summarizes the findings and contributions of this study. 

\section{Preliminary}\label{sec: prelim}
\begin{figure}[htbp]
    \centering
    \includegraphics[width=0.5\textwidth, height = 0.5\textheight, keepaspectratio]{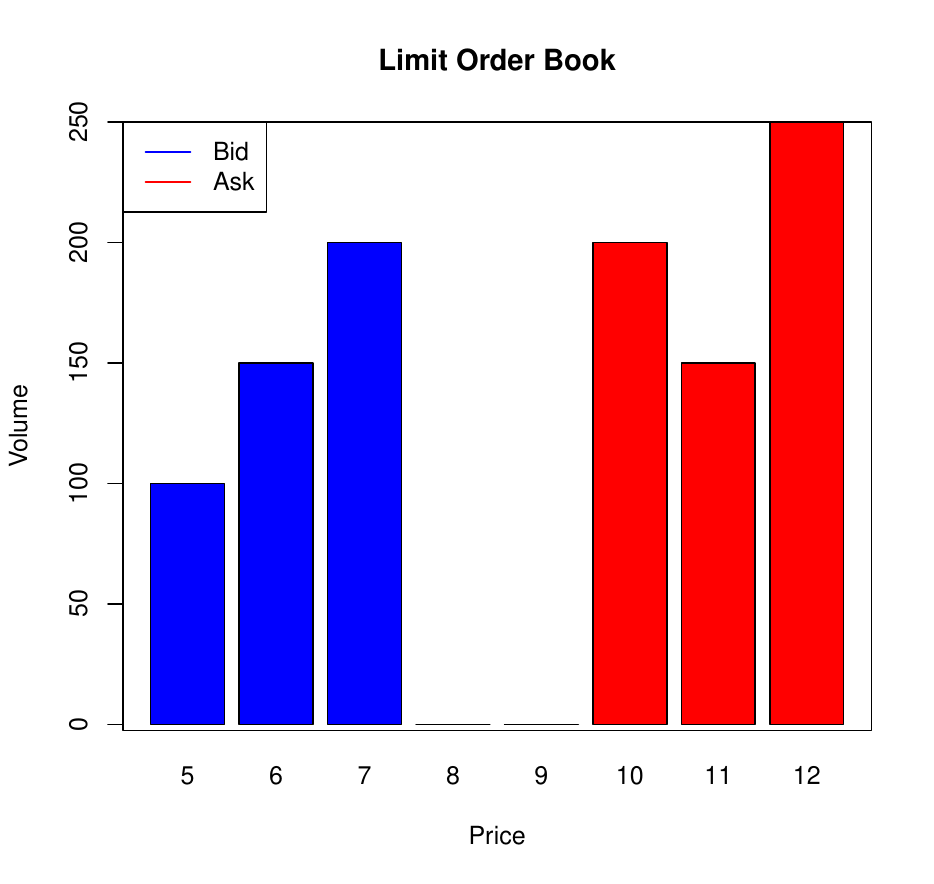}
    \caption{Example of a LOB with a simplified structure}
    \label{fig: LOB}
\end{figure}

 \cite{abergel2016limit} explain that a LOB is essentially a file in a computer that contains all orders sent to the market, with their characteristics, such as the order's sign, price, quantity, and timestamp. In academia, the usual approach involves simplifying the LOB's structure to facilitate its more efficient application.  Figure~\ref{fig: LOB} is an example of  a  LOB with a simplified structure.

 In Figure~\ref{fig: LOB}, the blue side depicts the bid prices (5, 6, and 7), representing the prices of limit buy orders, while the red side depicts the ask prices (10, 11, and 12), namely, the prices of the limit sell orders.  In the bid prices and ask prices, we also encounter the corresponding best bid price and best ask price. The best bid price is the highest among the limit buy orders, while the best ask price is the lowest among the limit sell orders. Hence, in this example, the best bid price is 7 and the best ask price is 10.

When discussing the best bid price and best ask price, there are two concepts we cannot ignore. One is the BAS, the other is the mid-price. The BAS is defined as the difference between the best ask price and the best bid price, while the mid-price is defined as the average of the bid price and the ask price. In the example in Figure~\ref{fig: LOB}, the BAS is $10 - 7 = 3$ and the mid-price is $(10 + 7)/2 = 8.5$.

In this paper, we define two new concepts, which contain the BAS and mid-price as special cases. Before delving into these new concepts, let us introduce a new definition:
\begin{definition}\label{def: TMOBAP and TMOBBP}
(TMOBAP and TMOBBP) The \textit{total market order book ask price} (TMOBAP) is the quotient of the value of a market order that is sufficient to purchase the entire volume of shares in the ask limit order book by the total number of shares available in the ask limit order book. Similarly, the \textit{total market order book bid price} (TMOBBP) is the quotient of the value of the market order that is sufficient to purchase the entire volume of shares in the bid limit order book by the total number of shares available in the bid limit order book.
\end{definition}
Referring to Definition~\ref{def: TMOBAP and TMOBBP}, the two new concepts we define are referred to as the TMOBBAS and the GMP, respectively. They are defined as follows:
\begin{align}
\textrm{TMOBBAS} &= \textrm{TMOBAP $-$ TMOBBP},\label{eq: TMOBBAS}\\
\textrm{GMP} &= (\textrm{TMOBAP + TMOBBP})/2.\label{eq: GMP}
\end{align}

Based on Definition~\ref{def: TMOBAP and TMOBBP}, it is evident that the values of the TMOBAP and TMOBBP are influenced by the levels of bid and ask prices, including the best bid and ask, as well as the subsequent tiered bids and asks. Consequently, the values of the TMOBBAS and GMP are also correlated with the depth of the bid and ask stacks, reflecting the range and distribution of these prices. In this study, we refer to this quantity as the “depth” and denote it by the symbol $d$. Let us continue to use Figure~\ref{fig: LOB} to provide a clearer explanation of these new terms. Based on Definition~\ref{def: TMOBAP and TMOBBP}, Equations~\eqref{eq: TMOBBAS} and \eqref{eq: GMP}, and referencing Figure~\ref{fig: LOB}, we can derive the following:
\begin{itemize}
\item If we only consider the best bid price and the best ask price, then the depth $d = 1$. The TMOBAP $= 10\cdot 200 / 200 = 10$, which is the best actual ask price. Similarly, we can see that TMOBBP $= 7\cdot 200/200 = 7$, which is  the  best actual bid price. In this scenario, the TMOBBAS = 10 - 7 = 3, and the GMP = $(10 + 7) / 2 = 8.5$.
\item If we consider  the  two largest bid prices (7 and 6) and the two smallest ask prices (10 and 11), then the depth $d = 2$. The TMOBAP $= (10\cdot 200 + 11\cdot 150)/(200 + 150) \approx 10.43$ and TMOBBP $= (7\cdot 200 + 6\cdot 150)/(200 + 150)\approx 6.57$. In this scenario, the TMOBBAS $\approx 10.43-6.57 = 3.86$ and GMP $\approx (10.43 + 6.57) / 2 = 8.5$.
\item If we consider the three largest bid prices (7, 6, and 5) and the three smallest ask prices (10, 11, and 12), then the depth $d = 3$. The TMOBAP $= (10\cdot 200 + 11\cdot 150 + 12\cdot 250)/(200+150+250)\approx 11.08$ and TMOBBP $= (7\cdot 200 + 6\cdot 150 + 5\cdot 100)/(200 + 150 + 100)\approx 6.22$. In this scenario, the TMOBBAS $\approx 11.08 - 6.22 = 4.86$ and GMP $\approx (11.08 + 6.22)/2 = 8.65$.
\end{itemize}

The outcomes of the computations detailed above indicate that as the depth $d$ increases, the value of TMOBAP increases, while the value of TMOBBP decreases, leading to an overall increase in the value of TMOBBAS. Nevertheless, the trend observed for the GMP is not consistent, and remains indeterminate:
\begin{itemize}
\item If the increase in TMOBAP is greater than the decrease in TMOBBP, then GMP will increase.
\item If the increase in TMOBAP equals the decrease in TMOBBP, then GMP will remain the same.
\item If the increase in TMOBAP is smaller than the decrease in TMOBBP, then GMP will decrease.
\end{itemize}

\section{Overview of the Datasets}\label{sec: data}
Obtaining the high-frequency LOB data with which to compute the TMOBBAS and GMP for real-world analysis is a challenging endeavor due to the value and sensitivity of this information. Fortunately, the Limit Order Book System -- The Efficient Reconstructor (LOBSTER)\footnote{The official website of LOBSTER is \url{https://lobsterdata.com/}.} provided us with free samples of high-frequency data with which to conduct academic research.

In our analysis, we adhere to the guidelines outlined in \cite{huang2011lobster} to select HFT data samples for Amazon (AMZN), Apple (AAPL) and Google (GOOG) from 9:30:00 AM to 4:00:00 PM on June 21, 2012. To mitigate the effects of the ``burn-in'' period,\footnote{
A ``burn-in'' period refers to an initial phase in the life of a financial instrument, system, or model, during which it may exhibit atypical behavior or performance due to various factors such as initialization, data calibration, or the settling of initial conditions. This period is often considered less representative of the instrument’s or system’s true long-term behavior, and as such, it is sometimes excluded from analysis to prevent skewed results. During the burn-in period, the following might occur: 
i) \textit{Initialization Effects:} Systems might have startup procedures that don’t reflect regular operations.
ii) \textit{Data Noise:} Early data might be noisy due to initial data collection or calibration issues.
iii) \textit{Model Calibration:} Financial models might require  adjusting  to market conditions.
iv) \textit{Adoption Rates:}  There might be a period of low trading volume for new financial instruments  as the market adjusts to their presence.}
we eliminate the initial 5\% of the timestamp data, along with the corresponding price and volume information. Table~\ref{table: summary} summarizes the cleaned datasets that we will use.
\begin{table}[htbp]
\tbl{Summary of our datasets}
{\begin{tabular}{l c c}
\hline
Ticker & Number of Timestamps & Range of Depth\\ \hline
AMZN & 248,201 & $1-10$\\
AAPL & 365,112 & $1-10$\\
GOOG & 132,410 & $1-10$\\ \hline
\end{tabular}}
\label{table: summary}
\end{table}

To analyze TMOBBAS or GMP at different depths, we must truncate our dataset to include a specific number of ask prices, bid prices, ask volumes and bid volumes. For instance, if we intend to examine the TMOBBAS and GMP of AAPL at a depth of 2, we should truncate the dataset for AAPL to include two ask prices, two bid prices, two ask volumes and two bid volumes. Table~\ref{table: dataset} is the part of the dataset for AAPL. It is evident that to explore the TMOBBAS and GMP at a depth of 2, we need to truncate to nine columns, including the timestamp, two ask prices, two ask volumes (sizes), two bid prices, and two bid volumes.
\begin{table}[htbp]
\tbl{Part of dataset for AAPL at depth 2}
{\begin{tabular}{c c c c c c c c c}
\hline
Time & Ask & Ask & Bid & Bid & Ask & Ask & Bid & Bid\\
     & price 1 & size 1 & price 1 & size 1 & price 2 & size 2 & price 2 & size 2\\ \hline
$\vdots$ & $\vdots$ & $\vdots$ & $\vdots$ & $\vdots$ & $\vdots$ & $\vdots$ & $\vdots$ & $\vdots$\\
35159.318815640 & 5865700 & 100 & 5863100 & 100 & 5866100 & 1900 & 5862900 & 100\\
35159.428838154 & 5865700 & 100 & 5863100 & 100 & 5865900 & 200 & 5862900 & 100\\
35159.432266639 & 5865700 & 100 & 5863100 & 100 & 5865900 & 200 & 5862900 & 100\\
35159.439130585 & 5865700 & 100 & 5863200 & 100 & 5865900 & 200 & 5863100 & 100\\
35159.450166122 & 5865700 & 100 & 5863200 & 100 & 5865900 & 200 & 5863100 & 100\\
$\vdots$ & $\vdots$ & $\vdots$ & $\vdots$ & $\vdots$ & $\vdots$ & $\vdots$ & $\vdots$ & $\vdots$\\
57599.913117637 & 5776700 & 300 & 5775400 & 410 & 5776800 & 200 & 5775300 & 1400\\ \hline
\end{tabular}}
\label{table: dataset}
\end{table}

The values in the ``Time'' column are the time difference between the current time and midnight in seconds\footnote{Beginning with Section~\ref{sec: rtn}, the figures will present time in a standard format, converting seconds into a more conventional unit. This adjustment is intended to enhance the clarity and prevent any confusion for the reader.}. For example, ``34200'' means 34,200 seconds from midnight, which, when converted to hours, is 9.5 hours, namely, 9:30:00 AM. Similarly, ``57600'' indicates 57,600 seconds from midnight, corresponding to 16 hours or 4:00:00 PM.

Regarding the price columns, ``Ask price 1'' is the best ask price, i.e., the lowest ask price, while ``Ask price 2'' is the second-best ask price. Similarly, ``Bid price 1'' is the best bid price, i.e., the highest bid price, and ``Bid price 2'' is the second-best bid price. The values in the price columns are in US dollars (USD) multiplied by 10,000. For example, ``5865700'' is 586.57 USD.

Lastly, the values in the size columns are the corresponding number of shares for the given ask price or bid price.

\section{Return Time-Series}\label{sec: rtn}
In Section~\ref{sec: prelim}, we presented the TMOBBAS and GMP as financial instruments. Assuming these represent the prices of certain assets  we are considering investing in, it is pertinent to discuss the concept of their returns. Subsequently, we will introduce the respective log-return calculations for the TMOBBAS and GMP.

\begin{figure}[htbp]
\centering
\subfloat[]{\includegraphics[width=0.49\textwidth]{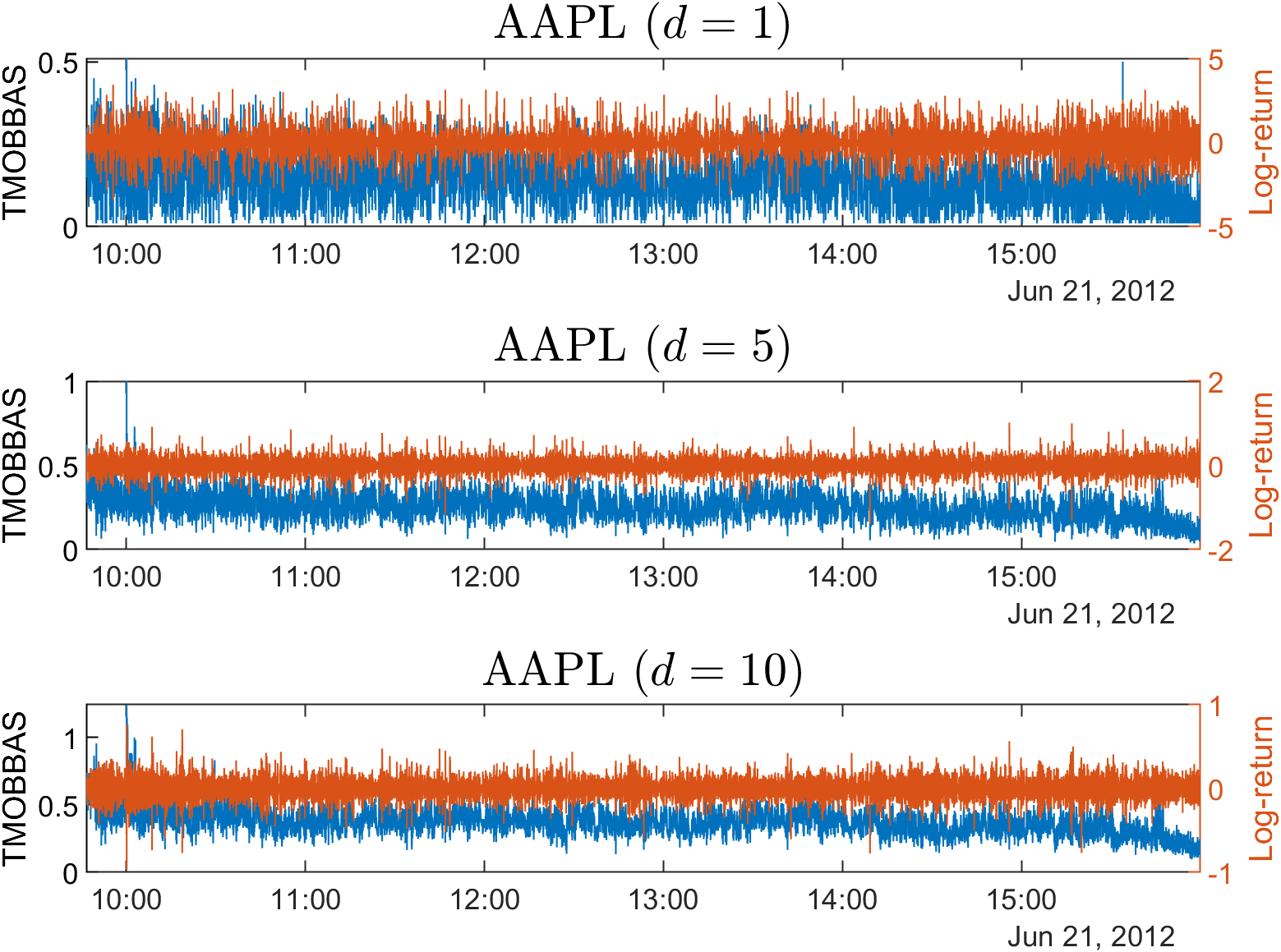}}
\label{figure: gmp_log_return_aapl}
\subfloat[]{\includegraphics[width=0.49\textwidth]{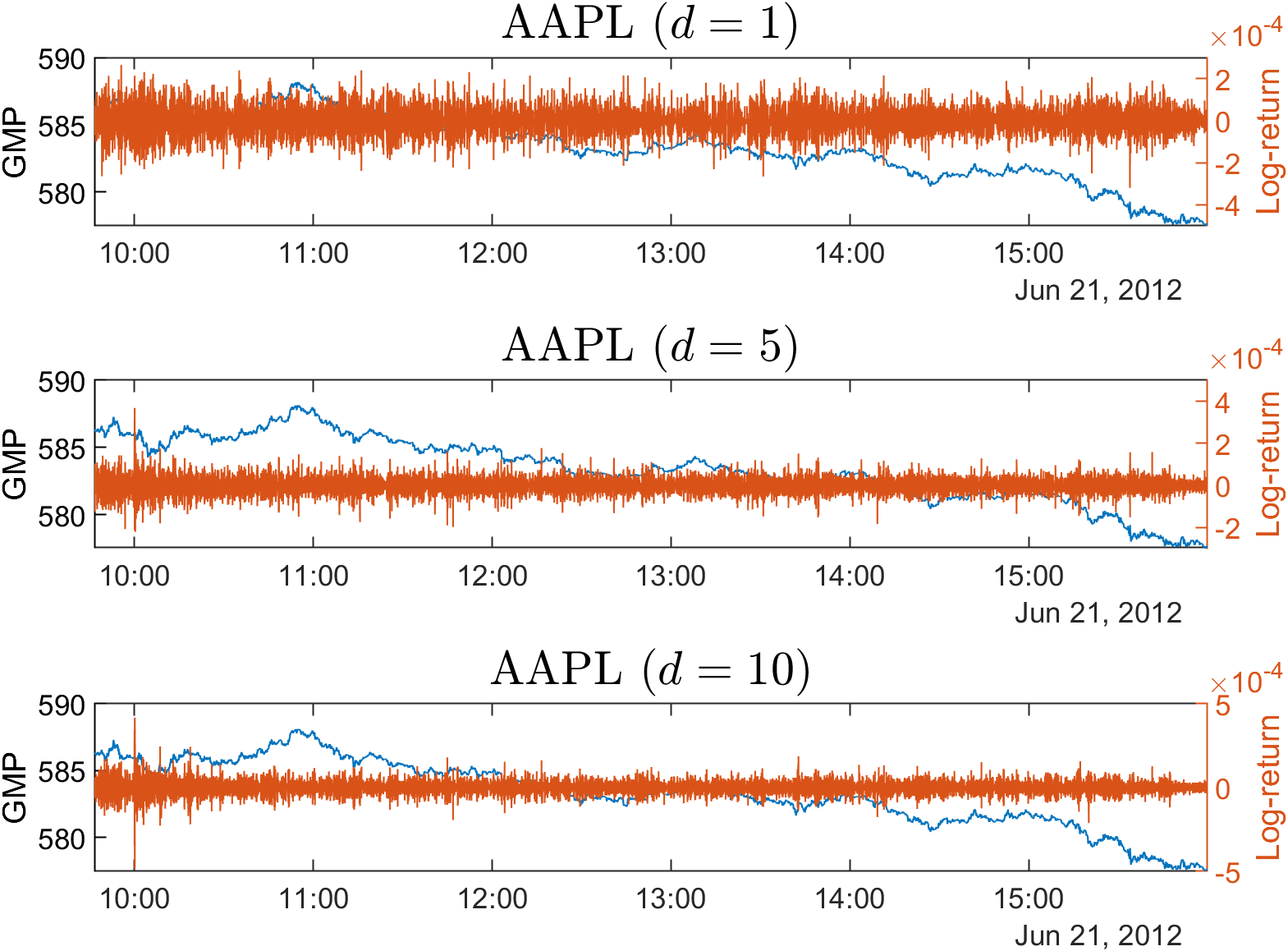}}
\label{figure: tmobbas_log_return_aapl}
\caption{The evolution of (a) TMOBBAS, (b) GMP, and their log-returns over time at depths of 1, 5, and 10 for AAPL (both two subgraphs contain two vertical axes. The left vertical axis indicates the value of TMOBBAS or GMP, and the right vertical axis indicates the value of the log-returns. The evolution of the TMOBBAS or GMP is represented by the blue line while the orange line represents the evolution of the log returns)}
\label{figure: gmp_tmobbas_aapl}
\end{figure}

Considering the extremely short duration between trading actions in HFT, which can span from milliseconds to microseconds, we will focus on the \textit{log returns}  when studying the TMOBBAS and GMP in our research.  Specifically, let $S_{t, d}$ denote the TMOBBAS or GMP of a given ticker at timestamp $t$ and depth $d$, where 9:30 AM $\le t\le$ 4:00 PM and $1\le d\le 10$. Then the \textit{log-return} of TMOBBAS or GMP, $r_{t, d}$ is defined as
\begin{equation}\label{eq: log_return}
r_{t, d} = \log\left(\frac{S_{t+\Delta t, d}}{S_{t, d}}\right) = \log(S_{t +\Delta t, d}) - \log(S_{t, d}),
\end{equation}
where $t+\Delta t$ ($\Delta t > 0$) is the timestamp for the next time step after timestamp $t$. To help us visualize the  price time series and return time series, for instance, Figure~\ref{figure: gmp_tmobbas_aapl} displays the evolution of the TMOBBAS and GMP over time at depths of 1, 5, and 10 for AAPL.\footnote{To conserve space, our primary focus in the main text will be on AAPL from this section on, due to its having the largest market capitalization up to 2024 among these three stocks. Analyses for AMZN and GOOG can be found in Appendix~\ref{appendix: figure} and Appendix~\ref{appendix: table}. Since the analyses for AMZN and GOOG are similar to those for AAPL, we will ignore the explanatory text.}

\subsection{The Non-Normality of the Returns}
We  defined log returns in Equation~\eqref{eq: log_return}. Moving forward, our investigation into TMOBBAS and GMP will primarily focus on analyzing these log returns as a core aspect of our research.

The initial focus of our investigation is to examine the distribution of returns. In financial research, it is commonly assumed that the distribution of returns follows a normal distribution \citep{bachelier1900theorie, markowitz1952portfolio, sharpe1964capital, fama1965behavior, samuelson1965proof}.

The question of whether a given distribution is normal can be addressed by:
\begin{itemize}
\item  comparing fits the distribution of a non-parametric kernel density distribution against the normal distribution.
\item via Quantile--Quantile (QQ) plots; or
\item measurements of the skewness and kurtosis (especially the kurtosis), central moments characterizing the behavior of the tails.
\end{itemize}
A QQ plot is a graphical tool that examines the correspondence between the quantiles of a sample’s distribution and those of a reference theoretical distribution. The plot’s axes are scaled to reflect a linear association between the quantiles, suggesting that the sample distribution aligns with the theoretical one. When the theoretical distribution is  normal, the plot is known as a normal QQ plot. In a normal QQ plot, the horizontal axis (representing the normal distribution) is calibrated in units of standard deviations.

\begin{figure}[htbp]
\centering
\subfloat[]{\includegraphics[width=0.8\textwidth, height = 0.3\textheight, keepaspectratio]{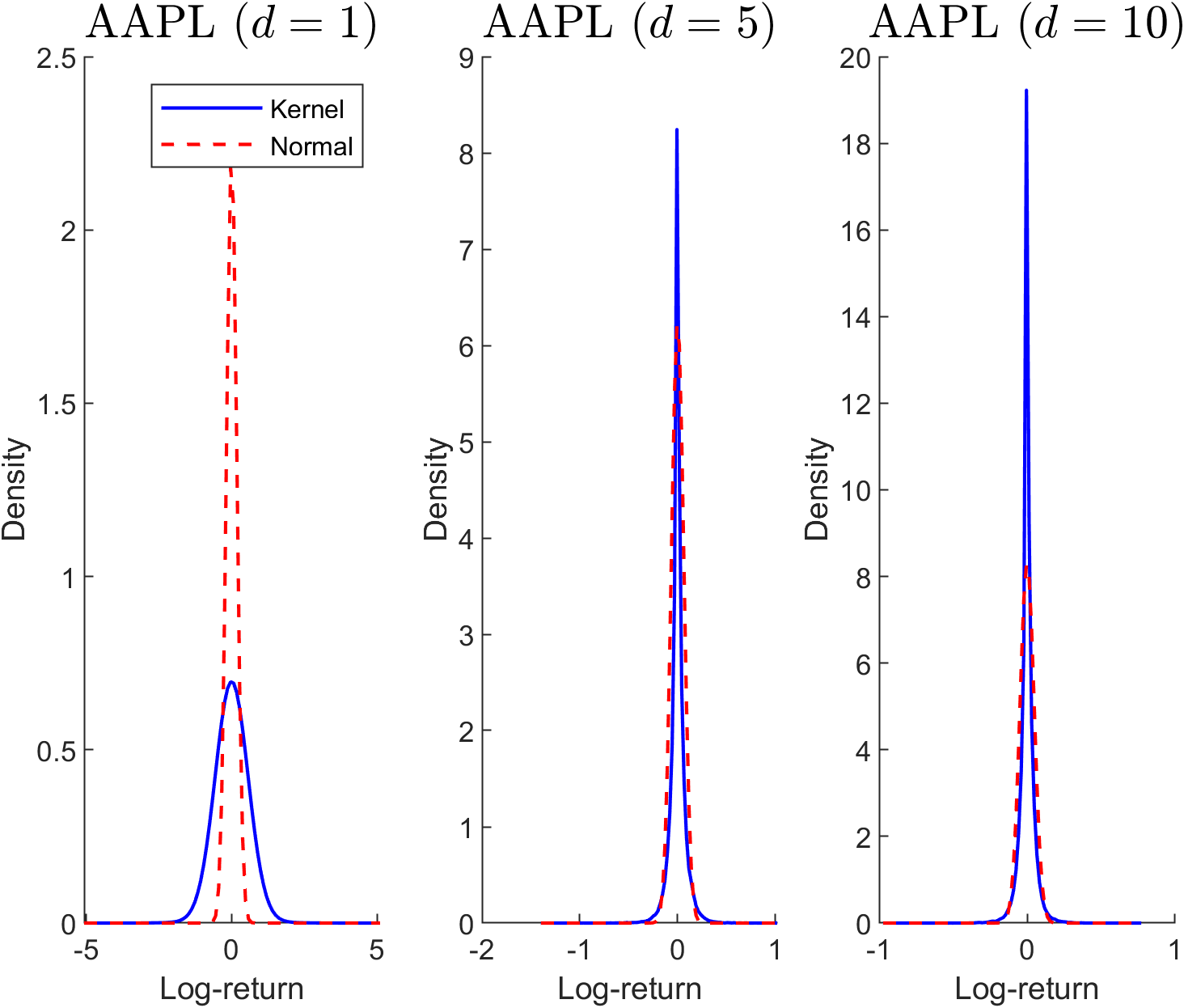}}
\label{figure: tmobbas_no_gaussian_aapl}
\subfloat[]{\includegraphics[width=0.8\textwidth, height = 0.3\textheight, keepaspectratio]{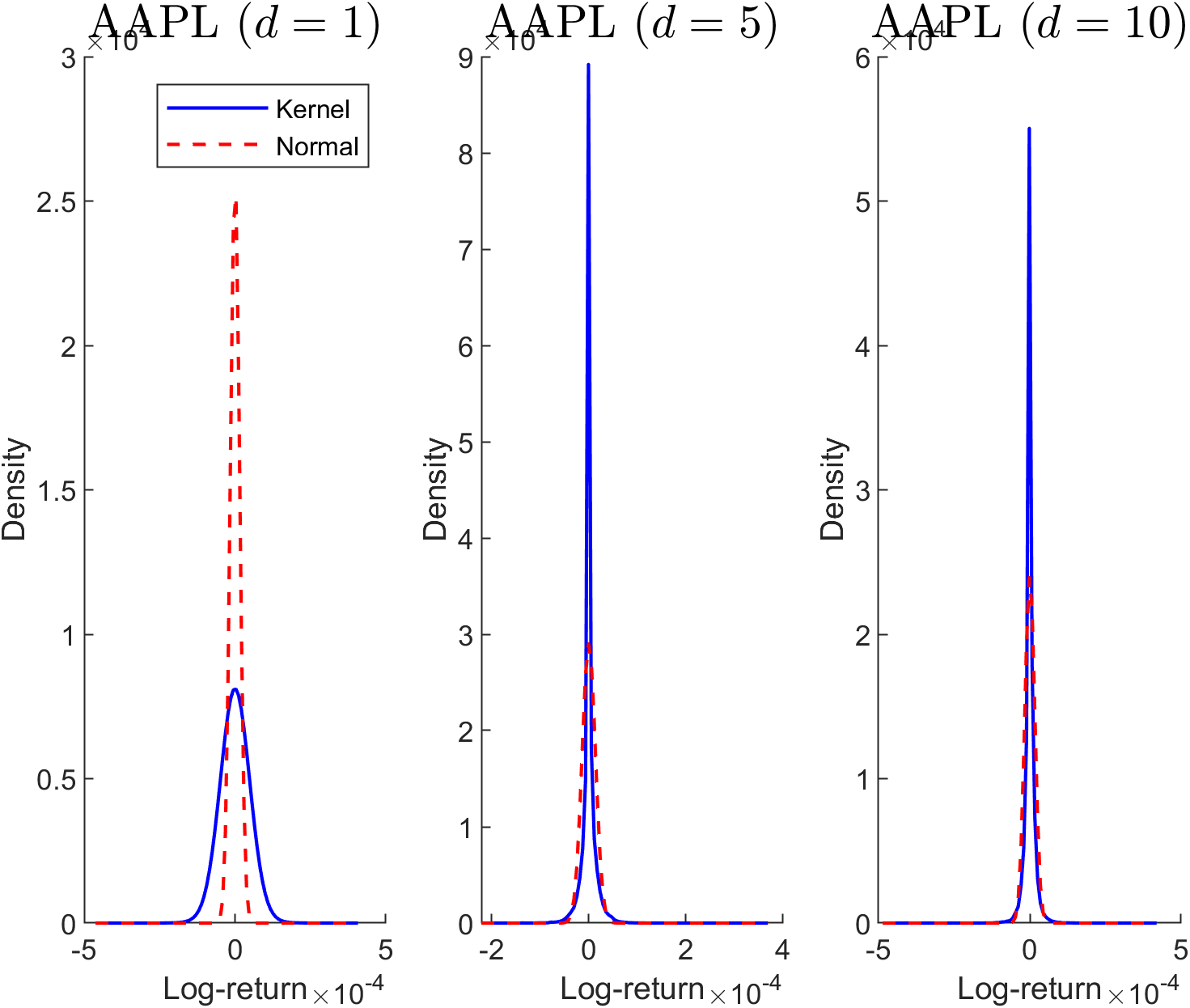}}
\label{figure: gmp_no_gaussian_aapl}
\caption{Comparison between the kernel density (represented by blue solid lines) of log-returns of (a) TMOBBAS, (b) GMP and the corresponding normal distribution with the same sample mean and standard deviation (represented by the red dashed lines) at depths of 1, 5, and 10 for AAPL}
\label{figure: no_gaussian_aapl}
\end{figure}

Figure~\ref{figure: no_gaussian_aapl} provides the kernel density -- normal density comparisons for the TMOBBAS and GMP for AAPL at depths of 1, 5, and 10. It is obvious that the returns are non-normal.

\begin{figure}[htbp]
\centering
\subfloat[]{\includegraphics[width=0.8\textwidth, height = 0.3\textheight, keepaspectratio]{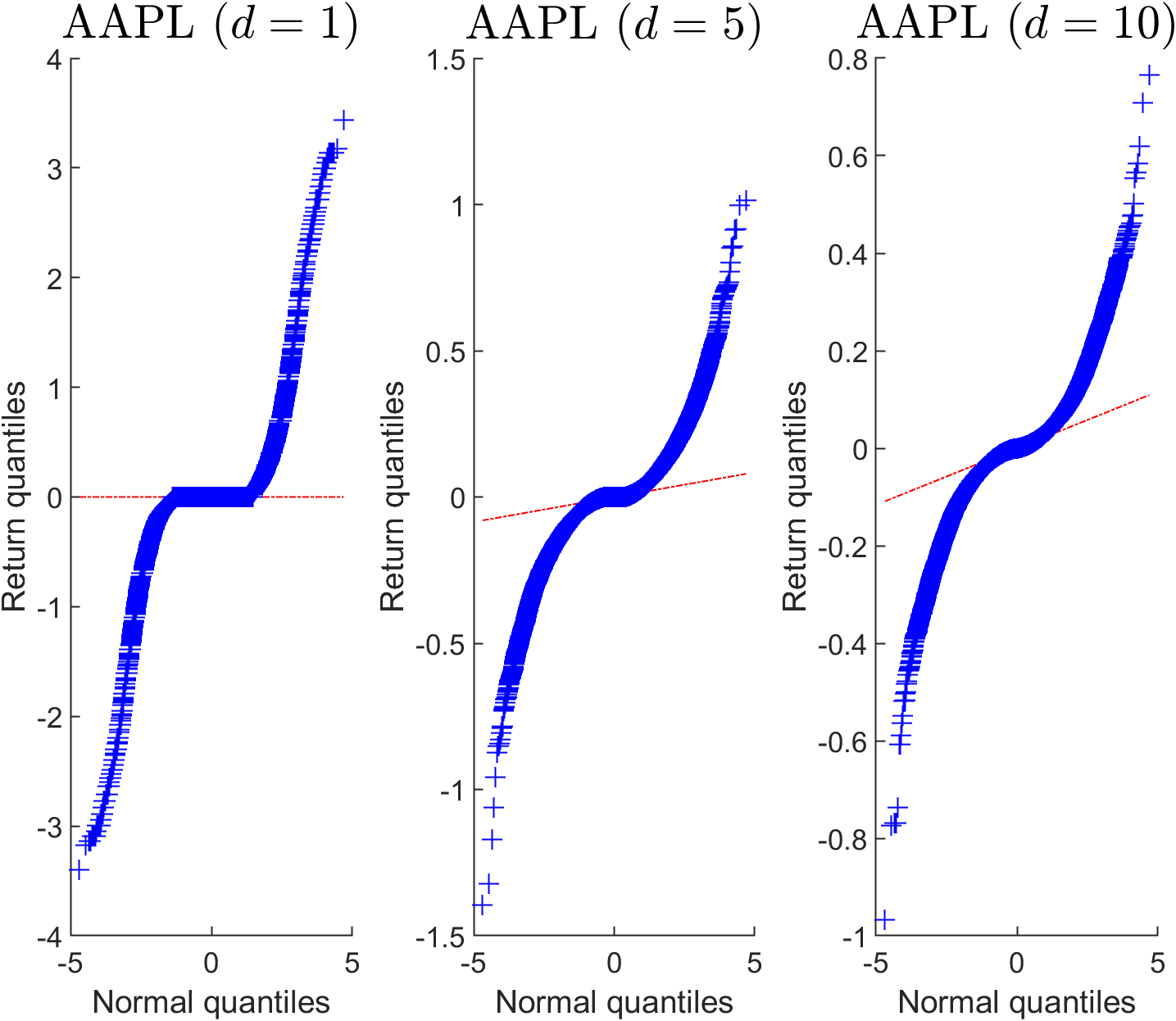}}
\label{figure: tmobbas_qq_plot_aapl}
\subfloat[]{\includegraphics[width=0.8\textwidth, height = 0.3\textheight, keepaspectratio]{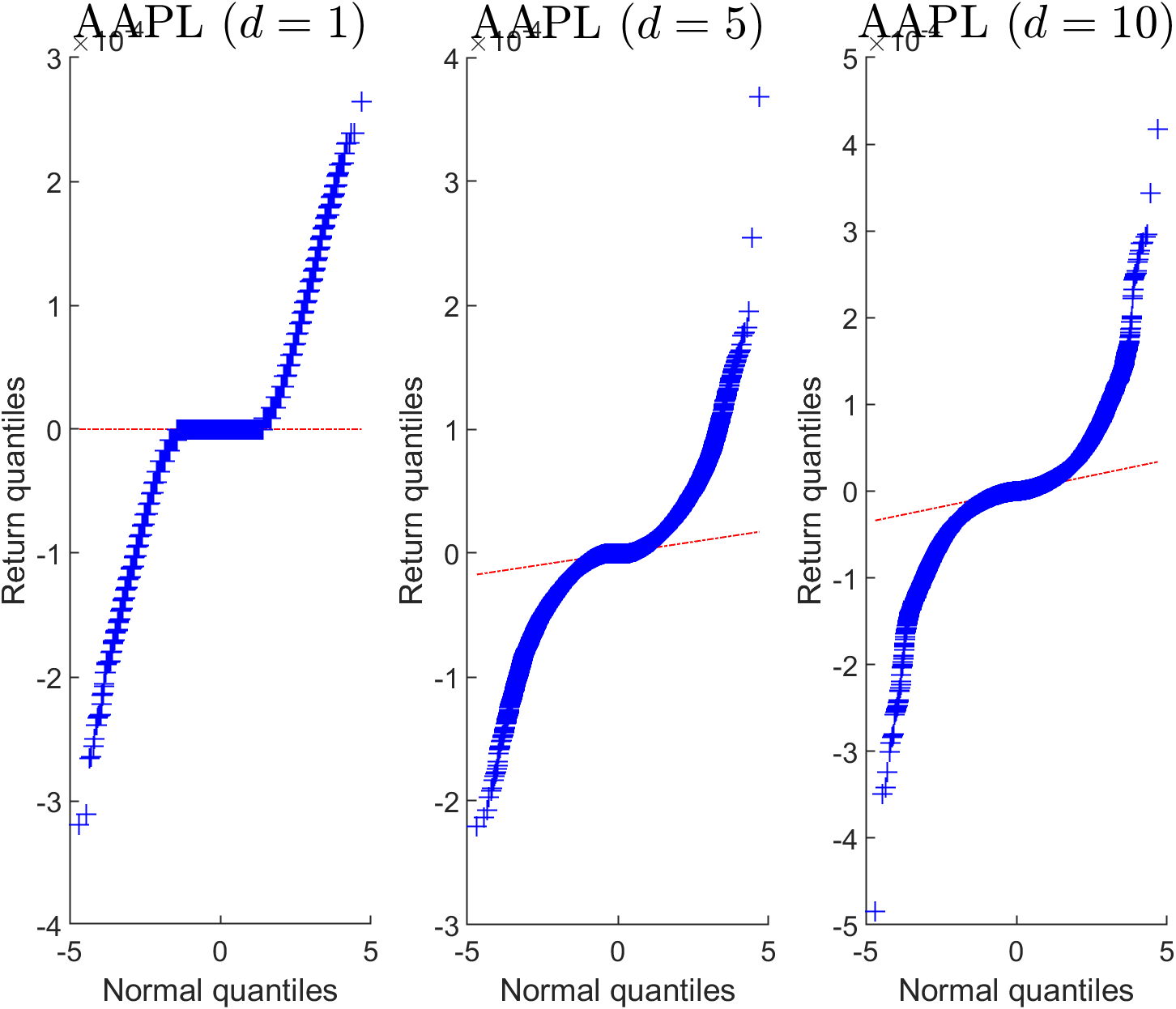}}
\label{figure: gmp_qq_plot_aapl}
\caption{Normal QQ plots of the return distributions for the (a) TMOBBAS and (b) GMP for AAPL at depths of 1, 5, and 10}
\label{figure: qq_plot_aapl}
\end{figure}

 Figure~\ref{figure: qq_plot_aapl} shows the normal QQ plots analyzed in Figure~\ref{figure: no_gaussian_aapl}. The  returns are decidedly non-normal, exhibiting what are referred to as \textit{fat tails}.\footnote{Fat-tailed distributions exhibit large negative and positive values with greater probability than a normal distribution.}

 A normal distribution has kurtosis $\kappa = 3$. The difference $\kappa - 3$ is referred to as the excess kurtosis. In addition to the traditional excess kurtosis, we also propose a robust measure of excess kurtosis for characterizing a distribution's tails, as suggested by \cite{hogg1972more, hogg1974adaptive}:
\begin{equation}\label{eq: robust_kurt}
\kappa_{\rm robust} - 2.59 = \frac{U_{0.05}-L_{0.05}}{U_{0.5}-L_{0.5}}-2.59,
\end{equation}
where
\begin{equation*}
\begin{aligned}
U_{\alpha} &= \frac{1}{\alpha}\int_{1-\alpha}^{1}F^{-1}(y)\textrm{ d}y,\\
L_{\alpha} &= \frac{1}{\alpha}\int_{0}^{\alpha} F^{-1}(y)\textrm{ d}y,
\end{aligned}
\end{equation*}
and $F^{-1}(y)$ is the $y$-th quantile of the sample distribution. The reason for subtracting $2.59$ in Equation~\eqref{eq: robust_kurt} is to ensure that the robust excess kurtosis of the normal distribution is zero.

\begin{figure}[htbp]
\centering
\subfloat[]{\includegraphics[width=0.49\textwidth, height = 0.4\textwidth, keepaspectratio]{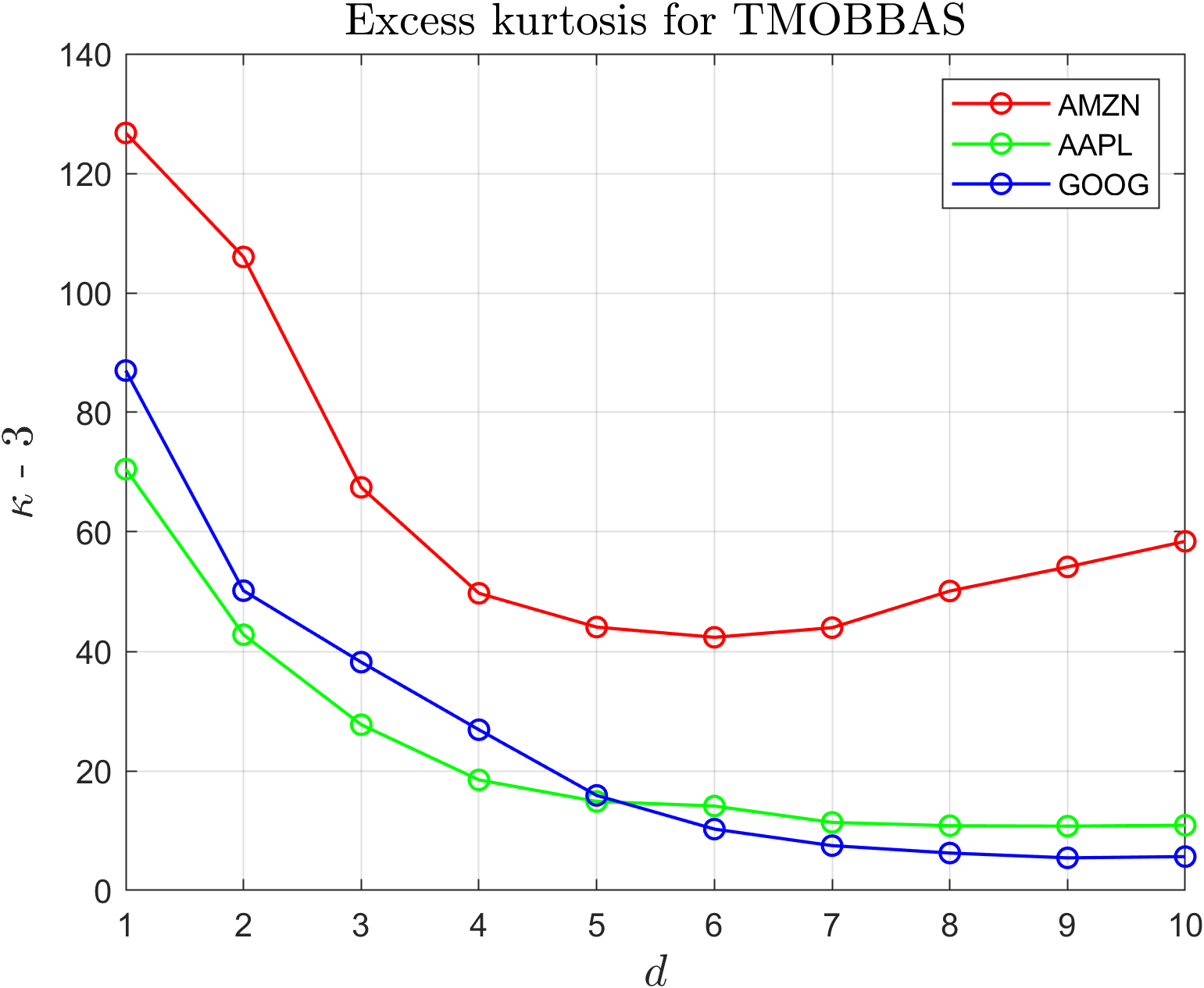}}
\label{figure: tmobbas_kurt}
\subfloat[]{\includegraphics[width=0.49\textwidth, height = 0.4\textwidth, keepaspectratio]{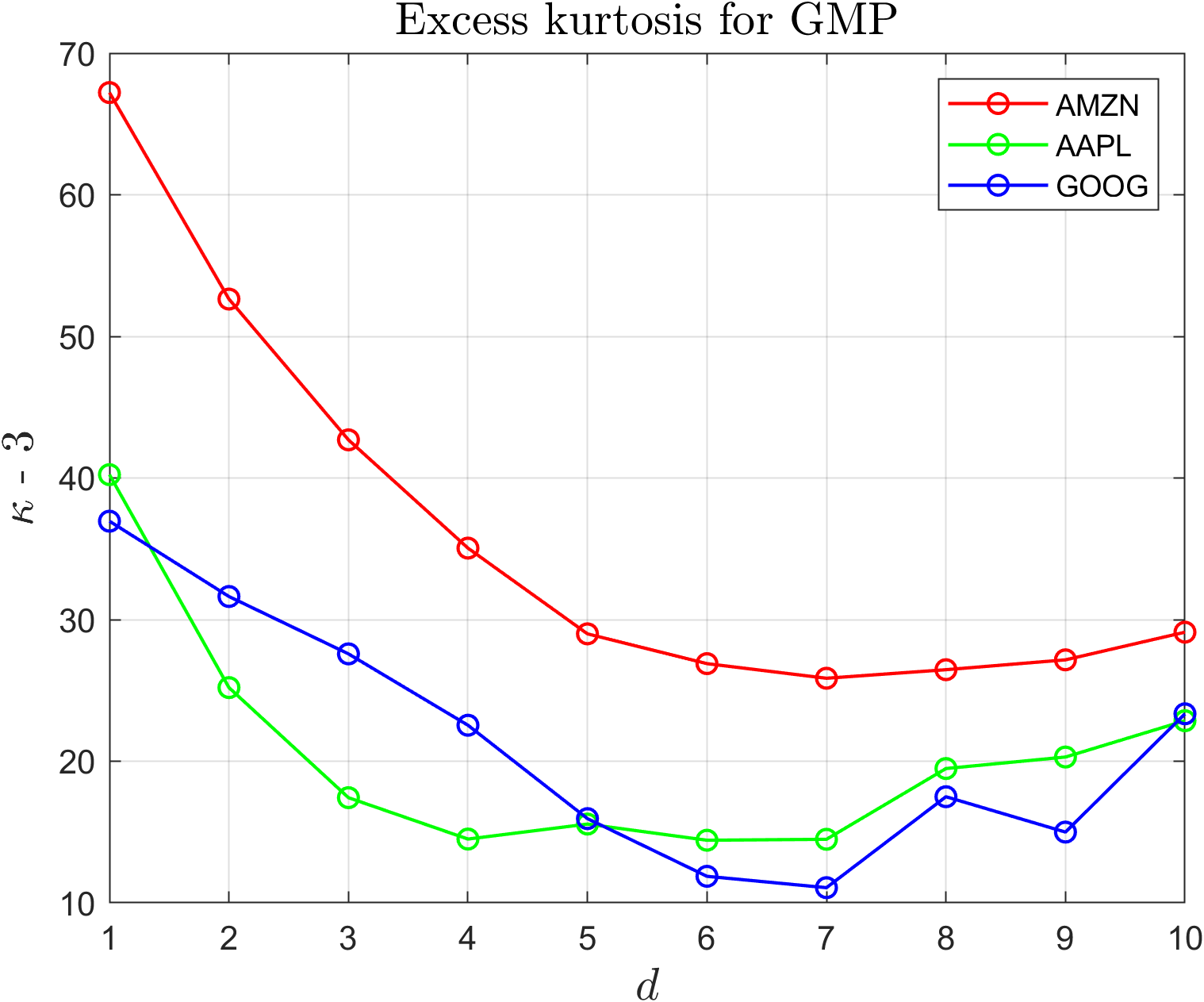}}
\label{figure: gmp_kurt}
\caption{The evolution of excess kurtosis for (a) TMOBBAS and (b) GMP over depth (AMZN, AAPL, and GOOG are represented by red, green, and blue line with dots, respectively)}
\label{figure: kurt}
\end{figure}

\begin{figure}[htbp]
\centering
\subfloat[]{\includegraphics[width=0.49\textwidth, height = 0.4\textwidth, keepaspectratio]{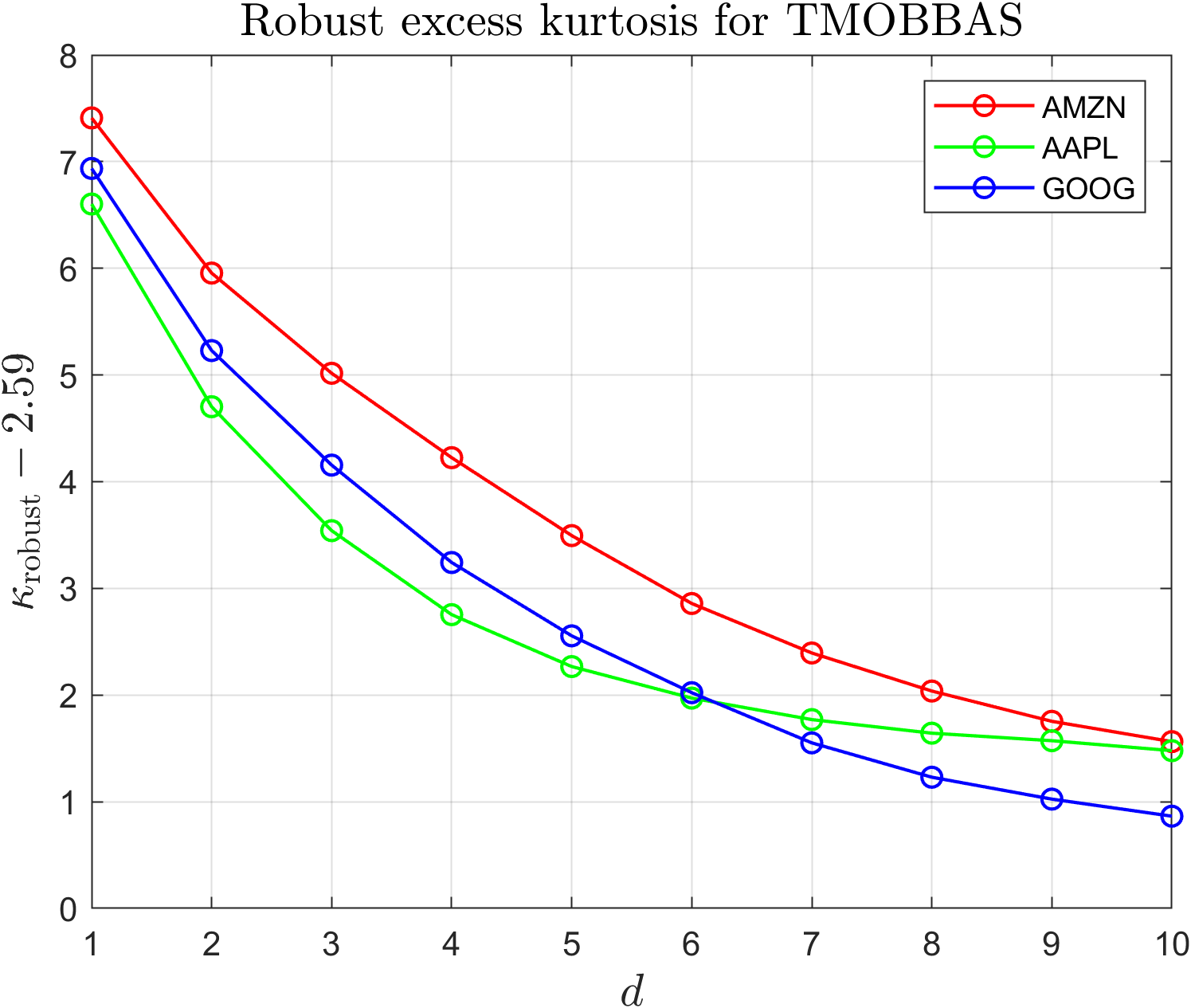}}
\label{figure: tmobbas_robust_kurt}
\subfloat[]{\includegraphics[width=0.49\textwidth, height = 0.4\textwidth, keepaspectratio]{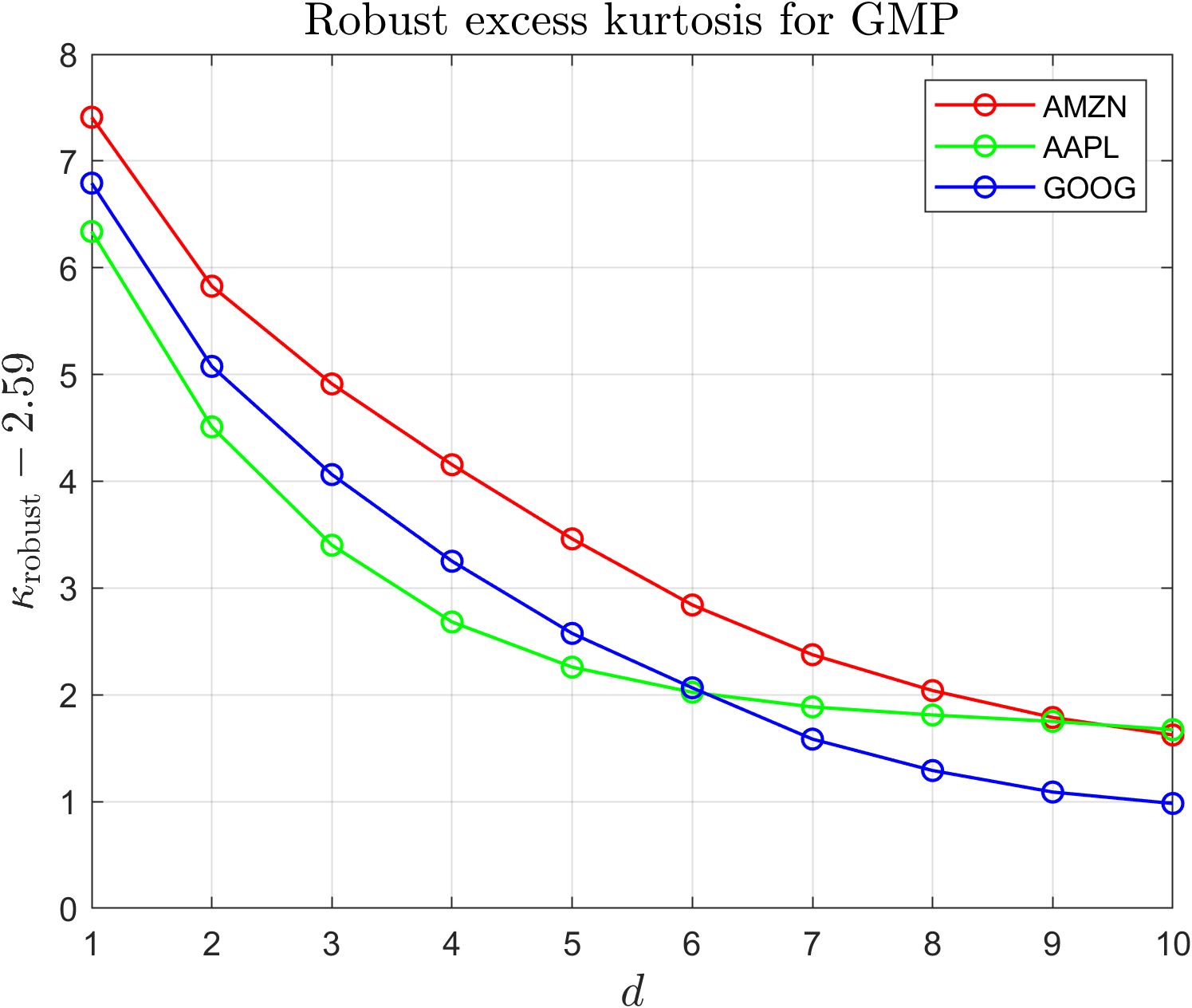}}
\label{figure: gmp_robust_kurt}
\caption{The evolution of robust excess kurtosis for (a) TMOBBAS and (b) GMP over depth (AMZN, AAPL, and GOOG are represented by red, green, and blue line with dots, respectively)}
\label{figure: robust_kurt}
\end{figure}

From Figures~\ref{figure: kurt} and \ref{figure: robust_kurt}, we can observe that both the excess kurtosis and the robust excess kurtosis for TMOBBAS and GMP at each depth  are  significantly greater than zero, indicating that the distributions of the log-returns of TMOBBAS and GMP for these three stocks  are  \textit{leptokurtic} (with positive excess kurtosis and positive robust excess kurtosis).

In synthesizing the analyses of the fits to the distributions, including the QQ plots and the calculation of both the excess kurtosis and the robust excess kurtosis, we find compelling empirical evidence to refute the notion that the returns of TMOBBAS and GMP follow a normal distribution. In light of this departure from normality, we turn our attention to investigating the tail behavior of these distributions, as discussed in \cite{rachev2000stable, shimokawa2007agent, lux2009stochastic, rogers2011asset}. Profiling the tail statistics aligns with the adage of “let the tail speak for itself,” as emphasized in \cite{embrechts1997modelling, mcneil2015quantitative}. In the subsequent subsections, we explore three techniques for characterizing the tails of the return distribution.

\subsection{Tail Behavior of Returns}

\subsubsection{Fitting a Generalized Pareto Distribution}
The initial approach we consider uses the generalized Pareto distribution (GPD) to model the tails of the returns. The cumulative distribution function (CDF) of the GPD is given by
\begin{align}\label{eq: GPD}
F_{\rm GPD}(x;\sigma, \xi) = 
\begin{cases}
1 - (1 + \xi x/\sigma)^{-1/\xi} & \textrm{if }\xi\ne 0,\\
1 - e^{-x/\sigma} &\textrm{if }\xi = 0,
\end{cases}
\end{align}
where $x\in[0, \infty)$ if $\xi\ge 0$, and $x\in[0, -\sigma/\xi]$ if $\xi < 0$. Here, $\sigma$ is the scale parameter and $\xi$ is the shape parameter.\footnote{We assume a location (modal) value of zero.} The sign of $\xi$ determines the nature of the tail of the distribution:
\begin{itemize}
\item if $\xi > 0$, the distribution has a power-law decay (i.e., is heavy-tailed);
\item if $\xi = 0$, the distribution decays exponentially; and
\item if $\xi < 0$, the support of the distribution is bounded.
\end{itemize}
As demonstrated by \cite{balkema1974residual} and \cite{pickands1975statistical},  the GPD can accurately approximate a broad category of tail distributions.

\begin{figure}[htbp]
\centering
\subfloat[]{\includegraphics[width=0.8\textwidth, height = 0.3\textheight, keepaspectratio]{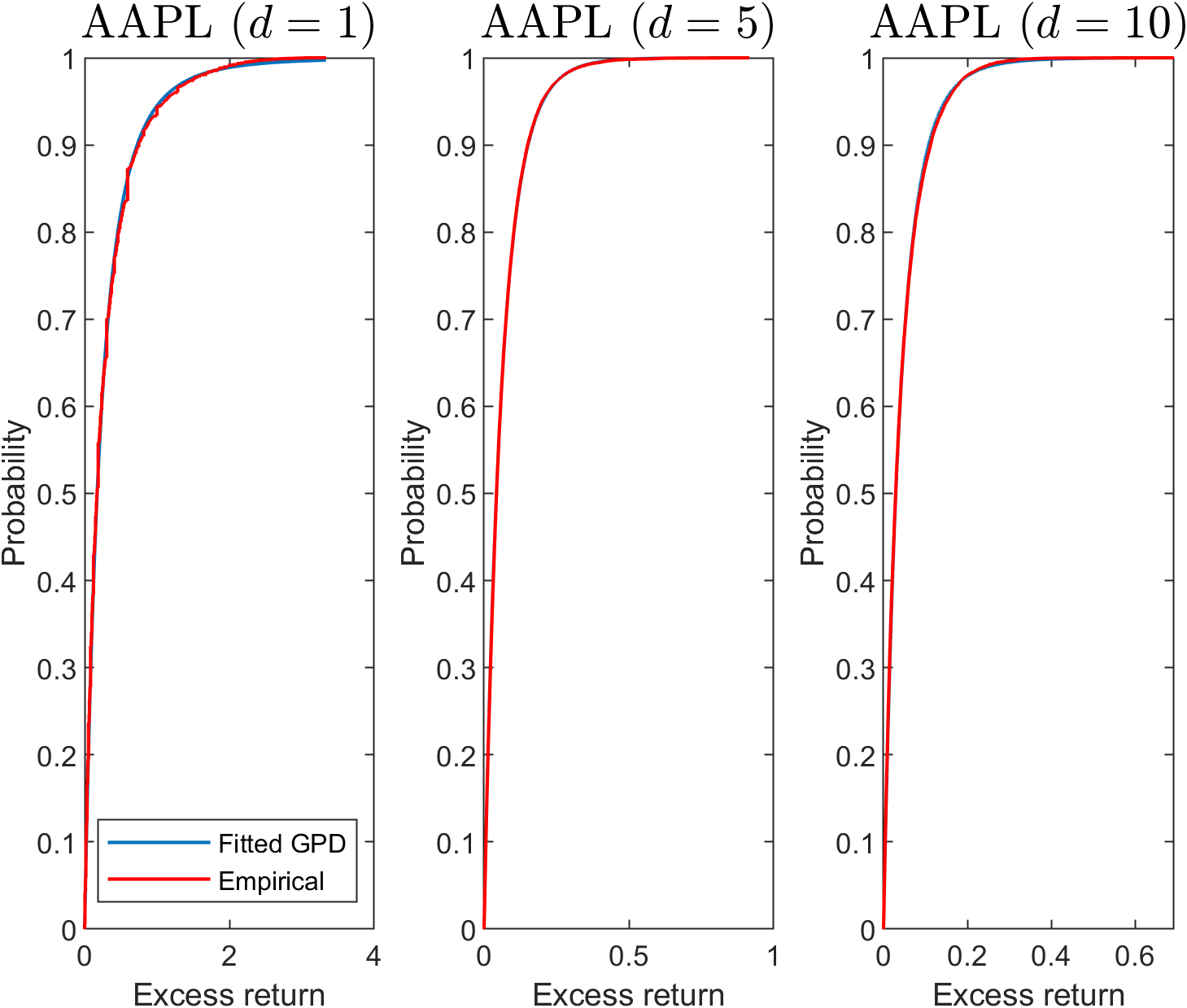}}
\label{figure: tmobbas_gpd_fit_aapl}
\subfloat[]{\includegraphics[width=0.8\textwidth, height = 0.3\textheight, keepaspectratio]{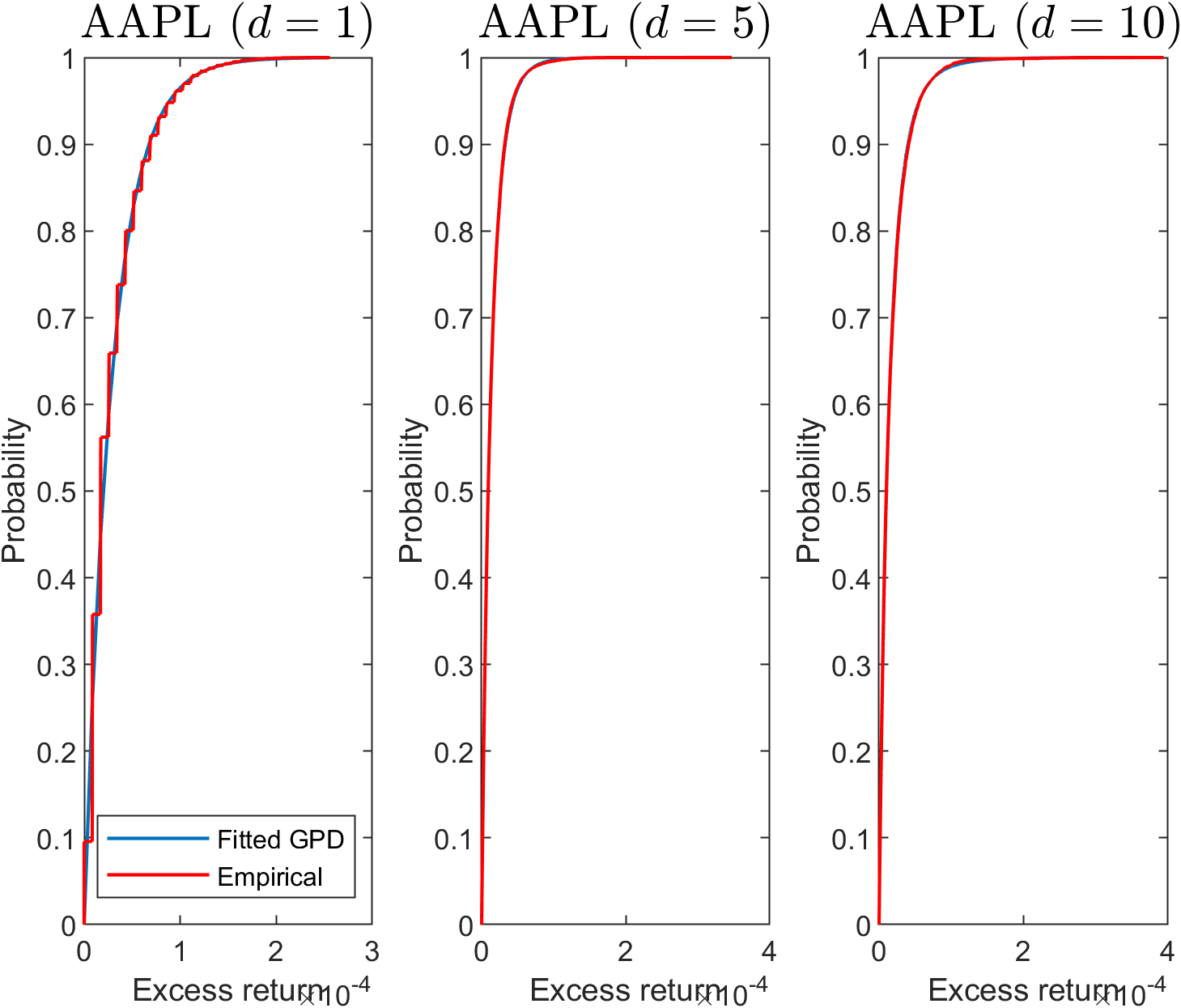}}
\label{figure: gmp_gpd_fit_aapl}
\caption{Comparison between empirical CDF (represented by orange solid lines) of tail and its fitted CDF of GPD (represented by blue solid lines) for (a) TMOBBAS and (b) GMP at depths of 1, 5, and 10 for AAPL}
\label{figure: gpd_fit_aapl}
\end{figure}

We define the positive (right) tail region to consist of the largest 5\% of the return data\footnote{We employ the mean excess plot to justify our selection of the largest 5\% of the return data to represent the right tail. Readers can refer to \cite{mcneil2015quantitative} for more details about the mean excess plot.} for the GMP at each depth. We used the MATLAB \texttt{gpfit()} function\footnote{More details about the \texttt{gpfit()} function can be found at \url{https://www.mathworks.com/help/stats/gpfit.html}.} to fit a GPD's CDF to the empirical CDF of the tail regions. The function returns an estimate of  and 95\% confidence intervals (CIs) for  the parameters $\sigma$ and $\xi$.  A graphical  comparison of the CDFs for the empirical data and GPD fit for TMOBBAS and GMP for AAPL at depths of 1, 5, and 10 are shown in Figure~\ref{figure: gpd_fit_aapl}.

The excess return is defined as the tail return value exceeding the threshold that marks the beginning of the 5\% upper tail region. As shown in Figure~\ref{figure: gpd_fit_aapl}, the fitting quality is consistently high, indicating that the GPD serves as a dependable model for the extreme return tails. The graphical outcomes for the shape parameter $\xi$ and its corresponding 95\% CI bounds are provided in Tables~\ref{table: tmobbas_gpd_fit_est_aapl} and \ref{table: gmp_gpd_fit_est_aapl}, which show that at the 95\% CI level, the estimated values of $\xi$ ($\hat{\xi}$) are positive, signifying that the tails exhibit ``heaviness''.\footnote{The similar results for AMZN and GOOG can be found in Appendix~\ref{appendix: table}.}

\begin{table}[htbp]
\tbl{The value of $\hat{\xi}$ and its 95\% CIs of TMOBBAS for AAPL at all 10 depths}
{\begin{tabular}{c c c | c c c}
\hline
Depth $d$ & $\hat{\xi}$ & 95\% CI & Depth $d$ & $\hat{\xi}$ & 95\% CI\\ \hline
1 & 0.271 & (0.251, 0.291) & 2 & 0.120 & (0.104, 0.137)\\
3 & 0.103 & (0.087, 0.119) & 4 & 0.060 & (0.045, 0.075)\\
5 & 0.078 & (0.063, 0.094) & 6 & 0.075 & (0.060, 0.090)\\
7 & 0.065 & (0.049, 0.080) & 8 & 0.091 & (0.075, 0.108)\\
9 & 0.111 & (0.094, 0.128) & 10 & 0.123 & (0.105, 0.140)\\ \hline
\end{tabular}}
\label{table: tmobbas_gpd_fit_est_aapl}
\end{table}

\begin{table}[htbp]
\tbl{The value of $\hat{\xi}$ and its 95\% CIs of GMP for AAPL at all 10 depths}
{\begin{tabular}{c c c | c c c}
\hline
Depth $d$ & $\hat{\xi}$ & 95\% CI & Depth $d$ & $\hat{\xi}$ & 95\% CI\\ \hline
1 & 0.024 & (0.009, 0.040) & 2 & 0.050 & (0.034, 0.065)\\
3 & 0.047 & (0.032, 0.062) & 4 & 0.072 & (0.057, 0.087)\\
5 & 0.102 & (0.087, 0.118) & 6 & 0.110 & (0.094, 0.126)\\
7 & 0.130 & (0.113, 0.147) & 8 & 0.157 & (0.140, 0.174)\\
9 & 0.174 & (0.156, 0.192) & 10 & 0.187 & (0.169, 0.205)\\ \hline
\end{tabular}}
\label{table: gmp_gpd_fit_est_aapl}
\end{table}

\subsubsection{The Hill Estimate for the Tail Index}
The tail index, denoted by $\alpha = 1/\xi$, is a parameter derived from the shape parameter $\xi$ of the GPD in \eqref{eq: GPD}. It characterizes the rate of decay of the power law tail: a higher value of $\alpha > 0$ indicates a faster decay of the tail and a quicker rise of the tail’s CDF. The Hill estimator, proposed by \cite{hill1975simple}, is a method that uses sample order statistics to estimate $\alpha$. Given a set of $n$ IID positive samples, $X_1, X_2,\cdots, X_n$, from the right tail of a random variable $X$ with support on $(-\infty, \infty)$, we arrange these samples in descending order: $X_{(1)}\ge X_{(2)}\ge\cdots\ge X_{(n)}$. The $k$-th largest value, $X_{(k)}$,  is  the $k$-th order statistic. The Hill estimator calculates $\alpha$ using the following formula:
\begin{equation}\label{eq: Hill}
\hat{\alpha}_{k, n}^{(H)} = \left[\frac{1}{k}\sum_{i=1}^k \ln X_{(i)} - \ln X_{(k+1)}\right]^{-1}.
\end{equation}
According to \cite{de1998comparison}, the estimator in \eqref{eq: Hill} converges in probability to the true value of $\alpha$ as $n$ and $k$ tend to infinity, under the condition that $k / n$ approaches zero:
\begin{align*}
\hat{\alpha}_{k, n}^{(H)}\overset{\mathbb{P}}{\longrightarrow}\alpha,\quad \textrm{as }n\to\infty, k\to\infty,\textrm{ such that }k/n\to 0,
\end{align*}
For practical applications, the ratio $k / n$ should be within the range of [0.01, 0.05], as recommended by \cite{mcneil2015quantitative}. The $(1-\theta)\times 100\%$ Wald CI for the Hill estimator is given by \cite{haeusler2007assessing} as:
\begin{equation*}
\textrm{CI}_n^{\rm (Wald)}(\theta, k) = \left[\left(1 + \frac{z_{\theta / 2}}{\sqrt{k}}\right)^{-1}\hat{\alpha}_{k, n}^{(H)}, \left(1 - \frac{z_{\theta / 2}}{\sqrt{k}}\right)^{-1}\hat{\alpha}_{k, n}^{(H)}\right],
\end{equation*}
where $z_p$ denotes the $p$-th quantile of the standard normal distribution. In this study, we computed the 95\% Wald CIs using $\theta = 0.05$.

\begin{figure}[htbp]
\centering
\subfloat[]{\includegraphics[width=0.8\textwidth, height = 0.3\textheight, keepaspectratio]{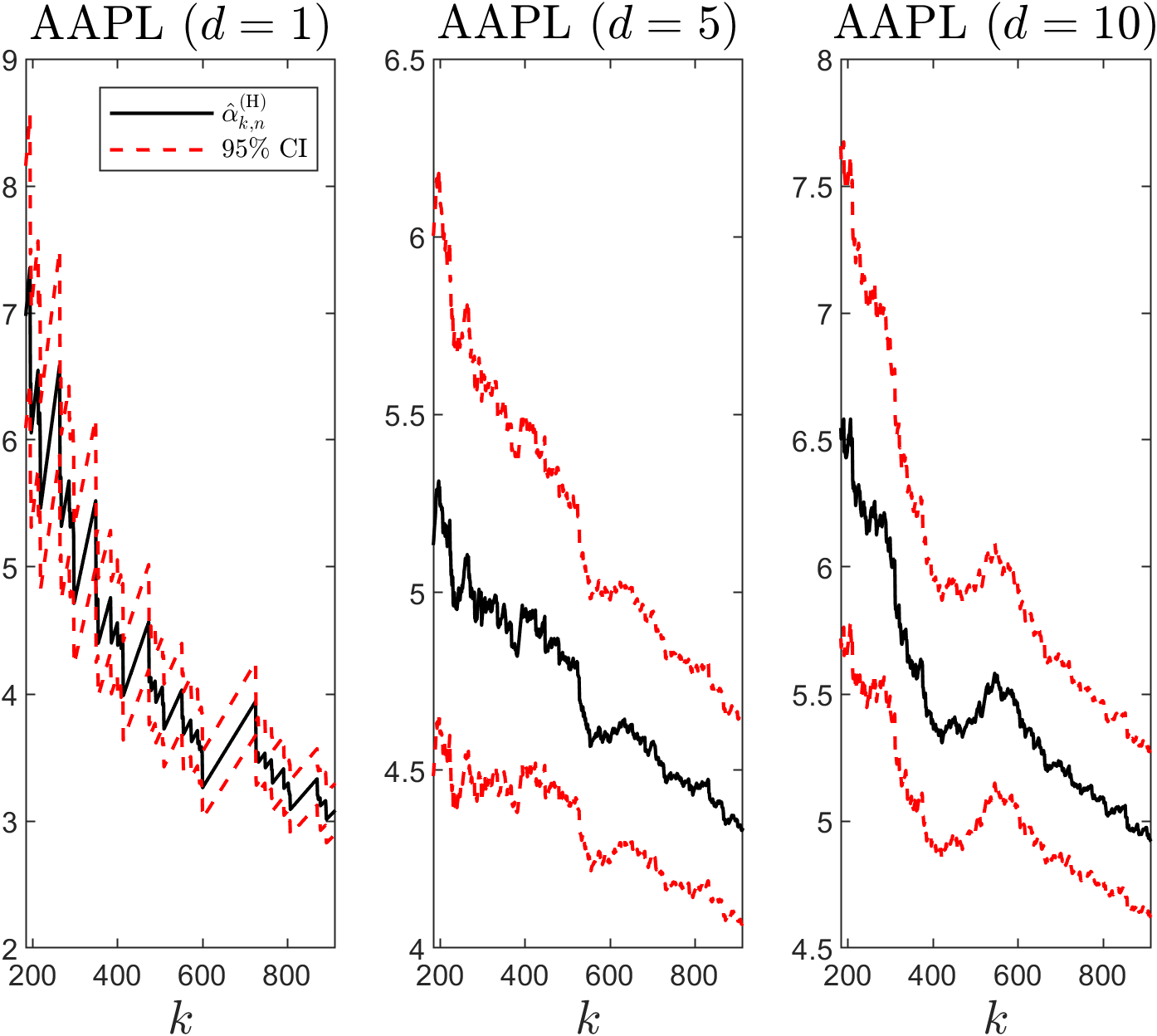}}
\label{figure: tmobbas_hill_plot_aapl}
\subfloat[]{\includegraphics[width=0.8\textwidth, height = 0.3\textheight, keepaspectratio]{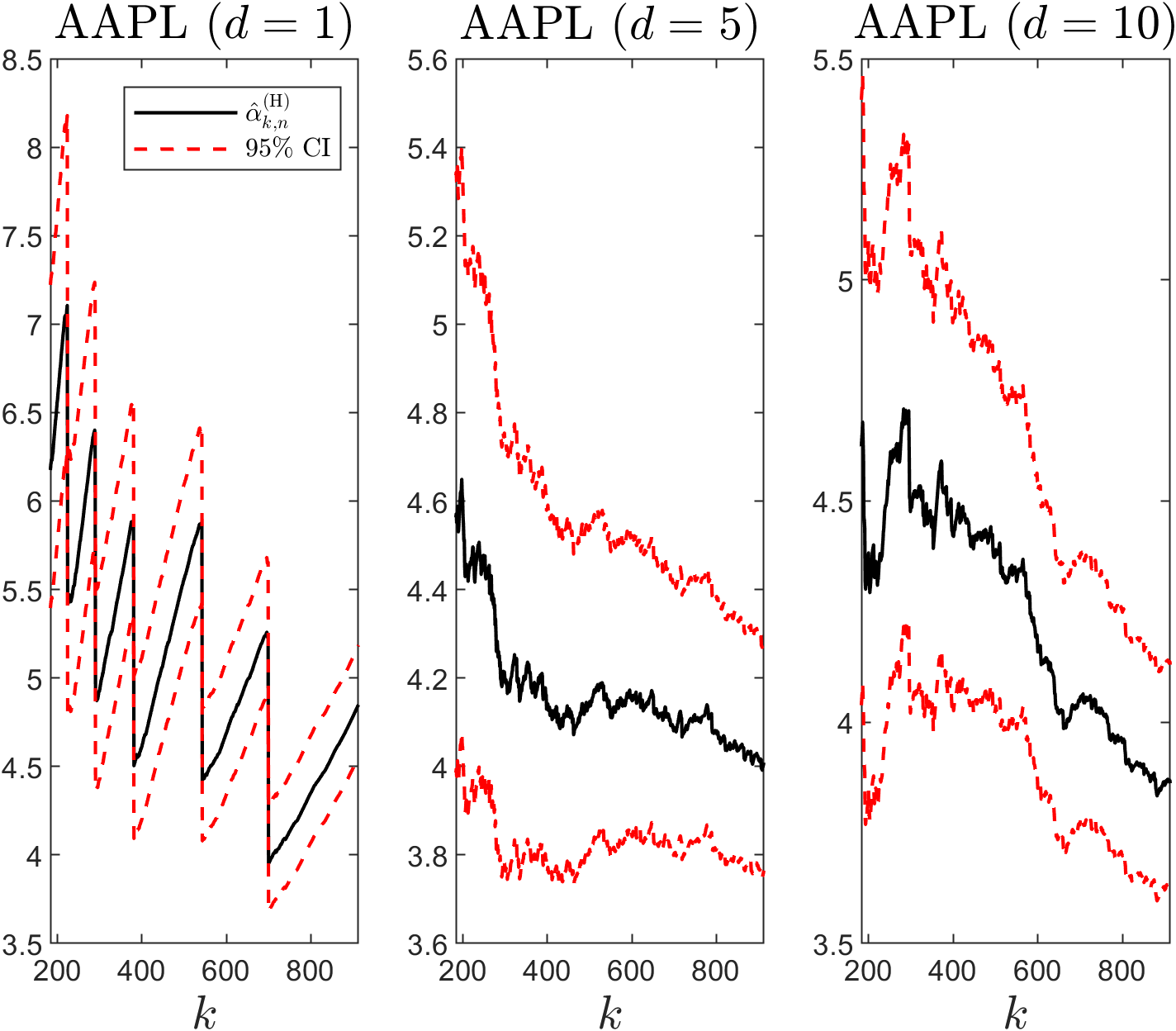}}
\label{figure: gmp_hill_plot_aapl}
\caption{The dependence of estimated tail index $\alpha$ (represented by black solid line) and its corresponding 95\% CI (represented by two red dashed lines) for (a) TMOBBAS and (b) GMP on the number of order statistics $k$ at depths of 1, 5, and 10 for AAPL}
\label{figure: hill_plot_aapl}
\end{figure}

Figure~\ref{figure: hill_plot_aapl} plots the estimated tail indices and the 95\% Wald CI as a function of $k$ of the TMOBBAS and GMP for AAPL and depths of 1, 5, and 10. Given that the dataset of 365,112 returns of TMOBBAS and GMP for AAPL, the highest 18,256 returns define the right tails. The recommendation of \cite{mcneil2015quantitative} would correspond to investigating the Hill estimate over order statistics $k\in[183, 913]$ (i.e, $k/n \in [0.01, 0.05]$). The results for depths of 1, 5, and 10 are representative of all ten depths and we can see that the estimated tail indices are significantly greater than zero and the 95\% CIs extend strictly to the right of zero, indicating that the tails are heavy.

\subsubsection{The ``Rank Minus 1/2'' Method}
The rank minus 1/2 method\footnote{ We thank one of the anonymous reviewers for introducing us to the novel test for the tail index discussed in this section.} is a novel strategy for estimating the tail index of a distribution; it was proposed by \cite{gabaix2011rank}. Unlike conventional methods like the Hill estimator, which  relies  solely on data from the extreme right tail, the rank minus 1/2 method takes into account the entire set of positive returns. This comprehensive approach enhances the method’s robustness and reduces its vulnerability to the effects of outliers. By using a larger  dataset segment, the rank minus 1/2 estimator seeks to deliver a more precise and dependable estimate of the tail index, a critical component in understanding the risk profiles of heavy-tailed distributions in disciplines such as finance, insurance, and extreme value theory.

To apply the rank minus 1/2 method, consider a collection of $n$ IID samples, $Z_1, Z_2,\cdots, Z_n$, from the positive returns. These samples are then organized in descending order: $Z_{(1)}\ge Z_{(2)}\ge\cdots\ge Z_{(n)} > 0$. To estimate the tail index $\zeta$\footnote{The tail index $\zeta$ here is decided by the power law: $\mathbb{P}(Z > s)\sim Cs^{-\zeta} (C, s > 0)$ and $f(s)\sim g(s)$ means $f(s) = g(s)\left(1 + o(1)\right)$.}, we employ the following log--log regression model:
\begin{equation}\label{eq: rank}
\log\left(t - \frac{1}{2}\right) = a - b\log\left(Z_{(t)}\right).
\end{equation}
Here, $t$ denotes the rank of each sample. For instance, $Z_{(1)}$ is the largest sample and therefore has a rank of 1, while $Z_{(n)}$ is the smallest of the largest samples and has a rank of $n$. Subsequently, we use the estimated coefficient $\hat{b}$ from the regression model~\eqref{eq: rank} to determine the tail index $\zeta$. In the case of a heavy-tailed distribution, the estimated $\hat{b}$ is expected to be positive.

Tables~\ref{table: tmobbas_rank_estimator_aapl} and \ref{table: gmp_rank_estimator_aapl} present the estimated coefficient $\hat{b}$, which  is  a proxy for the tail index $\zeta$, of the TMOBBAS and GMP for AAPL across all ten depth levels. It is evident that the $\hat{b}$ are positive, and the associated 95\% CIs for each $\hat{b}$ extends strictly to the right of zero. This indicates that the tail index of TMOBBAS and GMP at each depth  is  greater than zero, confirming that the distributions of TMOBBAS and GMP at all ten depths have heavy tails.\footnote{The similar results for AMZN and GOOG can be found in Appendix~\ref{appendix: table}.}
\begin{table}[htbp]
\tbl{The value of $\hat{b}$ and its 95\% CI of TMOBBAS for AAPL}
{\begin{tabular}{c c c | c c c}
\hline
Depth $d$ & $\hat{b}$ & 95\% CI & Depth $d$ & $\hat{b}$ & 95\% CI\\ \hline
1 & 0.923 & (0.917, 0.928) & 2 & 0.366 & (0.364, 0.369)\\
3 & 0.405 & (0.402, 0.407) & 4 & 0.473 & (0.471, 0.475)\\
5 & 0.496 & (0.494, 0.499) & 6 & 0.521 & (0.519, 0.524)\\
7 & 0.524 & (0.522, 0.526) & 8 & 0.522 & (0.520, 0.524)\\
9 & 0.520 & (0.518, 0.522) & 10 & 0.519 & (0.517, 0.510)\\ \hline
\end{tabular}}
\label{table: tmobbas_rank_estimator_aapl}
\end{table}

\begin{table}[htbp]
\tbl{The value of $\hat{b}$ and its 95\% CI of GMP for AAPL}
{\begin{tabular}{c c c | c c c}
\hline
Depth $d$ & $\hat{b}$ & 95\% CI & Depth $d$ & $\hat{b}$ & 95\% CI\\ \hline
1 & 1.145 & (1.140, 1.150) & 2 & 0.394 & (0.392, 0.397)\\
3 & 0.464 & (0.462, 0.467) & 4 & 0.495 & (0.492, 0.497)\\
5 & 0.507 & (0.505, 0.510) & 6 & 0.525 & (0.523, 0.527)\\
7 & 0.527 & (0.525, 0.529) & 8 & 0.526 & (0.524, 0.528)\\
9 & 0.520 & (0.518, 0.522) & 10 & 0.517 & (0.515, 0.518)\\ \hline
\end{tabular}}
\label{table: gmp_rank_estimator_aapl}
\end{table}

\section{Dynamics of Log-Returns}\label{sec: dynamics}
In the preceding section, we used the GPD to characterize the right tails of the logarithmic returns for both TMOBBAS and GMP. Specifically, we identified the top 5\% of the log-return data as the tail and used the GPD to model it. Our analysis included estimating the parameters $\xi$ and the tail index $\alpha$ along with their corresponding 95\% CIs for each depth. This revealed that the  log-return distributions  for both TMOBBAS and GMP have heavy tails, indicative of significant deviations from normality.

The method employed in the previous sections relied on a historical or static approach, which involved sequentially sampling the returns data from fixed historical periods. This provides a snapshot of the market activity and global events within a finite timeframe. Within this framework, we treated the dataset as a series of historical log returns, assuming they were IID. This simplification is widely adopted by practitioners, including traders and portfolio managers, for its simplicity and accessibility. However, it is imperative to acknowledge the common caveat found in fund prospectuses, that past performance does not guarantee future returns.

Unlike the historical approach, dynamic methods investigate  the informational content in historical data more deeply. These methods operate under the assumption that historical returns arise from a dynamic univariate distribution, where characteristics such as dependency may vary over time. Seeking to uncover the nature of this distribution, dynamic methods generate extensive predictive samples of asset returns,  focusing  on extreme events or tail behavior. The rationale behind using dynamic methods lies in their ability to adapt to changing market conditions and take into account evolving risk factors that  static models may not capture. By continuously updating the model parameters based on new information, dynamic methods offer improved accuracy in predicting future returns and mitigating potential risks. While historical methods remain crucial, especially for modeling new high-frequency spread indices and providing insights into risk management frameworks like Basel I and II, dynamic approaches offer heightened sensitivity to potential significant shifts in market performance. By incorporating the time-varying nature of market dynamics, dynamic methods provide more robust and reliable predictions, making them invaluable tools for investors and risk managers.

In this section, we adopt a time-series approach designed to identify the  best-fitting models for  time-series  data, considering both in-sample and out-of-sample performance. By leveraging Monte Carlo simulations, our goal is to generate a true sample of IID log returns  for subsequent periods. This methodological shift distinguishes our approach from historical returns, which inherently lack the IID property.

\subsection{ARMA(1,1)-GARCH(1,1) with Normal Inverse Gaussian Distribution}
Our dynamic framework incorporates several key components.
We use a versatile ARMA(1,1)-GARCH(1,1) model to fit the log-returns data,  integrating  an NIG distribution to model the innovations within the ARMA-GARCH framework, allowing the taking into account of tails that are heavier than those of a normal distribution. Subsequently, we generate a large sample of asset innovation values from the NIG distribution. By applying inverse ARMA-GARCH to this sample set, we derive a comprehensive set of  returns values. These values are then incorporated into our analysis to model spread indices across various depths of the order book, providing deeper insights into market dynamics and risk management strategies. Additionally, we aim to employ the same dynamic framework for analyzing the GMP data alongside TMOBBAS to enhance the comprehensiveness of our analysis.

 Suppose  a time-series of returns, \( r_t \), is stationary, a useful general model for describing it is the synthesis of the autoregressive moving-average (ARMA) model and the generalized autoregressive conditional heteroscedasticity (GARCH) model. The ARMA \citep{Engle1982} component explicitly models the behavior of the returns, whereas the GARCH \citep{Bollerslev1986} component explicitly models its variance.
Both models contain theoretically infinite parameters;  the finite number of parameters employed denotes the varieties of the models. The ARMA(\( p, q \)) model \citep{tsay2010analysis} is
\begin{equation}\label{ARMA_GARCH}
r_t = \phi_0 + \sum_{i=1}^{p} \phi_i r_{t-i} + a_t + \sum_{j=1}^{q} \theta_j a_{t-j}, \nonumber
\end{equation}
where each shock, \( a_t \), is a zero-mean random variable. The first two terms in Equation~\eqref{ARMA_GARCH} describe the autoregressive dependence of \( r_t \) on previous returns; the second two terms add the influence of a weighted (moving) average of shocks, \( a_t \). The GARCH(\( m, s \)) model relates \( a_t \) to, and provides a model for, the variance \( \sigma_t^2 \) of the series:
\begin{equation}\label{ARMA_GARCH_full}
a_t = \sigma_t \epsilon_t, \quad \sigma_t^2 = \alpha_0 + \sum_{i=1}^{m} \alpha_i a_{t-i}^2 + \sum_{j=1}^{s} \beta_j \sigma_{t-j}^2.
\end{equation}
 The  so-called innovations, \( \epsilon_t \), are zero-mean, unit-variance, independent, identically distributed random variables. The GARCH model is autoregressive in both \( \sigma_t^2 \) and \( a_t^2 \).

Identifying the daily variance as the volatility of the time series, Equation~\eqref{ARMA_GARCH_full} captures the property of conditional heteroscedasticity:  the  property that the volatility is not constant relative to that of prior days. With six parameters, the ARMA(1,1)–GARCH(1,1) model
\begin{align*}
r_t &= \phi_0 + \phi_1 r_{t-1} + a_t + \theta_1 a_{t-1}, \\
a_t &= \sigma_t \epsilon_t, \\
\sigma_t^2 &= \alpha_0 + \alpha_1 a_{t-1}^2 + \beta_1 \sigma_{t-1}^2,
\end{align*}
provides enough generality to model many time series of returns. However, providing a fit to a particular time series requires the specification of the distribution governing the random variables of the innovations. In the dynamic optimization method, we assume that the innovations, \( \epsilon_t \), are governed by the NIG distribution given by 
\begin{equation}\label{NIG_pdf}
f(x|\alpha, \beta, \mu, \delta) = \frac{\delta}{2\alpha K_1(\delta)} \exp \left( -\frac{\delta}{\alpha} \left( \sqrt{\beta^2 + (x - \mu)^2} - \sqrt{\beta^2 + \mu^2} \right) \right), \nonumber
\end{equation}
where \( K_1(\cdot) \) is the modified Bessel function of the second kind, and \( \alpha \), \( \beta \), \( \mu \), and \( \delta \) are shape and scale parameters.
The NIG distribution, with its ability to handle both heavy tails and skewness, provides a flexible and accurate fit for the financial log-return data of TMOBBAS and GMP. Unlike other heavy-tailed distributions, such as Student's $t$ (Std) or the generalized error distribution (GED), the NIG distribution is infinitely divisible, which is essential for option pricing and modeling returns over infinitesimally small time intervals.\footnote{See \cite{barndorff1997, prause1999processes, blasques2014_2}}

While the Std is widely used for modeling heavy tails, it has certain limitations for option pricing, as it can lead to divergent integrals when used in a Black--Scholes--Merton framework \cite{cassidy2010pricing}. Similarly, the GED is not infinitely divisible, restricting its utility in complex financial models and time series analyses \cite{cont2004financial}. NIG’s flexibility in capturing the skewness and leptokurtosis of returns makes it a superior choice for our analysis.

Moreover, to ensure robustness, we performed comparisons using alternative heavy-tailed distributions such as GED and Std. The models had very close values resulting from the AIC and BIC tests, with the NIG distribution offering the best balance for both heavy tails and asymmetry. Table~\ref{AIC_BIC} compares the AIC and BIC for these models, supporting the use of the NIG distribution and confirming its appropriateness for our dynamic model, offering a balance between modeling accuracy and computational feasibility.

\begin{table}[htbp]
\tbl{Comparison of AIC and BIC for different distributions}
{\begin{tabular}{c c c}
\hline
Model & AIC & BIC \\
\hline
GED & -3.7323 & -3.7315 \\
Std & -3.7848 & -3.7841 \\
NIG & -3.7749 & -3.7741 \\
\hline
\end{tabular}}
\label{AIC_BIC}
\end{table}

\subsection{The TMOBBAS Hill Index Estimations}
The dynamic model seeks to provide a statistically accurate larger sampling of returns for computing the tail index. This is achieved by fitting an ARMA(1,1)–GARCH(1,1) with an NIG-distribution model to the log-return time series of the TMOBBAS index of depths \( d=1,2,...,10 \), generating the model parameters, computing the shock series \( a_{t} \) and the variances \( \sigma_{t}^2 \) predicted from the fitted GARCH(1,1) model, generating \( 10,000 \) of the innovation series \( \epsilon_{t} \), and performing the inverse transformations \( \epsilon_{t} \) in the ARMA(1,1)–GARCH(1,1) model to generate a dynamic ensemble of predicted returns for the TMOBBAS index for each depth. The ensemble of returns \( \{r_{t}^{(s)}\}_{s=1}^{10} \) is the output from the dynamic model, which is then incorporated into our analysis to model spread indices for various depths of the order book, providing deeper insights into market dynamics and risk management strategies.

\begin{figure}[htbp]
    \centering
    \includegraphics[width=0.5\textwidth, height=0.5\textheight, keepaspectratio]{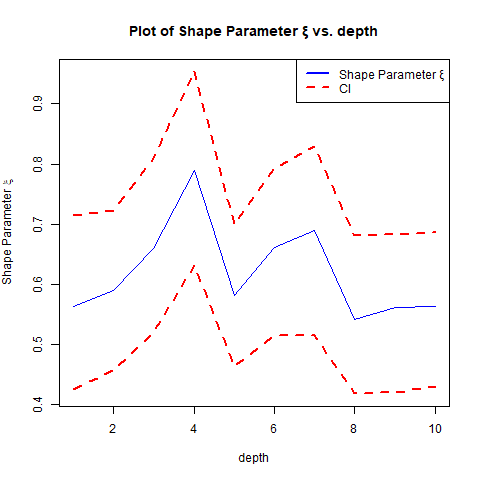}
    \caption{The dependence of estimated shape parameter \( \xi \) (represented by blue solid line) and its corresponding 95\% CI (represented by two red dashed lines) on the depth for AAPL}
    \label{Shape_AAPL_dyn}
\end{figure}

Figure~\ref{Shape_AAPL_dyn} illustrates the dependence of the estimated \( \xi \) and its corresponding 95\% CI on the depth.
From Figure~\ref{Shape_AAPL_dyn}, we observe that the estimated values of \( \xi \) are significantly greater than zero, and their corresponding 95\% CIs for each depth extend strictly to the right of zero. Thus, we should consider a power-law decay (i.e., heavy-tailed) distribution to fit the entire density of the  log returns  of the TMOBBAS and confirm our findings from the historical method.

\begin{figure}[htbp]
\centering
\subfloat[]{\includegraphics[width=0.33\textwidth, height=0.2\textheight, keepaspectratio]{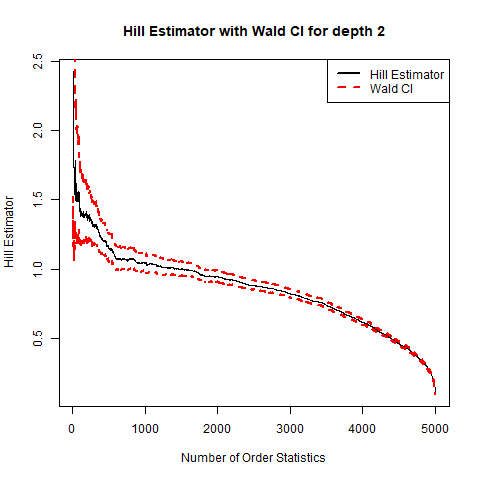}}
\label{tmobbas_hill_aapl_1}
\subfloat[]{\includegraphics[width=0.33\textwidth, height=0.2\textheight, keepaspectratio]{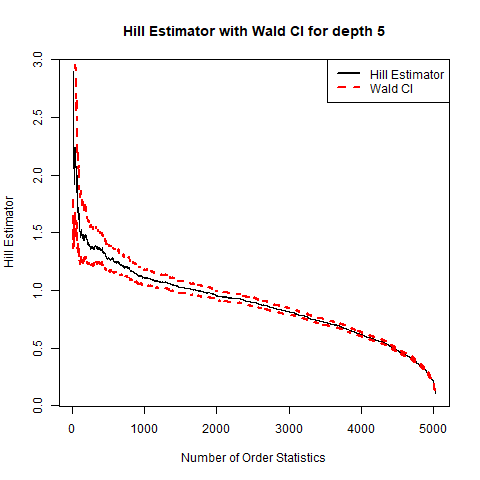}}
\label{tmobbas_hill_aapl_5}
\subfloat[]{\includegraphics[width=0.33\textwidth, height=0.2\textheight, keepaspectratio]{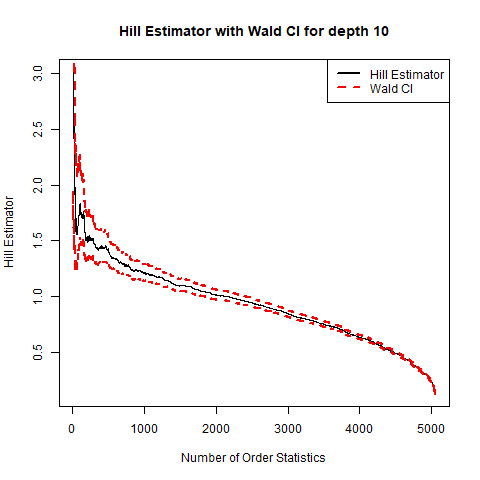}}
\label{tmobbas_hill_aapl_10}
\caption{The dependence of estimated tail index $\alpha$ (represented by black solid lines) and its corresponding 95\% CI (represented by two red dashed lines) on the number of order statistics at depths (a) 2, (b) 5, and (c) 10 for AAPL}
\label{Hill_AAPL_dyno}
\end{figure}

For estimating the Hill index for each depth, consistent with our historical method, we set \( k/n \approx 0.05 \) and \( \theta = 0.05 \).  Figure~\ref{Hill_AAPL_dyno} shows  that the estimated tail indices are significantly greater than zero, with the 95\% CIs extending strictly to the right of zero. This indicates that the tails of the distribution should be considered heavy-tailed, corroborating our findings from the historical method.

Furthermore, in Figure~\ref{Hill_AAPL_dyno}, we observe a decreasing trend in the evolution of the estimated tail index \( \alpha \) and its associated CIs as a function of the number of order statistics  as  we include more extreme values (higher order statistics), the variability in the estimates of the tail index decreases, leading to a more conservative estimate. In other words, a larger number of order statistics results in a more robust estimation, yielding a stable and lower estimate of the tail index.

\begin{figure}[htbp]
    \centering
    
\includegraphics[width=0.5\textwidth, height=0.5\textheight, keepaspectratio]{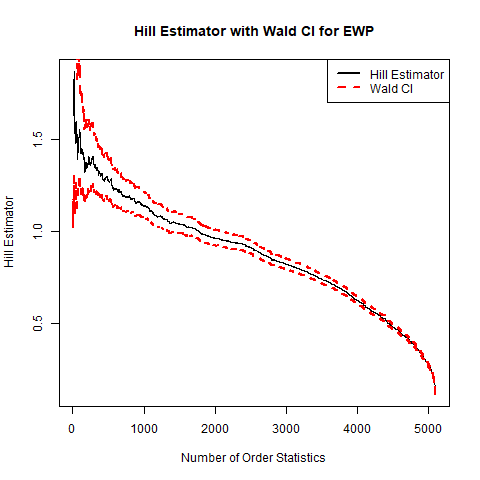}

    \caption{The dependence of estimated tail index $\alpha$ (represented by black solid line) and its corresponding 95\% CI (represented by two red dashed lines) on the number of order statistics for EWP for AAPL}
    \label{Hill_AAPL_dyno_EWP}
\end{figure}

Meanwhile, we examined the tail behavior of an equally weighted portfolio (EWP) containing the  log returns  from all ten depths. Adopting an EWP approach facilitated the aggregation of  log returns  across multiple depths, offering a holistic view of the  market behavior  while minimizing the influence of factors that might be specific to one particular depth. As illustrated in Figure~\ref{Hill_AAPL_dyno_EWP}, our analysis reaffirms the heavy-tailed nature of the distribution, indicating significant deviations from normality. This observation underscores the robustness of our method in capturing heavy-tailed behavior across the TMOBBAS index, further enhancing our understanding of the dynamics of tail risk within the market.

These findings provide valuable insights into the tail behavior of the log-returns for the TMOBBAS index, particularly when comparing results obtained from historical methods with dynamic approaches. Our examination of the Hill index for each depth, reaffirms,  consistently with our historical method, the heavy-tailed nature of the distribution.

\subsection{The GMP Hill Index Estimations}
In this subsection, we extend our analysis to the GMP, applying the same method previously employed  to analyze  the TMOBBAS. By leveraging the dynamic framework described earlier, we aim to provide insights into the tail behavior and dynamics of the  log returns  of the GMP for various depths of the order book.

The dynamic model used an ARMA(1,1)–GARCH(1,1) model, an NIG distribution to fit the time series of the GMP index for depths $d=1,2,...,10$. This generated a statistically accurate ensemble of predicted log-returns for the GMP at depths of 2, 5, and 10, facilitating the modeling of spread indices and offering deeper insights into market dynamics and risk management strategies.

\begin{figure}[htbp]
\centering
\subfloat[]{\includegraphics[width=0.33\textwidth, height=0.2\textheight, keepaspectratio]{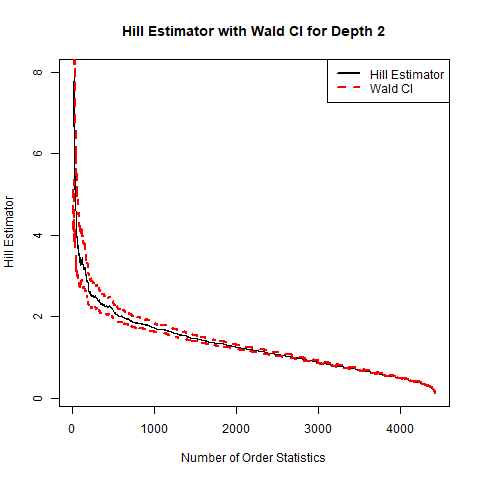}}
\subfloat[]{\includegraphics[width=0.33\textwidth, height=0.2\textheight, keepaspectratio]{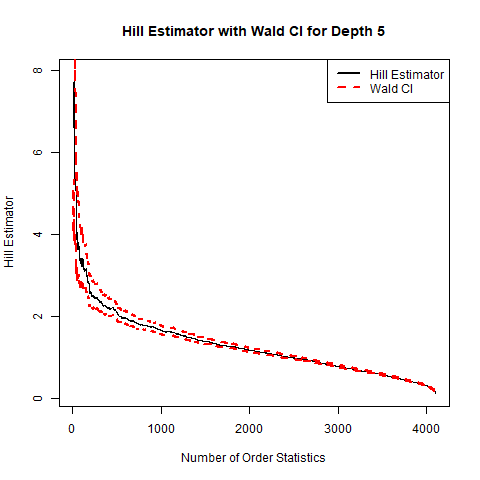}}
\subfloat[]{\includegraphics[width=0.33\textwidth, height=0.2\textheight, keepaspectratio]{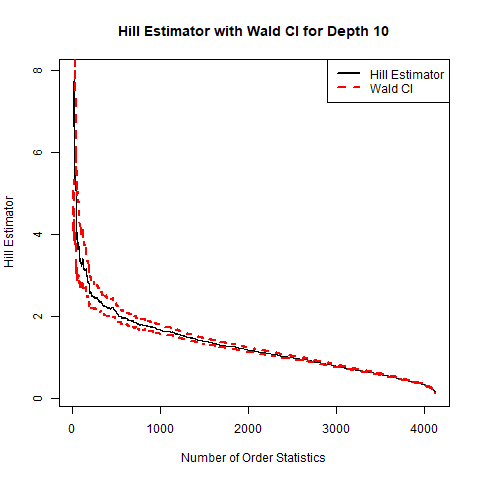}}
\caption{The  dependence of estimated tail index $\alpha$ (represented by black solid lines) for GMP and its corresponding 95\% CI (represented by two red dashed lines) on the number of order statistics at depths of (a) 2, (b) 5, and (c) 10 for AAPL}
\label{Hill_AAPL_dyno_GMP}
\end{figure}

We estimated the Hill index for each depth of the GMP,  which is  consistent with our method applied to the TMOBBAS. Figure~\ref{Hill_AAPL_dyno_GMP} presents the dependence of the estimated tail index $ \alpha $ and its corresponding 95\% CIs on the number of order statistics at each depth for GMP.

\begin{figure}[htbp]
    \centering
    
    \includegraphics[width=0.5\textwidth, height=0.5\textheight, keepaspectratio]{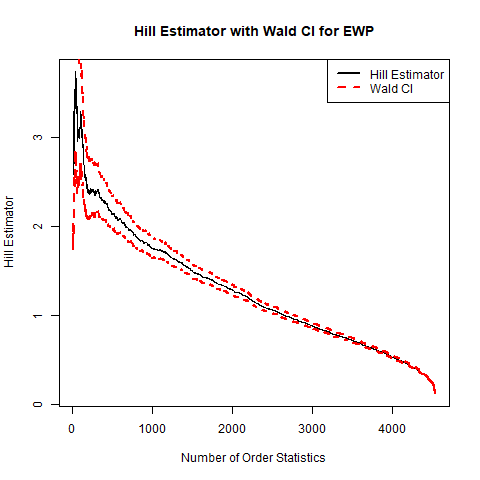}

    \caption{The dependence of the estimated tail index $\alpha$ (represented by black solid line) of GMP and its corresponding 95\% CI (represented by two red dashed lines) on the number of order statistics for EWP for AAPL}
    \label{GMP_Hill_AAPL_dyno_EWP}
\end{figure}

Additionally, we examined the tail behavior of an EWP consisting of the  log returns  from all ten depths of the GMP. This approach allowed us to gain a comprehensive understanding of  this market's behavior while mitigating the influence of factors specific to some particular depth. Figure~\ref{GMP_Hill_AAPL_dyno_EWP} presents the dependence of the estimated tail index $ \alpha $ and its corresponding $95\%$ CIs on the number of order statistics for the GMP EWP.

By conducting a thorough analysis of the GMP using the new method we already employed, we aim to enhance our understanding of market dynamics and tail risk within the context of the order book. This analysis complements our investigation of the TMOBBAS and GMP, providing valuable insights into the behavior of both indices and informing effective risk management strategies in financial markets.

Furthermore, by comparing the analyses of the tail behavior of GMP and TMOBBAS, we observe that while there are similar trends in the shape of their tail indices, there are also notable differences, suggesting distinct risk profiles for each index. It is worth noting that the Hill index estimated at depths of 2, 5,  and  10 of GMP shows very similar values, with minimal variation. This contrasts with the behavior observed for TMOBBAS, where the Hill index varies significantly between depths, ranging from 0 to 3. The consistent behavior of the Hill index as a function of depth for GMP suggests a more stable tail behavior  than  TMOBBAS.

This may be due to the  market's nature and the order  book's  characteristics. The GMP may exhibit a more uniform distribution of  log returns  across depths, leading to consistent tail behavior. On the other hand, the TMOBBAS may experience more variability in trading activity and liquidity across depths, resulting in a fluctuating tail behavior.

This analysis complements our investigation of TMOBBAS and GMP, providing valuable insights into the behavior of both indices and informing effective risk management strategies in financial markets.

\section{Option Pricing and Implied Volatility} \label{sec: option pricing}
The primary objective of our paper is to introduce and analyze the new high-frequency spread indices, TMOBBAS and GMP, which capture the dynamics of variations in the spread across different depths of the order book. Our findings from previous sections show that the  log returns  and spreads at different depths exhibit heavy-tailed characteristics. In addition, the global spreads and mid-prices at various depths can significantly vary, impacting liquidity.
This observation underscores the potential challenges  market participants face, particularly high-frequency traders, who are sensitive to fluctuations in liquidity. In response to this concern,  an opportunity arises  to explore innovative risk management strategies. One such approach involves the  adopting  insurance instruments akin to portfolio insurance, aimed at hedging against the risk of low liquidity.

Thus, there is a compelling rationale for  developing  an option pricing model tailored specifically for spreads at different levels. By constructing an equally weighted index out of these spreads and offering derivative pricing on this index, we can  give market participants  a robust mechanism for hedging against the adverse effects of low liquidity.
This will  address the traders' practical needs and  contribute to the ongoing discourse on risk management in financial markets.

This section focuses on pricing options on the TMOBBAS and GMP indices and deriving their implied volatilities. Measures  of implied volatility play a crucial role in this context by encapsulating the current perspective of the market on the future risk, as implied by the observed transaction prices of option contracts. Given the
dynamic nature of the high-frequency market and the presence of temporal dependence, as manifested in volatility clustering, traditional measures of historical volatility may be inadequate for capturing the prevailing risk environment. Implied volatility offers a forward-looking outlook, integrating market expectations and sentiments, thereby furnishing valuable insights for traders and investors.

Our approach  employs a double-subordinated process to model  the TMOBBAS and GMP indices. This method has been chosen to enhance the accuracy and reliability of the pricing, particularly in capturing the dynamic nature of the high-frequency market and addressing the challenges associated with the heavy-tailed distributions of asset returns. This choice has been informed by empirical evidence suggesting that the returns of speculative assets, such as the TMOBBAS and GMP indices, are heavy-tailed distributions and asymmetric, challenging the assumption of normality underlying traditional option pricing models like the Black--Scholes--Merton (BSM) formula. A double-subordinated process enables the variance of the normal distribution to vary over time, thereby effectively capturing heavy-tailed phenomena. Specifically, one of the subordinated processes models the intrinsic time, enhancing our ability to capture the intricate dynamics of both the TMOBBAS and GMP indices and provide more accurate pricing for options based on them.\footnote{For a further discussion of the use of subordinators in financial modeling, see \cite{Sato_1999}, \cite{Schoutens_2003}, and \cite{shirvani2021multiple}.}

\subsection{Double Subordination Model for TMOBBAS and GMP}
In this subsection, we define a double-subordinated process for modeling the TMOBBAS and GMP indices and pricing options based on their dynamics. By outlining the model parameters and processes, we lay the groundwork for our analysis, aiming to capture the nuanced behavior of these indices and provide accurate pricing for options.\footnote{For more details on L\'evy subordinator processes, refer to \cite{duffie2010dynamic}.}

The double-subordination framework involves a L\'evy subordinator process, denoted by $\mathbb{X}=(X(t), t \ge 0)$, defined on a stochastic basis $(\Omega,\mathcal{F},\ \mathbb{F}=\left({\mathcal{F}}_t,t\ge 0\right), \mathbb{P})$. A process $\mathbb{X}$ is considered a L\'evy process if $X(0) = 0$ almost surely under $\mathbb{P}$,  and it  has independent increments, has stationary increments, and is continuous in probability. \footnote{For more details on L\'evy subordinator processes, refer to \cite{Sato_1999} and \cite{Schoutens_2003}.} A L\'evy process $\mathbb{T} = (T(t),\ t \ge 0,\ T(0) = 0)$ with non-decreasing trajectories (i.e., non-decreasing sample paths) is called a \textit{L\'evy subordinator}.
Since $T\left(0\right)=0$, the trajectories of $\mathbb{T}$ take only non-negative values. In the BSM option pricing model, the price dynamics of the underlying asset is given by
\begin{equation}
\label{Black_Eq1}
S_t^{\textrm{(BSM)}} = e^{X_t^{\textrm{(BSM)}}},\quad t\in [0,\tau],\nonumber
\end{equation}
where the log-process is
\begin{equation}
\label{Black_Eq2}
X_t^{\textrm{(BSM)}} = X_0 + \mu_1 t + \sigma_1 B_{t},\quad \mu_1 \in \mathbb{R},\ \sigma_1 > 0,\ X_0 = \ln(S_0),\ S_0 > 0,\nonumber
\end{equation}
and $ \mathbb{B} = (B_t,\ t \geq 0)$ is a standard Brownian motion. To accommodate the non-normality of the asset returns, \cite{Mandelbrot_1967} and \cite{Clark_1973} proposed the use of a subordinated Brownian motion, where the price process $S_t^{\textrm{(ss)}}$ and the log-price process are defined by

\begin{equation}
S_t^{\textrm{(ss)}} = e^{X_t^{\textrm{(ss)}}},\quad t\in [0,\tau],\nonumber
\end{equation}
\begin{equation}
X_t^{\textrm{(ss)}} = X_0+\mu_2 t+\sigma_2 B_{T(t)},\quad \mu_2 \in \mathbb{R},\ \sigma_2 > 0,
\end{equation}
where $\mathbb{T} = (T(t),\ t \geq 0,\ T(0) = 0 )$ is a L\'evy subordinator.

Various studies have demonstrated that single-subordinated log-price models commonly fail to capture the heavy-tailedness observed in financial returns.\footnote{ 
For example, see \cite{Lundtofte_2013} and \cite{shirvani2021equity}.} \cite{shirvani2021multiple} defined and investigated the properties of various  multiple subordinated  log-return processes designed to model leptokurtic asset returns. They showed that multiple-subordinated log-return processes can have heavier tails than single-subordinated models and that they are capable of capturing skewness and kurtosis. Therefore, a double subordination framework may be more appropriate for modeling the rather extreme behavior of both the TMOBBAS and GMP indices.

To apply double subordination to modeling the TMOBBAS and GMP index price processes, let $S_t$ denote the price process with the dynamics
\begin{gather}
	S_t = e^{X_t}, \quad t \in [0,\tau], \label{DSS} \nonumber \\
	X_t = X_0 + \mu_3 t + \gamma U(t) + \rho T(U(t)) + \sigma_3 B_{T(U(t))}, \quad t \ge 0, \nonumber \\
	\mu_3 \in \mathbb{R},\ \sigma_3 > 0,\ X_0 = \ln(S_0),\ S_0 > 0. \label{DSX}
\end{gather}
where the components of the triple ${B_s, T(s), U(s)}, s \ge 0$
are independent processes generating the stochastic basis
$(\Omega,\mathcal{F},\ \mathbb{F}=\left({\mathcal{F}}_t,t\ge 0\right), \mathbb{P})$, which represents the real world.

Consider the case where the subordinators $T(t)$ and $U(t)$ are inverse Gaussian (IG) L\'{e}vy processes; in this case, \cite{shirvani2021multiple} referred to $T(U(t))$ as the \textit{double inverse Gaussian subordinator} and to $X_t$ as a \textit{normal double inverse Gaussian (NDIG) log-price process}, with the characteristic function (CF) of $X_1$ given by

\begin{multline}
\label{chf_double_IG_Log_price}
    \varphi_{X_1}(v) =\\
    \exp \left\{
        i v \mu_3 + \frac{\lambda_U}{\mu_U}
        \left[ 1-\sqrt{
            1-\frac{2\mu_U^2}{\lambda_U}
            \left( \frac{\lambda_T}{\mu_T}
            \left( 1-\sqrt{1-\frac{\mu_T^2}{\lambda_T}(2iv\rho -\sigma_3^2 v^2) }  \right)+iv\lambda
            \right)
            }
        \ \right]
    \right\},
\end{multline}
with $v \in \mathbb{R}$.

To price European contingent claims on both the TMOBBAS and GMP indices, we follow \cite{shirvani2023bitcoin} and assume that the log-price process $X_t$ follows an NDIG model. An equivalent martingale measure $\mathbb{Q}$ is derived from the risk-neutral measure $\mathbb{P}$, ensuring that the discounted price process $e^{-rt} S_t$ is a martingale \citep[][Chapter 6]{duffie2010dynamic}. Using the martingale-corrected moment matching (MCMM) approach, the parameters of the process are estimated, and an appropriate drift term is added to ensure the martingale property.\footnote{\cite{yao2011note} constructed a martingale measure using the MCMM approach for the geometric L\'{e}vy process model and showed that this measure is an equivalent martingale measure if there is a continuous Gaussian part in the L\'{e}vy process. If $X_t$ is a pure-jump L\'{e}vy process, they pointed out that this measure cannot be equivalent to a physical probability. However, pricing European options under this measure is still arbitrage-free.}

The price of a European call option $\mathcal{C}$ with underlying asset $\mathcal{S}$ is 
\begin{equation}
\label{call_price}
    C (S_0,r,K,\tau )=e^{-r\tau} \ \mathbb{E}_{\mathbb{Q}} \left[ \mathrm{max}  (S^{(\mathbb{Q} )}_\tau - K,0 ) \right],\nonumber
\end{equation}
where $\tau$ denotes the maturity, $K$ is the strike price, and $S_t^{ (\mathbb{Q} )}$ is the price dynamics of $\mathcal{S}$ under $\mathbb{Q}$. \cite{shirvani2023bitcoin}  derived  the price dynamics of $\mathcal{S}$ under $\mathbb{Q}$ and the CF of the log-price $\ln S^{ (\mathbb{Q} )}_t$, respectively, given by

\begin{align}
\label{Eq3}
    S^{(\mathbb{Q})}_t &= \frac{b_t S_t}{M_{X_t} (1)} = S_0 e^{\left( r - K_{X_1}(1) \right) t + X_t},\quad t \in [0,\tau], \\
\label{chf_Log_price}
    \varphi_{\ln S^{(\mathbb{Q})}_t} (v) &= S^{iv}_0 \exp \left\{ \left[ iv (r-K_{X_1}(1))+\psi_{X_1} (v) \right] t \right\},\quad t \in [0,\tau].
\end{align}

The efficient computation of option prices relies on \cite{Carr_1998}'s option pricing formula, using the fast Fourier transform (FFT) and necessitating access to the CF of the log-price, represented by

\begin{equation}
\label{Carr_call}
C(S_0, r, k, \tau) = \frac{e^{-r\tau-ak}}{\pi} \int_0^{\infty} e^{-ivk} \frac{{\varphi}_{\ln S^{(\mathbb{Q})}_\tau}(v - i(a+1))}{a^2 + a - v^2 + i(2a + 2)v} dv,
\end{equation}
where $a>0$ and $E(S^{(\mathbb{Q})}_t)^{a}<\infty$. \footnote{Limits for $a$ ensuring stable option prices are determined through numerical experiments, see \cite{shirvani2023bitcoin}.}

For pricing options on both the TMOBBAS and GMP indices and deriving their implied volatility, we employ a double-subordinated process to model these indices. The method of moments and the empirical CF are used to estimate the parameters of the model, following \cite{Paulson_1975} and \cite{Yu_2003}.\footnote{Exploiting the fact that the probability density function (PDF) is the Fourier transform of the CF, the objective is to minimize the squared differences of the first four moments and the empirical and theoretical CFs. This involves estimating the model parameters through optimization subject to constraints based on the statistical properties of the data. See \cite{shirvani2023bitcoin}.}

\subsection{Option Pricing and Implied Volatility of TMOBBAS} 
To explore the realm of option pricing and implied volatility for the TMOBBAS index, we employ the NDIG L\'{e}vy model. This model facilitates the pricing of plain vanilla European options on the TMOBBAS index, providing valuable insights into market dynamics and risk management strategies. Let's explore the details.

We apply the NDIG L\'{e}vy model to the pricing of plain vanilla European options on the TMOBBAS index. Let $\mathcal{C}$ be a European call option, where the underlying risky asset $\mathcal{S}$ follows the log-price process given in Equation~\eqref{DSX}. 
The dynamics of $\mathcal{S}$ on $\mathbb{Q}$ are given by Equation~\eqref{Eq3}, and the CF of the  log price  is given by Equation~\eqref{chf_Log_price}. We evaluate the integral Equation~\eqref{Carr_call} using the FFT for a range of strike levels and maturity horizons. 

\begin{figure}[htbp]
\centering
\subfloat[]{\includegraphics[width=0.49\textwidth]{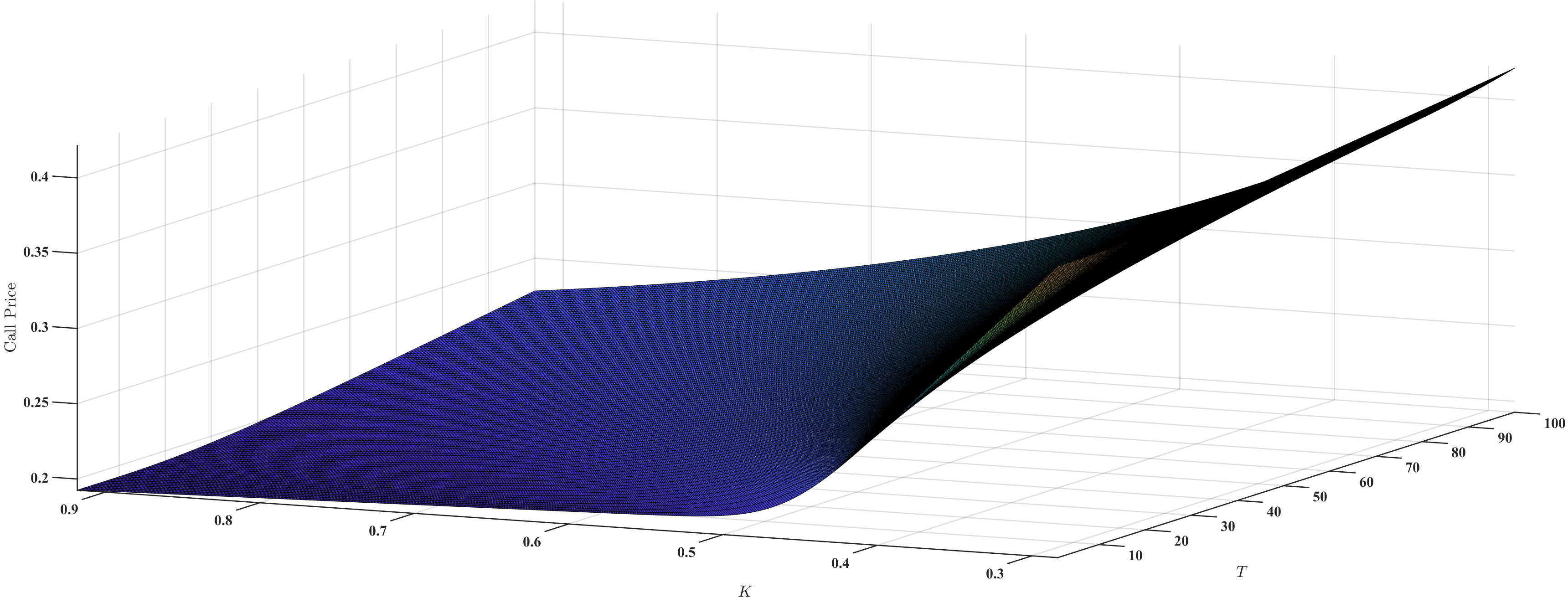}}
\label{Call_price_NDIG_5}
\subfloat[]{\includegraphics[width=0.49\textwidth]{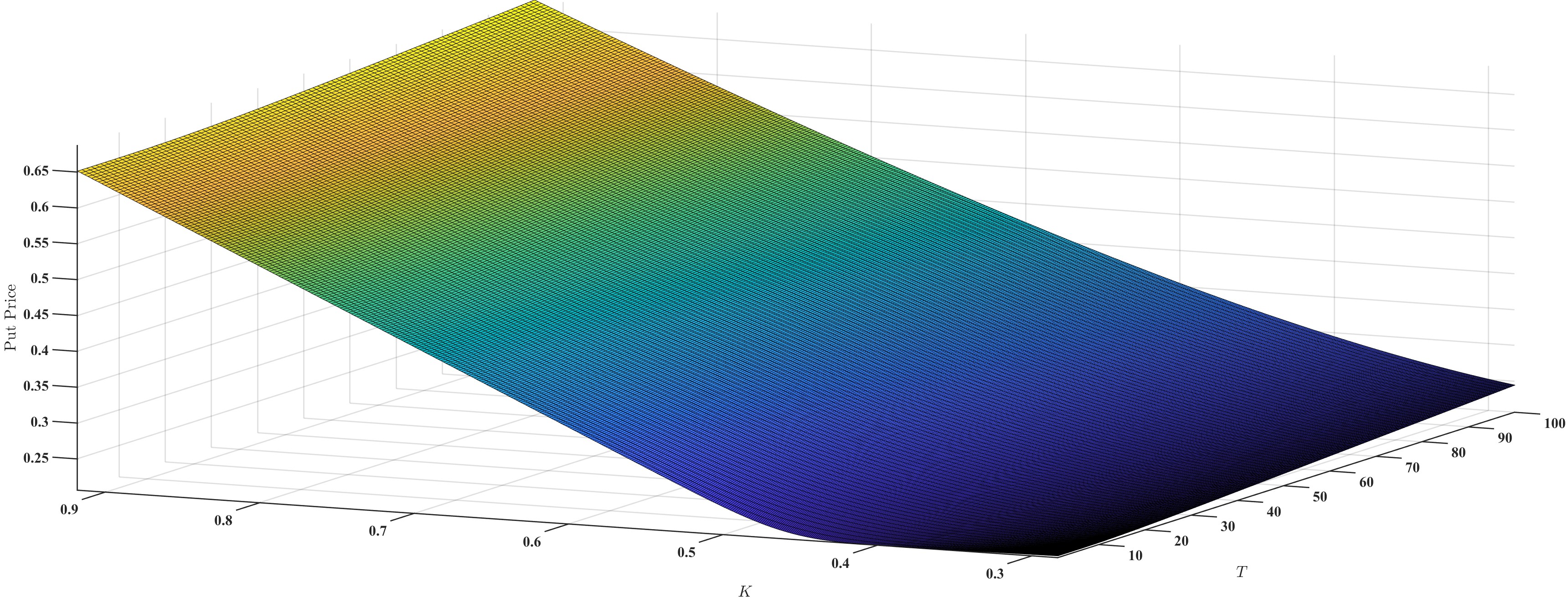}}
\label{Put_price_NDIG_5}
\subfloat[]{\includegraphics[width=0.49\textwidth]{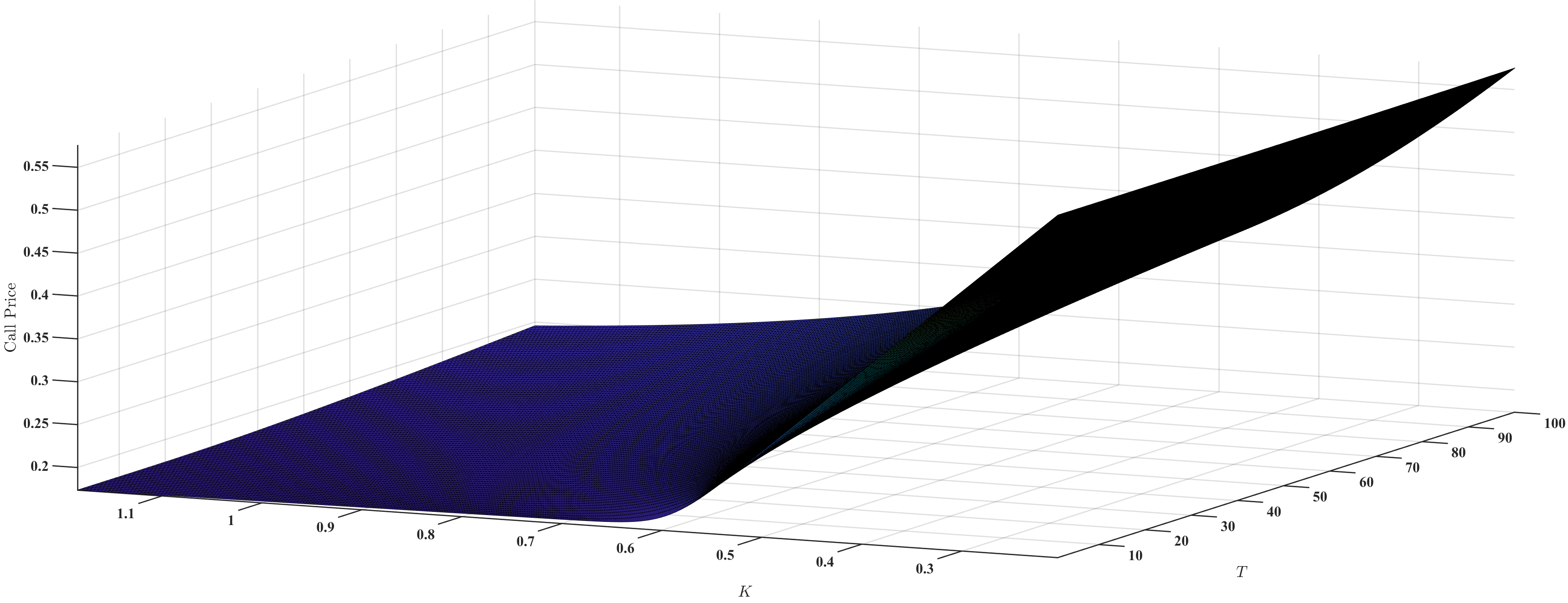}}
\label{Call_price_NDIG_10}
\subfloat[]{\includegraphics[width=0.49\textwidth]{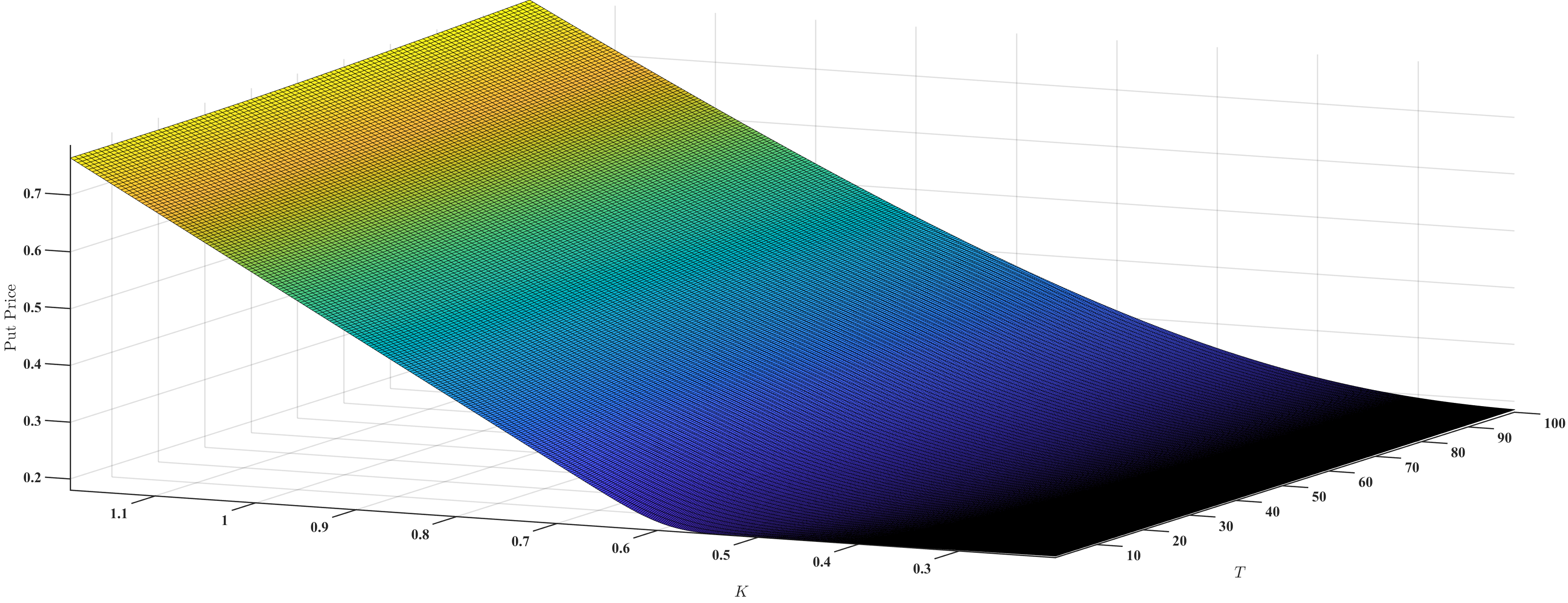}}
\label{Put_price_NDIG_10}
\caption{NDIG-based price surfaces for TMOBBAS index (a) at depth of 5 for call , (b) at depth of 5 for put, (c) at depth of 10 for call,  and (d) at depth of 10 for put as functions of the time to maturity and strike price.
}
\label{option_surfaces}
\end{figure}

Figure~\ref{option_surfaces} depicts the resulting prices for call and put options plotted against the time to maturity $\tau$ and the strike price $K$ at depths of 5 and 10. Since the call and put surfaces at different depths of the TMOBBAS index are similar, we only plot the TMOBBAS index at depths of 5 and 10. It's worth noting that for close-to-the-money values ($K \approx S$), where the strike price is close to the current asset price, the call and put option prices increase with $\tau$, reflecting increased future uncertainty. This behavior is consistent with market expectations, as higher uncertainty about the future value of the underlying asset leads to higher option prices. This phenomenon is often observed in financial markets, where options with longer maturities tend to command higher premiums due to the additional time value and increased uncertainty associated with longer time horizons.

\begin{figure}[htbp]
\centering
\subfloat[]{\includegraphics[width=0.49\textwidth]{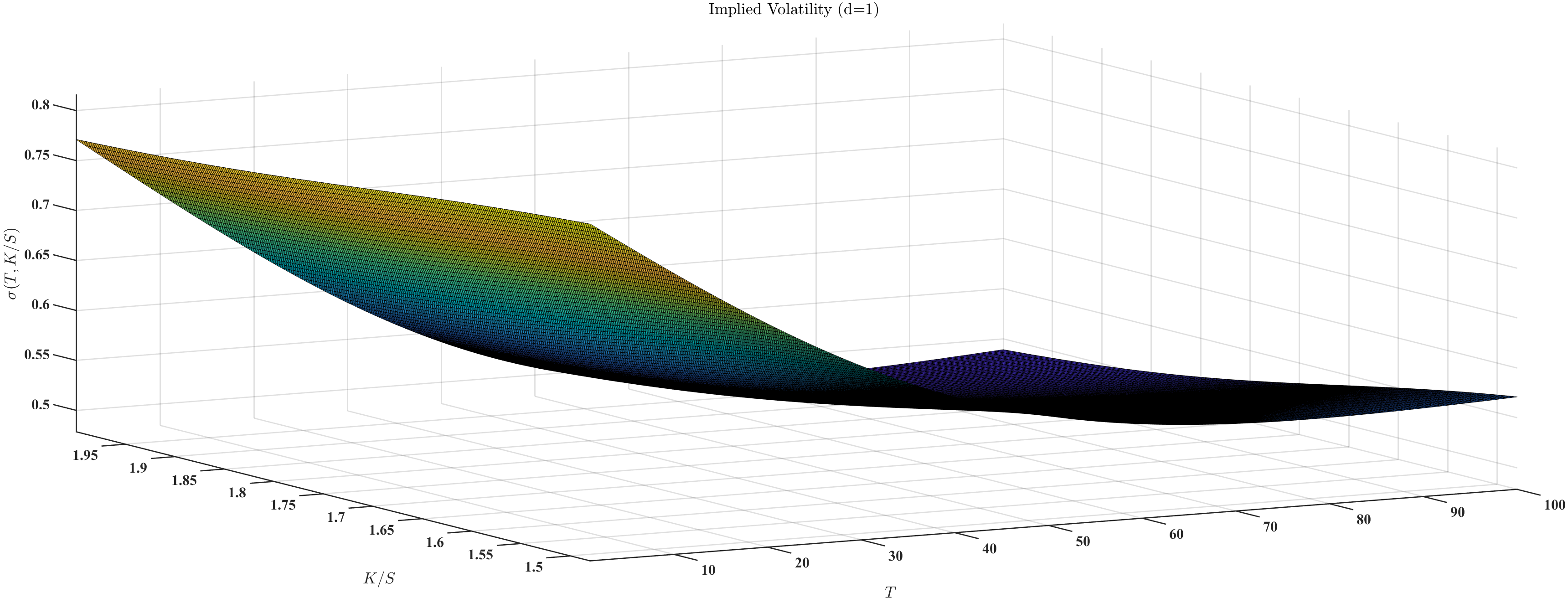}}
\subfloat[]{\includegraphics[width=0.49\textwidth]{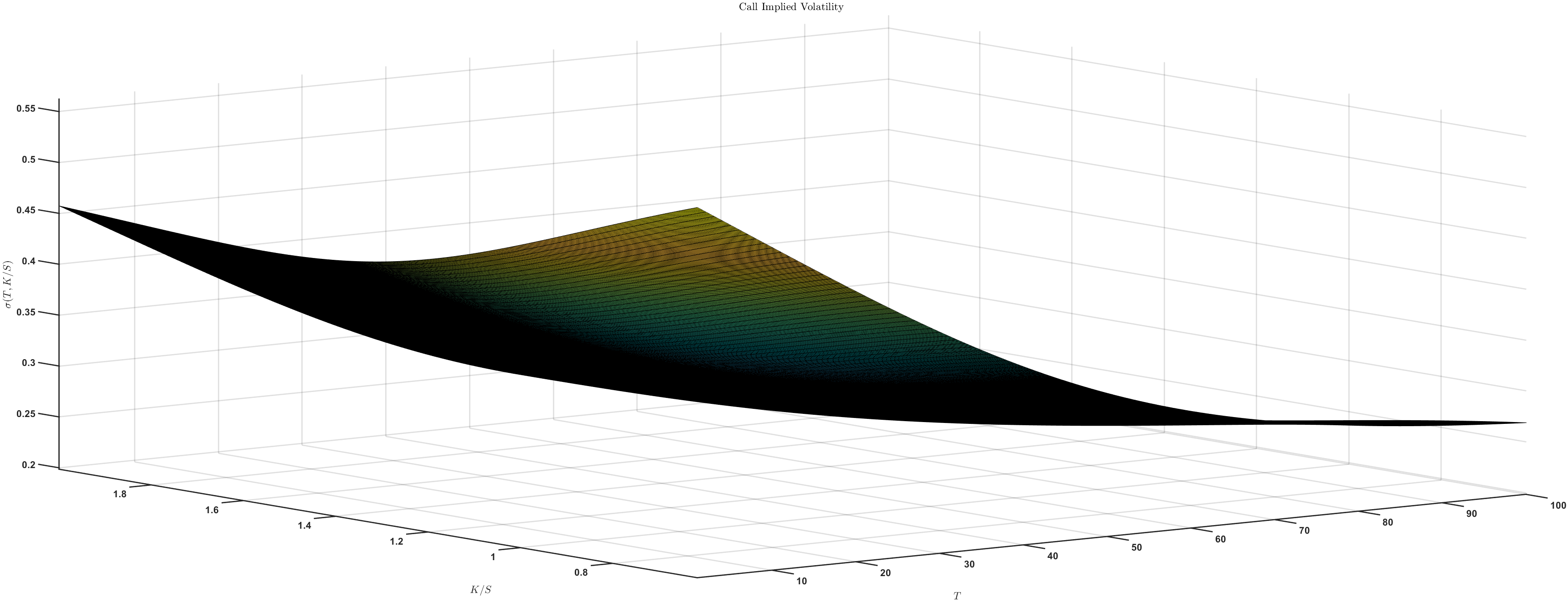}}

\subfloat[]{\includegraphics[width=0.49\textwidth]{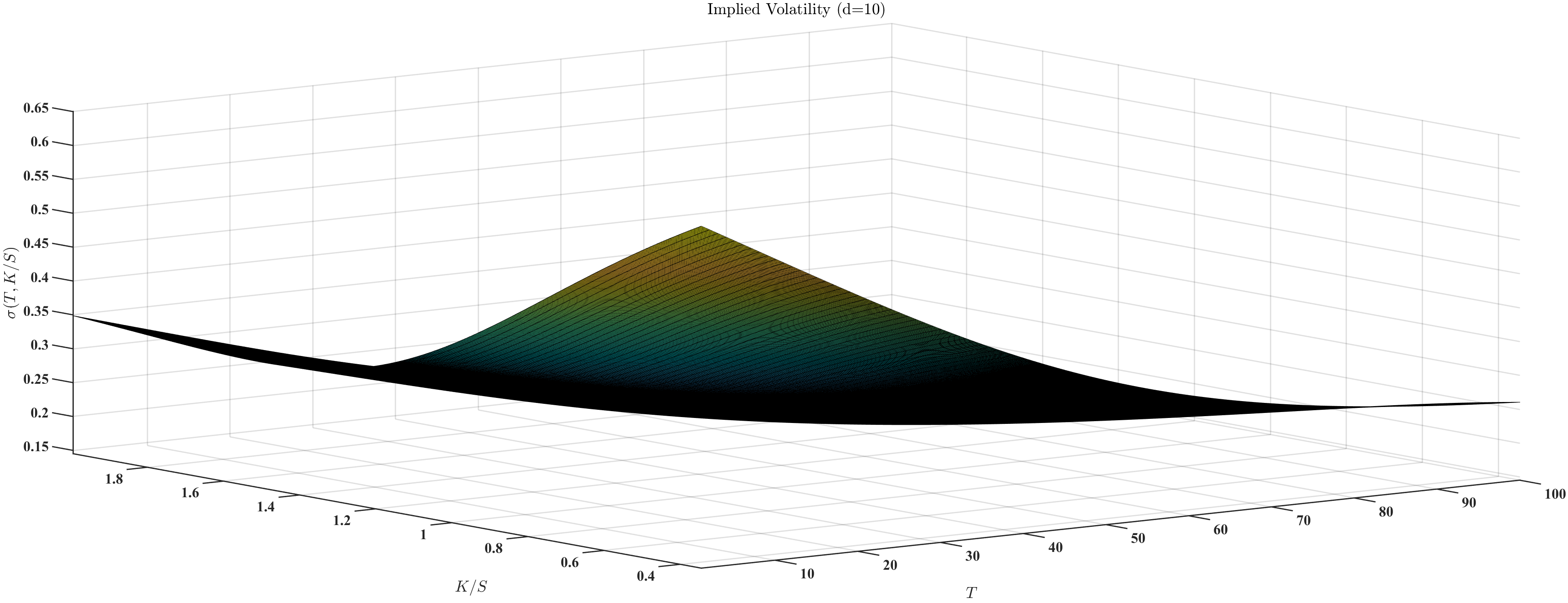}}
\subfloat[]{\includegraphics[width=0.49\textwidth]{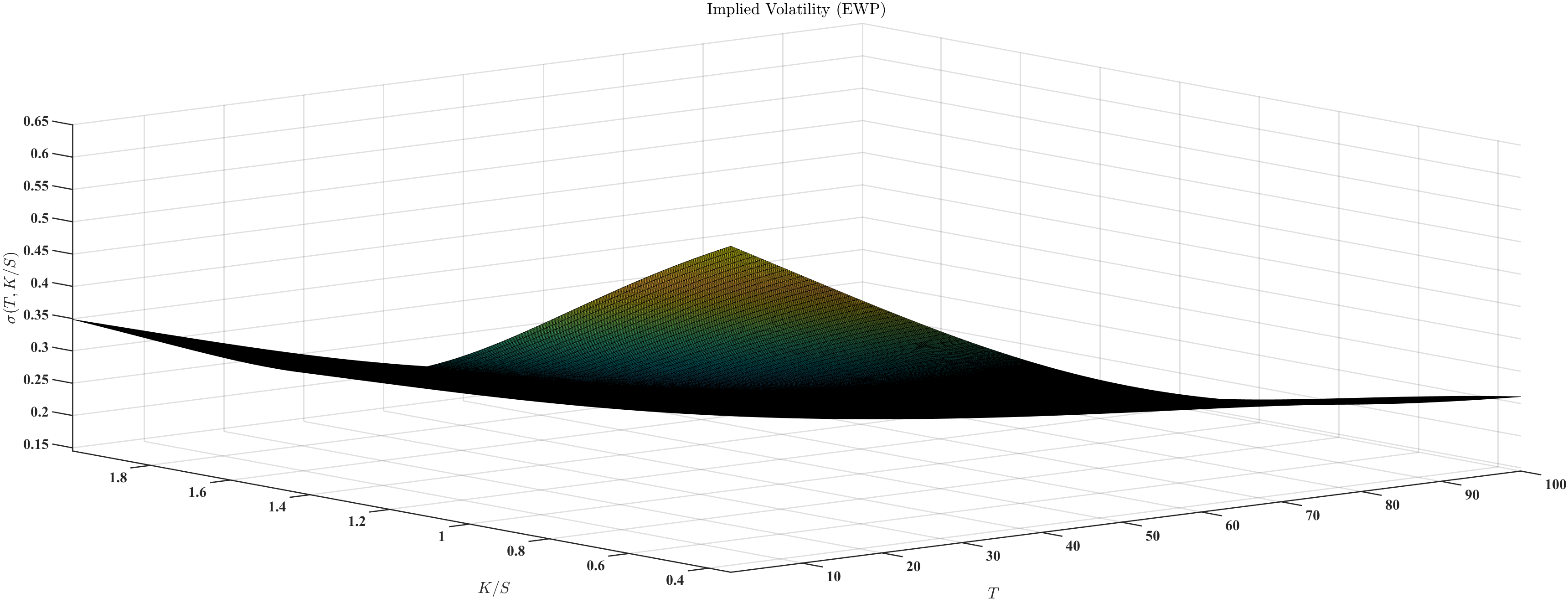}}
\caption{Implied volatility surface at depths of (a) 1, (b) 5, (c) 10, and (d) EWP for AAPL}
\label{IV_dyn}
\end{figure}

Figures~\ref{IV_dyn} also plot the implied volatility surface computed from call price options for the TMOBBAS index at depths of 1, 5, 10, and EWP as a function of $T$ and moneyness, $K/S_0$.  As is typically observed, at constant values of $T$, the implied volatility (future uncertainty) increases as strike prices move away from the value $K/S_0$ (the volatility ``smile''). At constant values of $K/S_0$, the implied volatility decreases as time to maturity increases. 

We also observe a decrease in implied volatility as depth increases, as illustrated in Figure~\ref{IV_dyn}, which presents an intriguing trend in the context of option pricing dynamics for the TMOBBAS index. Market depth, representing the level of liquidity and trading activity,  is a  crucial role in determining option prices and their associated implied volatilities.

The decrease in implied volatility with increasing depth suggests a relationship between market depth and option pricing stability. Deeper markets typically exhibit higher liquidity, characterized by  larger  buy and sell orders across various price levels. This increased liquidity contributes to a more stable and efficient pricing environment, reducing the uncertainty surrounding future price movements. As a result, options priced in deeper markets tend to reflect lower implied volatility levels, as market participants perceive less risk and uncertainty in price fluctuations.

Moreover, deeper markets often feature narrower bid--ask spreads, indicating tighter pricing and improved market efficiency. The reduced spread suggests buyers and sellers are more closely aligned in their pricing expectations, further contributing to the lower implied volatility observed in options priced based on the TMOBBAS index. 

Additionally, our analysis reveals a striking similarity between the implied volatility surface of the EWP encompassing all depths of the TMOBBAS index and the implied volatility of the TMOBBAS index with a depth of $10$. This suggests that the implied volatility of the EWP, which aggregates returns from all depths, closely mirrors the implied volatility behavior of the TMOBBAS index with the greatest market depth. Essentially, the EWP serves as a holistic representation of market sentiment and volatility by amalgamating information from diverse depths of the TMOBBAS index. Consequently, the implied volatility surface of the EWP offers a comprehensive perspective on market volatility, integrating insights from various depths.

Moreover, the notable resemblance between the implied volatility of the EWP and the TMOBBAS index with depth $10$ indicates the predominant influence of deeper market levels on overall market volatility. This underscores the critical role of deep liquidity levels in shaping market sentiment and option pricing dynamics. The convergence of implied volatility behavior between the EWP and the TMOBBAS index with the greatest depth underscores the importance of considering depth-specific factors in understanding and predicting market volatility.


\subsection{Option Pricing and Implied Volatility of GMP} 
 This subsection presents  the results of analyzing option pricing and implied volatility for the GMP index, leveraging the NDIG L\'{e}vy model. Our objective is to gain insights into the pricing dynamics of plain vanilla European options on the GMP index and assess implied volatility trends. 

\begin{figure}[htbp]
\centering
\subfloat[]{\includegraphics[width=0.49\textwidth]{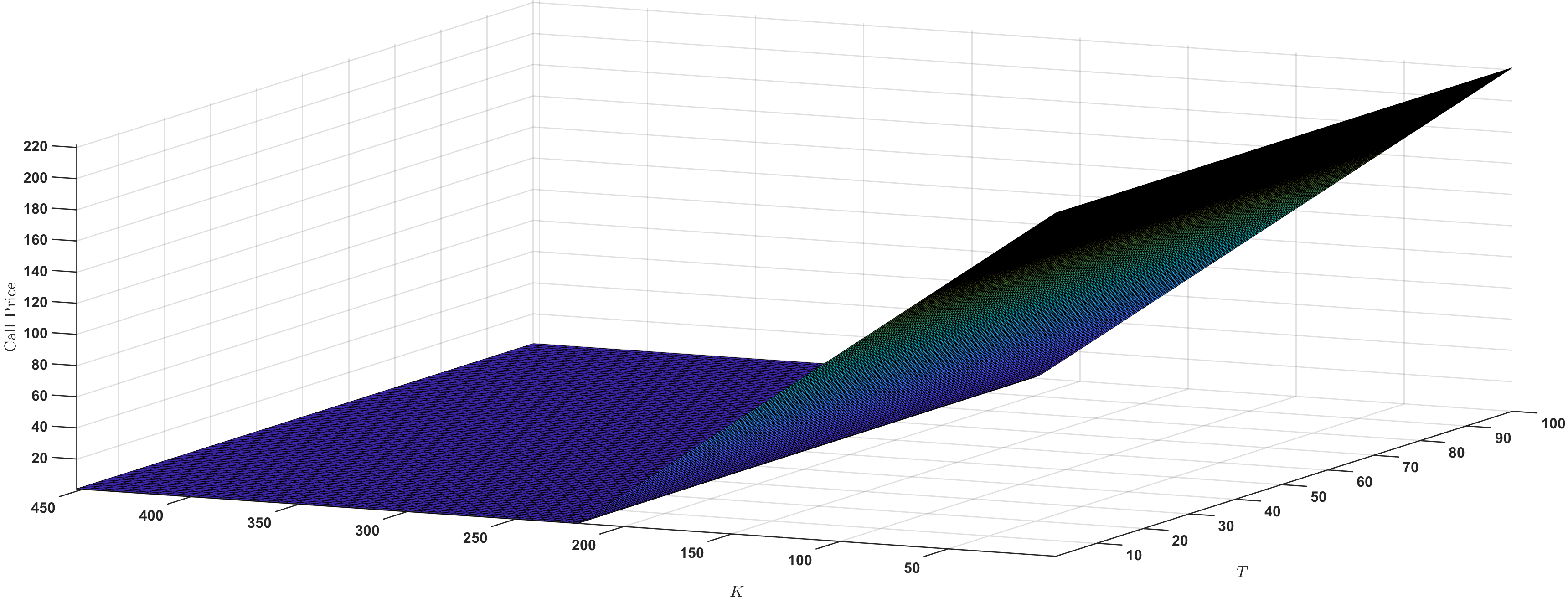}}
\label{GMP_Call_price_NDIG_5}
\subfloat[]{\includegraphics[width=0.49\textwidth]{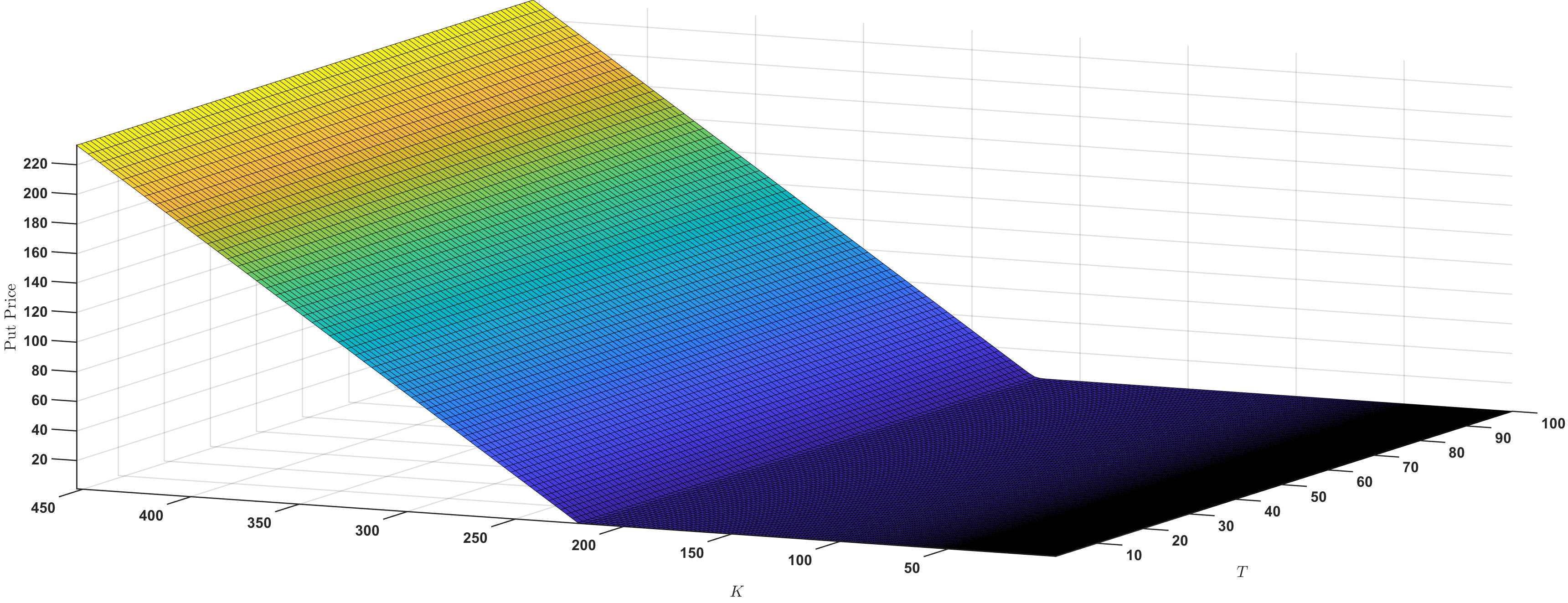}}
\label{GMP_Put_price_NDIG_5}
\subfloat[]{\includegraphics[width=0.49\textwidth]{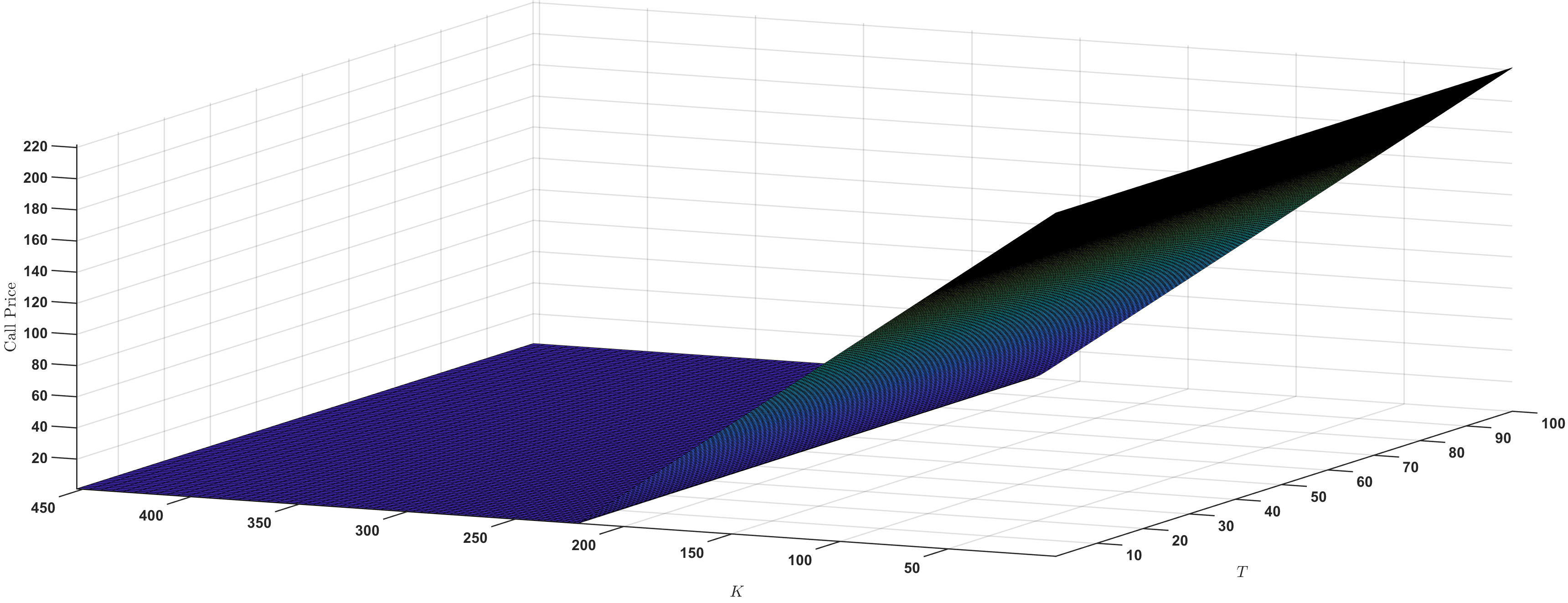}}
\label{GMP_Call_price_NDIG_10}
\subfloat[]{\includegraphics[width=0.49\textwidth]{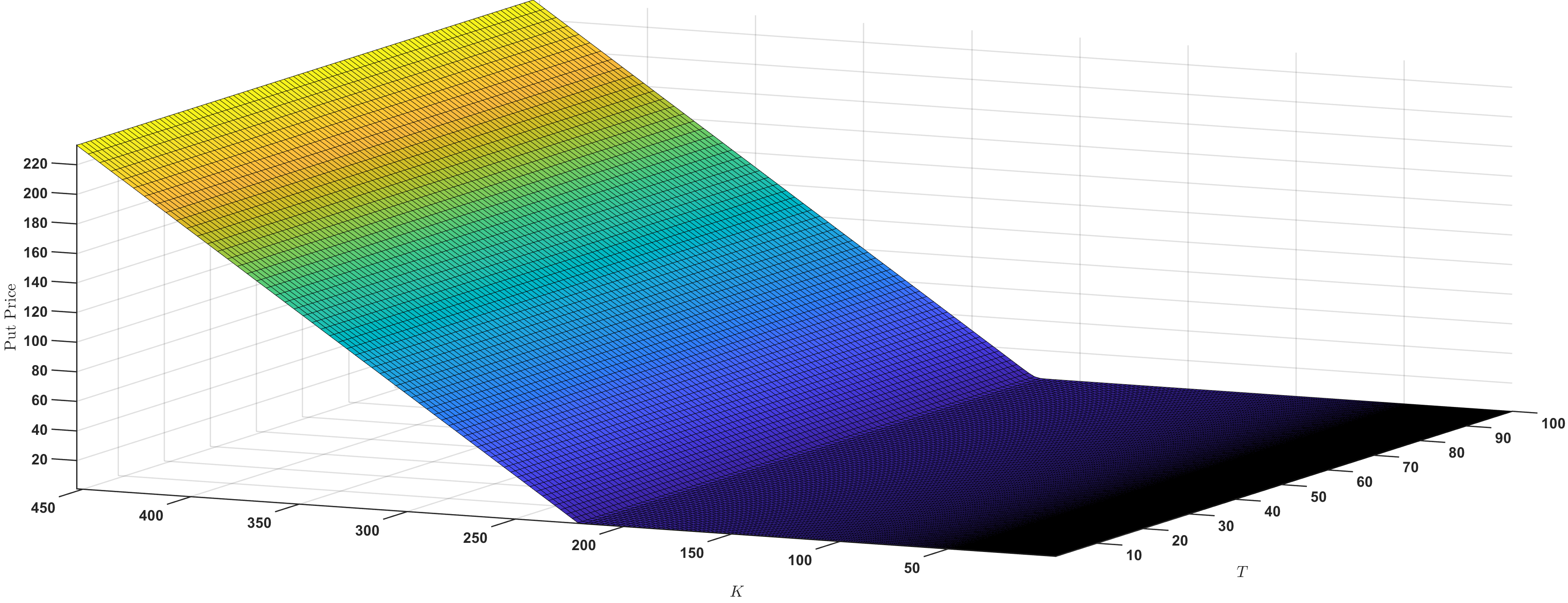}}
\label{GMP_Put_price_NDIG_10}
\caption{NDIG-based price surfaces for GMP index (a) at depth of 5 for calls, (b) at depth of 5 for puts, (c) at depth of 10 for calls,  and (d) at depth of 10 for puts as functions of the time to maturity and strike price.
}
\label{GMP_option_surfaces}
\end{figure}

Figure~\ref{GMP_option_surfaces} presents the pricing surfaces for call and put options plotted against the time to maturity $\tau$ and the strike price $K$, focusing on depths of 5 and 10 of the GMP index. Notably, these surfaces offer valuable insights into the option pricing behavior specific to the GMP index. 

Observing the patterns in the pricing surfaces reveals intriguing dynamics.  For  near-the-money options ($K \approx S$), where the strike price closely aligns with the current asset price, both call and put option prices demonstrate an increasing trend with $\tau$. This trend reflects the market's anticipation of heightened uncertainty, leading to elevated option premiums for longer-maturity contracts. Such observations align with established market phenomena, where longer-maturity options command higher premiums due to increased time value and augmented uncertainty associated with extended time horizons.

\begin{figure}[htbp]
\centering
\subfloat[]{\includegraphics[width=0.49\textwidth]{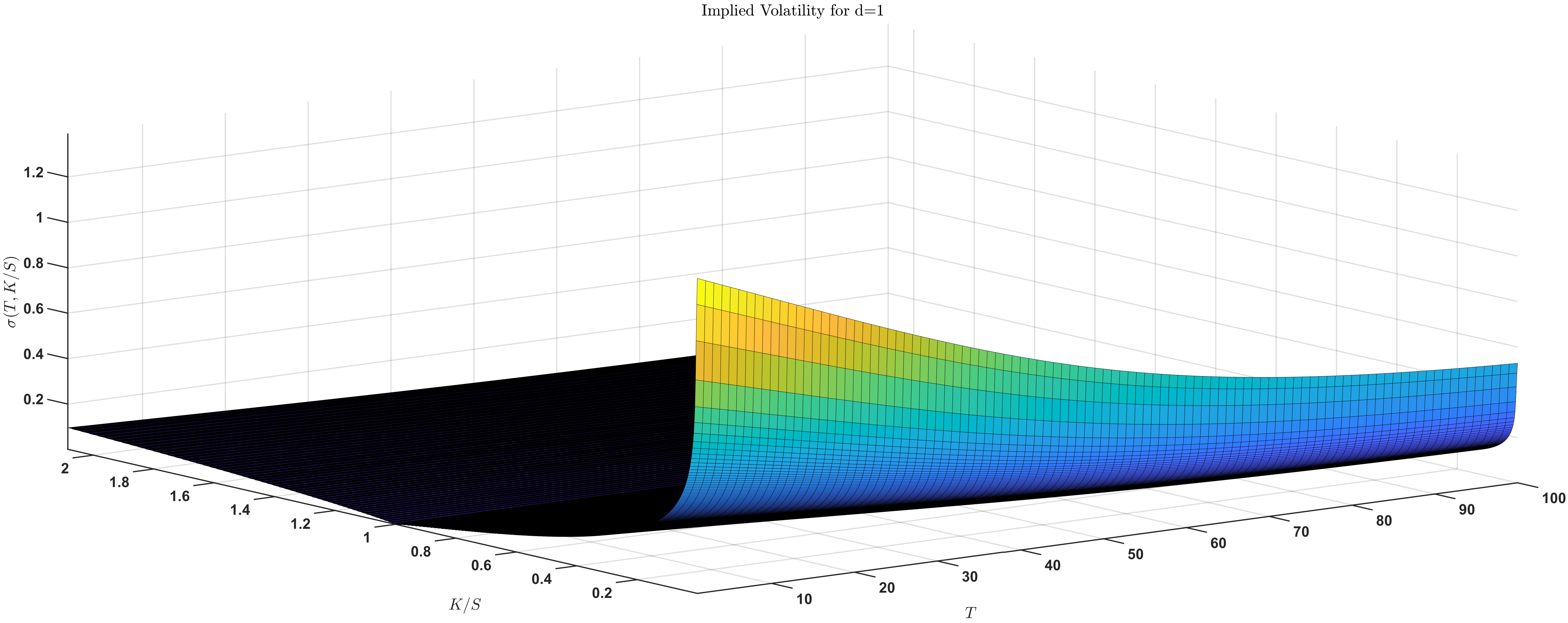}}
\subfloat[]{\includegraphics[width=0.49\textwidth]
{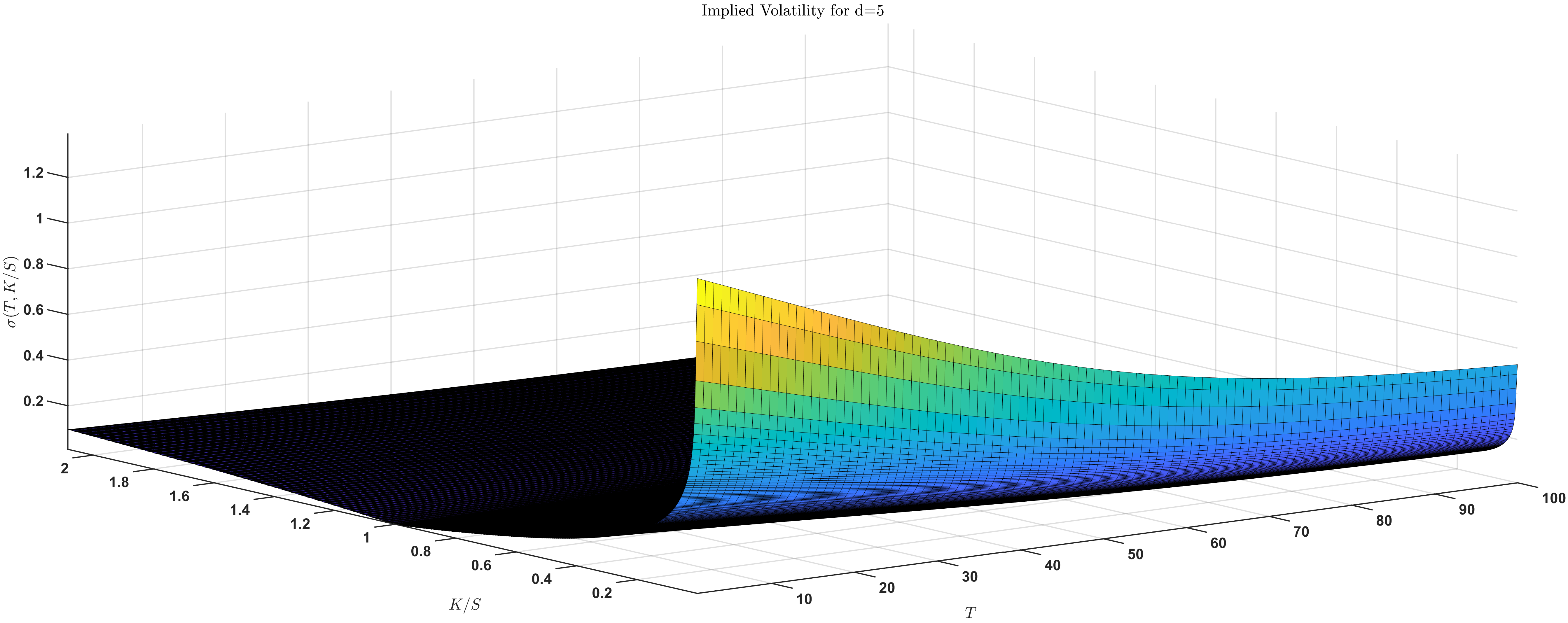}}

\subfloat[]{\includegraphics[width=0.49\textwidth]{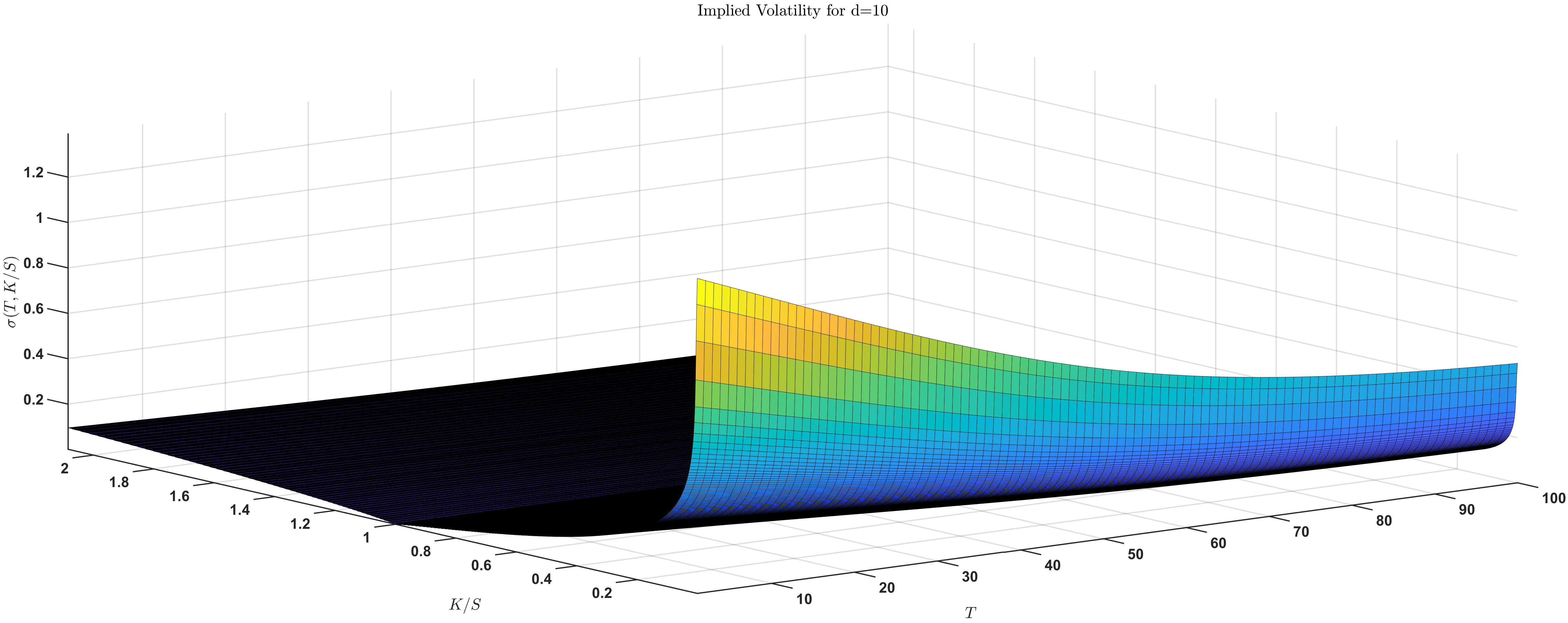}}
\includegraphics[width=0.49\textwidth]{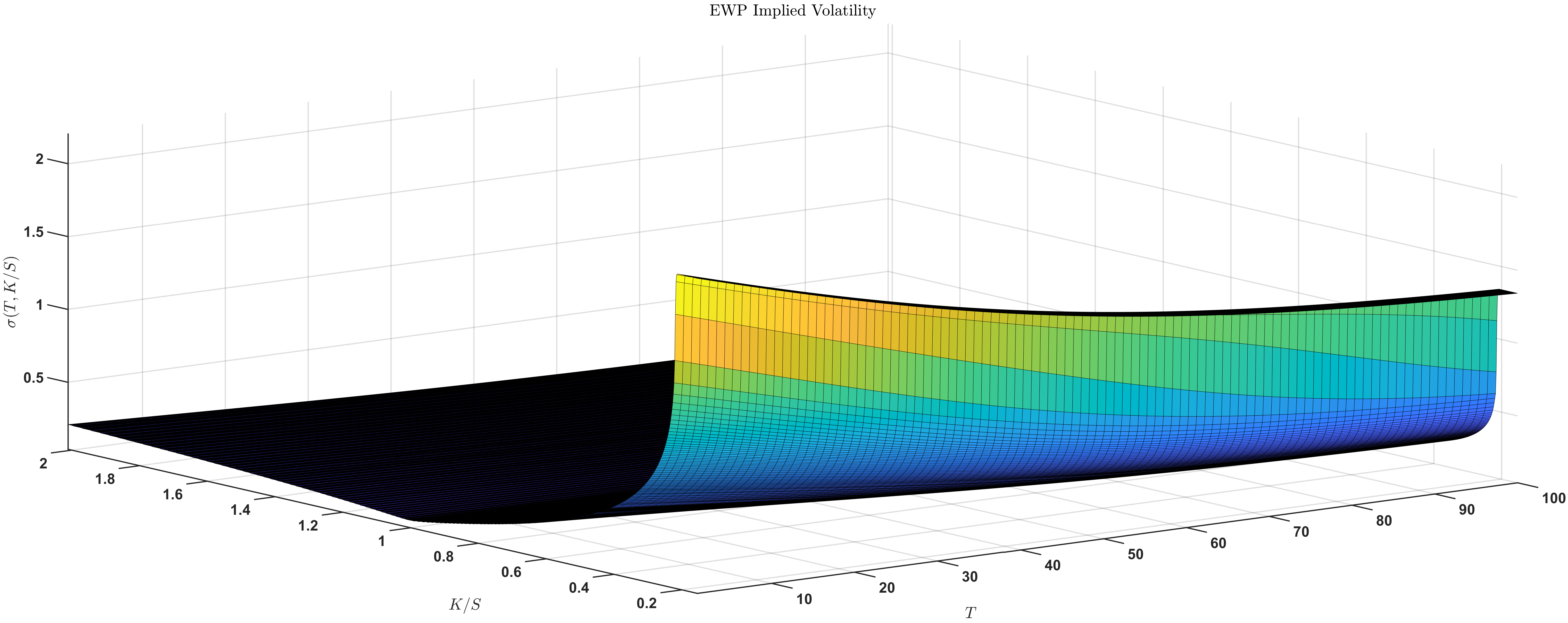}
\caption{GMP Implied volatility surfaces at depths of (a) 1, (b) 5, (c) 10, and (d) EWP for AAPL}
\label{IV_dyn_GMP}
\end{figure}

Figure~\ref{IV_dyn_GMP} presents the implied volatility surface computed  the price of call  options for the GMP index at depths of 1, 5, 10, and EWP as a function of $T$ and moneyness, $K/S_0$. Notably, the implied volatility exhibits a distinct ``smile'' pattern, characterized by a sharp increase in volatility as the moneyness approaches zero.

Moreover, there is notable minimum volatility observed close to zero moneyness, particularly evident as moneyness approaches 1 from both in-the-money and out-of-the-money options. This suggests that market participants perceive lower risk and uncertainty for options with moneyness near 1, reflecting a more stable pricing environment.

Interestingly, the implied volatility surface for GMP exhibits a consistent shape across different depths, with minimal variation as the depth increases. Unlike the TMOBBAS index, where the implied volatility decreases with increasing depth, the implied volatility of GMP remains relatively stable across all depths. This uniformity in the implied volatility suggests a robust and stable pricing environment for options on the GMP index, regardless of the market depth.

Comparing the implied volatility of the GMP with that of the TMOBBAS across all depths reveals that the  GMP's implied volatility  is consistently lower. This disparity in implied volatility underscores the differences in market dynamics and risk perceptions between the two indices, with the GMP exhibiting lower perceived risk and volatility than the TMOBBAS.


Regarding the implied volatility surface of the EWP for the GMP, our analysis indicates that it  is  slightly higher than the implied volatility of each depth of GMP.  This suggests that while the overall level of volatility may vary, the underlying volatility dynamics captured by the implied volatility surface exhibits uniform characteristics across different depths of the GMP. Also, despite this difference in overall volatility levels, we observe a remarkable similarity in the shapes of the implied volatility surfaces.

Similar to the individual depths of GMP, the implied volatility surface of the EWP also displays the characteristic volatility ``smile,'' with a sharp increase in volatility as the moneyness approaches zero. Additionally, the minimum volatility is observed close to zero moneyness, as the options move both in-the-money and out-of-the-money.


\section{Evaluation of Risk--Return Performance}\label{sec: rachev}
In this section, we will focus on measuring the risk--return performance for the TMOBBAS and the GMP. To do this, we will employ the Rachev ratio \citep{rachev2008advanced} as the metric. Unlike \textit{reward-to-variability} ratios such as the well-known Sharpe ratio \citep{sharpe1966mutual} and Sortino ratio \citep{sortino1994performance}, the Rachev ratio is a reward-to-risk ratio designed to assess the potential for extreme positive returns relative to the risk of extreme losses in a non-Gaussian setting. Intuitively, it reflects the potential for significant gains compared to the risk of significant losses, at a rarity frequency $q$ (quantile level) defined by the researcher.

The formula to compute the Rachev ratio for the TMOBBAS or GMP is
\begin{equation}
\textrm{RR}_{(\beta, \gamma)}(X) = \frac{\textrm{AVaR}_{\beta}(-X)}{\textrm{AVaR}_{\gamma}(X)},\label{eq: RRdef}
\end{equation}
where
\begin{itemize}
\item $X$ denotes the log-return of the TMOBBAS or GMP at a given time interval,
\item $\textrm{AVaR}_{\gamma}(X) = \gamma^{-1}\int_{0}^{\gamma}\textrm{VaR}_{u}(X)\textrm{ d}u$ is the average Value-at-Risk (VaR) at the level $\gamma\in(0,1]$, and
\item $\textrm{VaR}_{u}(X)= \inf\left\{m\in\mathbb{R}: \mathbb{P}[X+m < 0]\le u\right\}$ is the VaR at the level $u$.
\end{itemize}
The formula to compute $\textrm{AVaR}_{\gamma}(X)$ may seem a bit challenging. However, according to \cite{follmer2004stochastic}, we know that
\begin{equation}
\textrm{AVaR}_{\beta}(X) = \frac{\mathbb{E}[(q-X)^{+}]}{\beta}-q,\label{eq: AVaRformula}
\end{equation}
where 
\begin{itemize}
\item $\mathbb{E}[\cdot]$ represents the expectation,
\item $q$ equals to the $\beta$-quantile of $X$, and
\item $(q-X)^{+} = \max(q-X, 0)$.
\end{itemize}

\begin{figure}[htbp]
\centering
\subfloat[]{\includegraphics[width=0.49\textwidth, height=0.4\textheight, keepaspectratio]{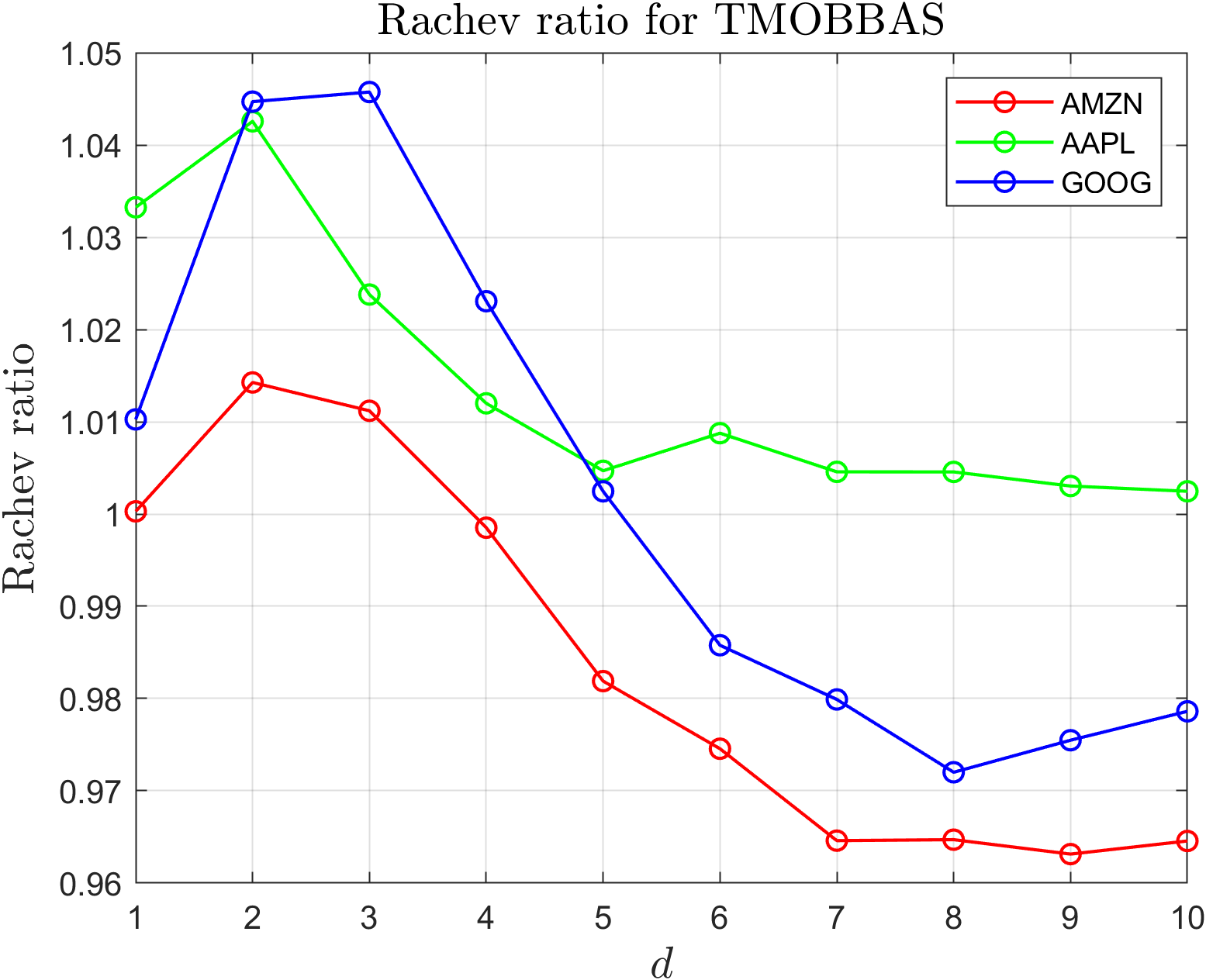}}
\label{figure: RR_TMOBBAS}
\subfloat[]{\includegraphics[width=0.49\textwidth, height=0.4\textheight, keepaspectratio]{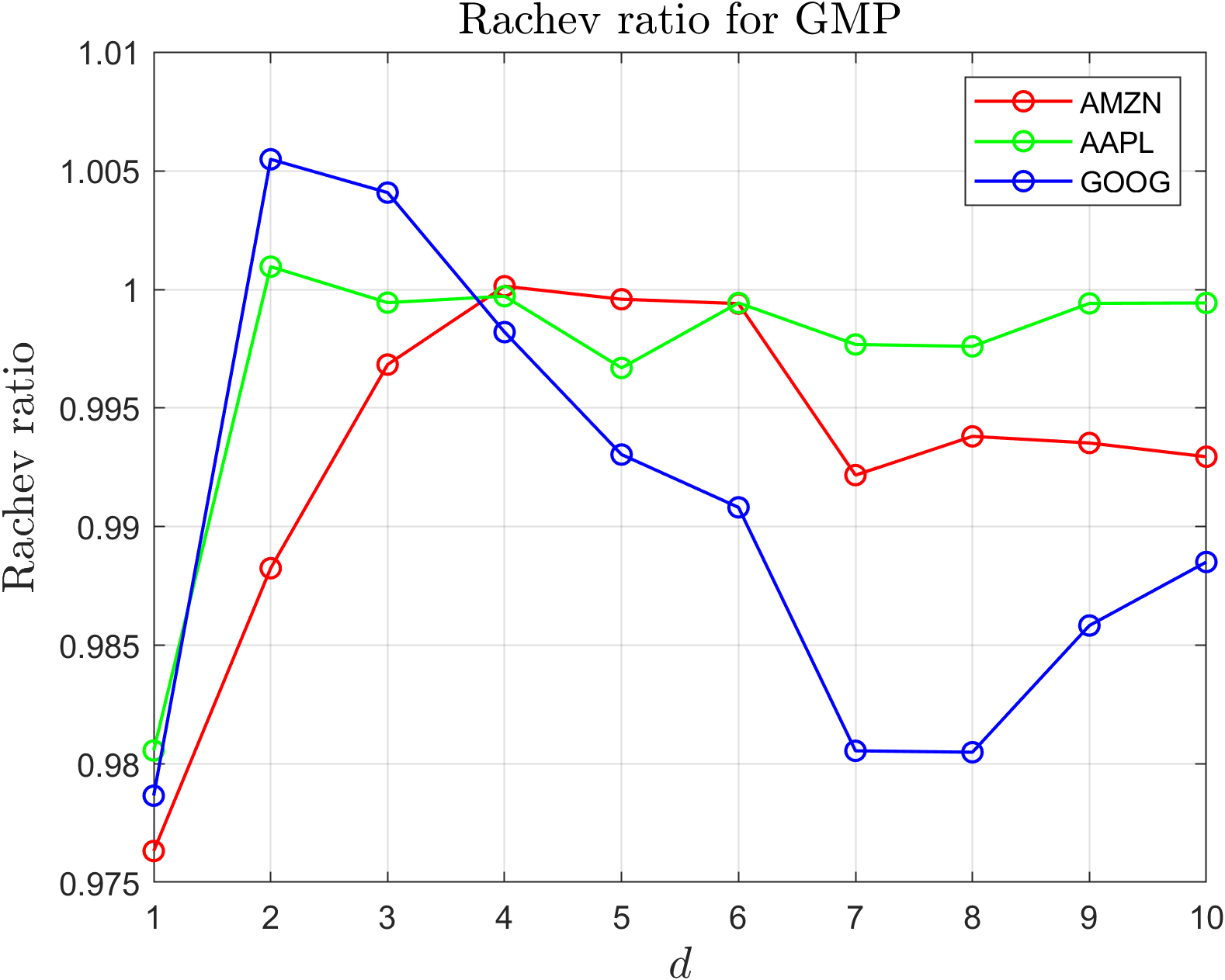}}
\label{figure: RR_GMP}
\caption{The dependence of the Rachev ratio of (a) TMOBBAS and (b) GMP for AMZN, AAPL, and GOOG on the depth (AMZN, AAPL, and GOOG are represented by red, green, and blue line with dots, respectively)}
\label{figure: RR}
\end{figure}

In this study, we set $\beta = \gamma = 0.05$ to explore the relationship between excess profit and excess loss when ``investing'' in the TMOBBAS and GMP. Next, we merge Equations~\eqref{eq: RRdef} and \eqref{eq: AVaRformula}, applying them to our dataset for AMZN, AAPL, and GOOG. Figure~\ref{figure: RR} illustrates the dependence of the Rachev ratio of TMOBBAS and GMP for AMZN, AAPL, and GOOG on the depth.

From Figure~\ref{figure: RR} (a), we can observe that
\begin{itemize}
\item The general trend shows that as the depth increases, the Rachev ratio initially rises and then declines.
\item For AAPL, the Rachev ratio remains consistently above 1 across different depths, suggesting that excess losses can always be offset by excess profits when investing in TMOBBAS at varying depths of AAPL. In particular, the highest Rachev ratio occurs at a depth of 2.
\item In the case of AMZN, as the depth increases to 5 or beyond, the Rachev ratio drops below 1, indicating that excess losses cannot be balanced by excess profits when investing in TMOBBAS at such depths. Similarly to AAPL, the Rachev ratio for AMZN peaks at a depth of 2.
\item For GOOG, as the depth increases to 6 or beyond, the Rachev ratio falls below 1, suggesting that excess losses cannot be balanced by excess profits when investing in TMOBBAS at such depths. Notably, the Rachev ratio for GOOG achieves relatively high values at depths 2 and 3.
\end{itemize}

From Figure~\ref{figure: RR} (b), we can observe that 
\begin{itemize}
\item The general trend shows that as the depth increases, the Rachev ratio initially rises and then declines.
\item For AAPL, the Rachev ratio increases first, reaching its peak at a depth of 2, approximately 1.001. Then, this value remains close to, but less than, 1 for the next eight depths. This suggests that the excess loss can almost be balanced by the excess profit when investing in GMP for AAPL at various depths.
\item For AMZN, the Rachev ratio also increases initially, reaching its peak at a depth of 4, around 1. Afterward, the overall trend of the Rachev ratio  decreases  until it reaches a value of approximately 0.993 at a depth of 10. This suggests that the excess loss cannot almost be balanced by the excess profit when investing GMP for AMZN at various depths.
\item For GOOG, the Rachev ratio initially increases, reaching its peak at a depth of 2, approximately 1.005. It then decreases to its lowest point at a depth of 8, around 0.981, before increasing again to about 0.988 at a depth of 10. Overall, we can see that the excess loss can be balanced by the excess profit when investing in GMP for GOOG at depths of 2, 3, and 4. However,  excess loss cannot be balanced by excess profit for other depths.
\end{itemize}

\section{Assessing Systemic Risk Contagion from Log-Returns of TMOBBAS to Individual Spread Levels}\label{sec: CoETL}
In this section, we evaluate the tail risk dependencies between the  log returns  of the TMOBBAS and the individual spread levels across different order book depths using systemic risk measures. The goal is to understand how extreme events in the individual spreads propagate through the spread hierarchy and impact the global spread index, represented by the mean of the GMP across depths 1 through 10.

We adopt a conditional risk assessment framework using CoVaR (Conditional Value at Risk), CoETL (Conditional Expected Tail Loss), and CoCVaR to capture these dynamics. Specifically, the CoVaR measures quantify how distress in the log-return of the TMOBBAS impacts the tail risk of individual spreads and vice versa, providing insight into the bidirectional nature of systemic risk within the order book. Similar to the previous section, we define the CoVaR using the quantiles of the conditional distribution, as outlined in \cite{Mainik14}. CoETL further extends this by capturing the expected tail loss beyond the CoVaR threshold, offering a more granular view of the tail dependencies between these variables.

Let \( Y \) and \( X \) denote two random variables representing the  log returns  of TMOBBAS and individual spreads, respectively. Following \cite{Mainik14}, the CoVaR at level \( q \), denoted as \(\text{CoVaR}_{q}\), is defined as

\begin{equation*}
\text{CoVaR}_{q} := F^{-1}_{Y|X\leq F^{-1}_{X}(q)}(q) = \text{VaR}_{q}\left(Y|X\leq \text{VaR}_{q}(X)\right),
\end{equation*}
where \( F_{Y|X} \) is the conditional distribution of \( Y \) given \( X \), and \( \text{VaR}_{q}(X) \) is the Value-at-Risk of \( X \) at level \( q \).

In addition to the CoVaR, we compute the corresponding Conditional Expected Tail Loss (CoETL), which  measures  the average loss given that both variables fall below their respective CoVaR values. CoETL is defined as

\begin{equation*}
\text{CoETL}_{q} := \mathbb{E}\left(Y|Y\leq\xi_{q},X\leq\text{VaR}_{q}(X)\right).
\end{equation*}

These measures help capture the magnitude of potential tail losses, conditional on distress in the other variable, and are critical in assessing the systemic risk transmission within the order book.

To conduct this analysis, we leverage the time series of  log returns  of the TMOBBAS and the returns for individual spreads \(R(t; d)\) across different depths \(d = 1, \ldots, 10\),  and  the total spread index \(I(t)\). For a given time \( T \), we model the joint dynamics of the system \(D(t) = (R(t; 1), \ldots, R(t; 10), I(t))\), \( t = 1, \ldots, T\), using a multivariate model to capture both linear and nonlinear dependencies between the spreads. Using this model, we generate 10,000 Monte Carlo scenarios \( D(T + 1, s) \), \( s = 1, \ldots, S \), representing potential future realizations of the spread series.

Based on the simulated scenarios, we compute the CoVaR for \((R(T + 1, d) , I(T + 1))\), which measures the tail risk of individual spreads conditioned on the TMOBBAS being in distress, and captures how the tail risk of individual spreads influences the overall system. 


The results, illustrated in Table~\ref{tab:CoES_CoETL}, reveal key insights into the nature of systemic risk across different depths of the order book. At a stress level of 99\%, we observe that deeper spreads (greater depth values) exhibit significantly larger values of the Conditional Expected Shortfall (CoES) and the Conditional Extreme Tail Loss (CoETL) compared to shallower spreads. This indicates that shocks in deeper levels have more substantial impacts on the overall system, as evidenced by values of the CoETL that grow in magnitude as the depth increases. Specifically, for $d = 6$, the CoETL reaches $-1.286$ at 99\%, the largest observed impact, highlighting the pronounced systemic risk linked to deeper order book levels.

Similarly, for the stress level of 95\%, deeper order book levels still carry a significant impact. The CoES increases as the depth $d$ increases, and for $d = 9$, the CoES is approximately $-0.058$. These findings demonstrate the asymmetric nature of risk propagation, with deeper spreads being more vulnerable to and imposing greater systemic shocks.

Moreover, the difference between the CoETL at  99\% and 95\% stress levels  further emphasizes the heavy-tailed nature of the risk at deeper spreads. This asymmetry in systemic risk highlights that deeper levels are more sensitive to shocks, which may amplify the risk, particularly during periods of high volatility.

In contrast, at shallower levels (e.g., $d = 1$), the CoETL values are lower (approximately $-0.091$ for 99\% and $-0.089$ for 95\%), suggesting a relatively smaller systemic influence. These findings emphasize that while shallower levels of the order book are more visible and reactive to market dynamics, they have less  systemic effect than  deeper levels.

\begin{table}[htbp]
\tbl{CoES and CoETL values at different depths for $\alpha = 0.95$ and $\alpha = 0.99$ for AAPL}
{\begin{tabular}{c c c c c}
\hline
Depth $d$ & CoES (95\%) & CoETL(95\%) & CoES(99\%) & CoETL(99\%) \\ \hline
1     & -0.002   & -0.089    & -0.007   & -0.091    \\ 
2     & -0.004   & -0.109    & -0.014   & -0.114    \\ 
3     & -0.003   & -0.153    & -0.012   & -0.157    \\ 
4     & -0.004   & -0.228    & -0.016   & -0.238    \\ 
5     & -0.007   & -0.338    & -0.029   & -0.346    \\ 
6     & -0.021   & -1.217    & -0.086   & -1.286    \\ 
7     & -0.034   & -0.387    & -0.138   & -0.438    \\ 
8     & -0.050   & -0.310    & -0.172   & -0.384    \\ 
9     & -0.058   & -0.428    & -0.183   & -0.493    \\ 
10    & -0.046   & -0.325    & -0.132   & -0.419    \\ \hline
\end{tabular}}
\label{tab:CoES_CoETL}
\end{table}

This analysis suggests that  deeper order book levels play a critical role in propagating and magnifying systemic risk during market distress, particularly at extreme quantiles. Consequently, effective risk management and monitoring require close attention to these deeper levels, as their systemic impact can be substantial, especially during times of increased market uncertainty.

\section{Conclusion}\label{sec: conclusion}
In this study, we expanded the concepts of BAS and mid-price to two new constructs: TMOBBAS, a development of the BAS,  and GMP, a development of the mid-price. Through a detailed analysis involving kernel density estimation, QQ-plots, and kurtosis measurement, we demonstrated that the return distributions of both TMOBBAS and GMP deviate significantly from normality.

Furthermore, we ascertained the presence of heavy tails in the return distributions by fitting a GPD, calculating the Hill estimator, and applying the rank minus 1/2 method. This  underscores the non-trivial characteristics of financial returns in high-frequency data.

Acknowledging the temporal dependence within the return time-series and the need to capture the heavy-tailed nature of these returns, we adopted an ARMA(1,1)-GARCH(1,1) model with an NIG distribution to model the dynamics of the return series. This constitutes a robust approach to handling the complexities inherent in financial market data.

Given the relationship between the  TMOBBAS and the GMP concepts  and market liquidity, we addressed the issue of hedging against the risk posed by low liquidity. To this end, we applied a novel double subordination option pricing model to price European options, offering a practical solution to manage and mitigate liquidity-related risks in trading strategies.

Furthermore, we used the Rachev ratio to evaluate the risk--return performance of the TMOBBAS and GMP for different depths of the order book, offering deeper insights into liquidity risks. We also analyzed the tail risk interdependencies between the TMOBBAS and individual spread levels, revealing how extreme events propagate through the order book, amplifying systemic risks.

In summary, this study provides a comprehensive framework for understanding the risks associated with HFT and liquidity, offering theoretical insights and practical tools for risk management.

\section*{Data Availability Statement}
Publicly available datasets were analyzed in this study. This data can be found here: LOBSTER (\url{https://lobsterdata.com/}).

\section*{Author Contributions}
YH: Data curation, Formal analysis, Investigation, Methodology, Software, Validation, Visualization, Writing - original draft, Writing - review \& editing; AS: Data curation, Formal analysis, Methodology, Software, Validation, Writing - original draft, Writing - review \& editing; BS: Writing - review \& editing; SR: Conceptualization, Investigation, Methodology, Project administration, Resources, Supervision, Visualization, Writing - review \& editing; FF: Resources, Writing - review \& editing


\section*{Disclosure Statement}
The authors declare that the research was conducted in the absence of any commercial or financial relationships that could be construed as a potential conflict of interest.

\section*{Funding}
The authors declare that no financial support was received for the research, authorship, and publication of this article.

\bibliographystyle{apacite}
\bibliography{paper_clean}
\clearpage

\appendix

\section{Figure}\label{appendix: figure}
The figures in the appendices present the outcomes of the analyses conducted for AMZN and GOOG. As these analyses parallel those performed for AAPL, we  only furnish  the results, omitting the explanatory narratives for brevity and consistency.

\subsection{Static Method}

\subsubsection{The Evolution of TMOBBAS, GMP and Their Log-Returns over Time}
\begin{figure}[htbp]
\centering
\subfloat[]{\includegraphics[width=0.49\textwidth]{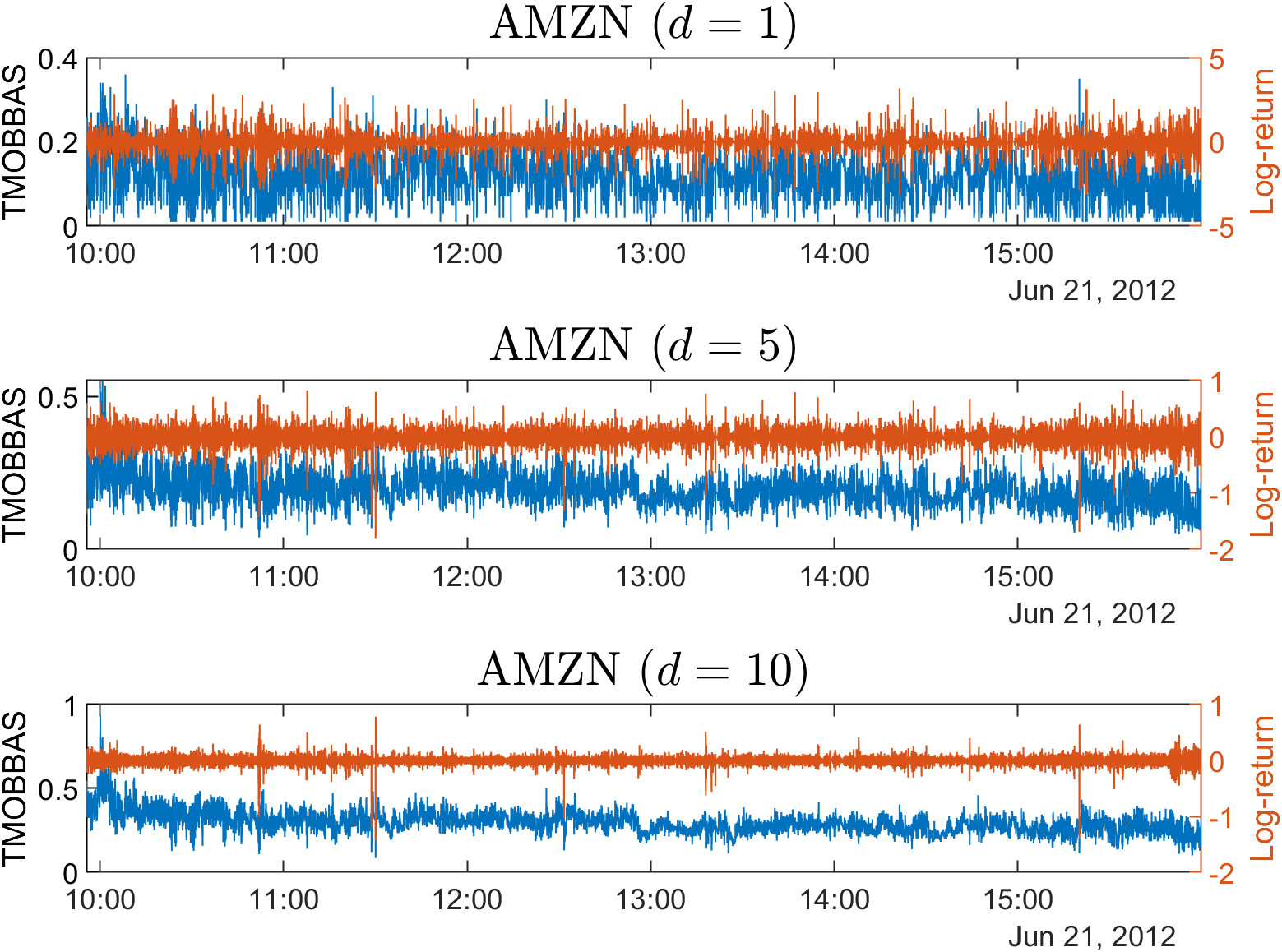}}
\label{figure: TMOBBAS_amzn}
\subfloat[]{\includegraphics[width=0.49\textwidth]{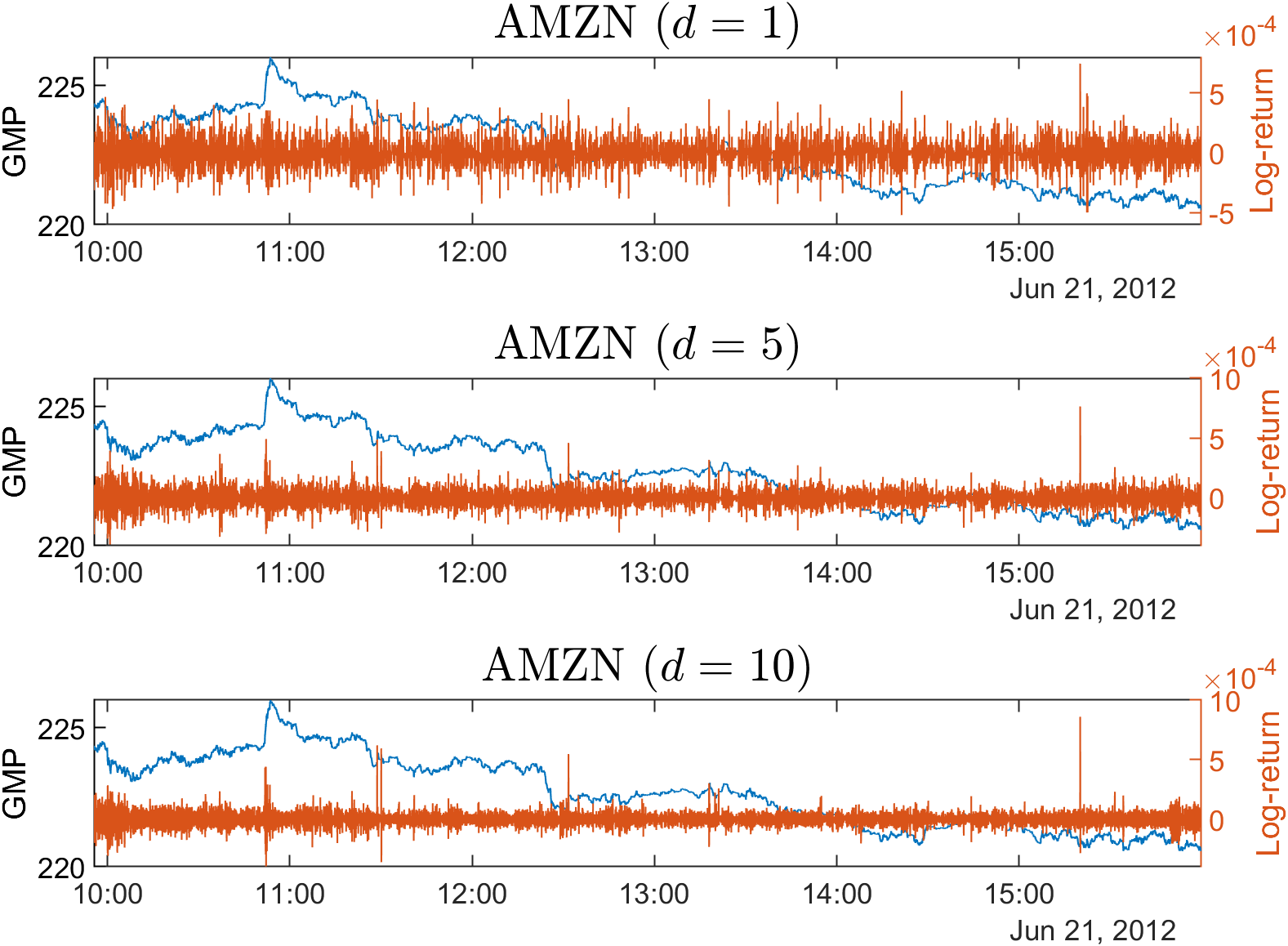}}
\label{figure: GMP_amzn}
\caption{The evolution of (a) TMOBBAS, (b) GMP and their log-returns over time at depths of 1, 5, and 10 for AMZN (each graph contains two vertical axes. The left vertical axis is for TMOBBAS (or GMP), and the right vertical axis is for the log-return. The evolution of TMOBBAS (or GMP) is represented by the blue line while the orange line the evolution of the log returns)}
\label{figure: TMOBBAS and GMP_amzn}
\end{figure}

\begin{figure}[htbp]
\centering
\subfloat[]{\includegraphics[width=0.49\textwidth]{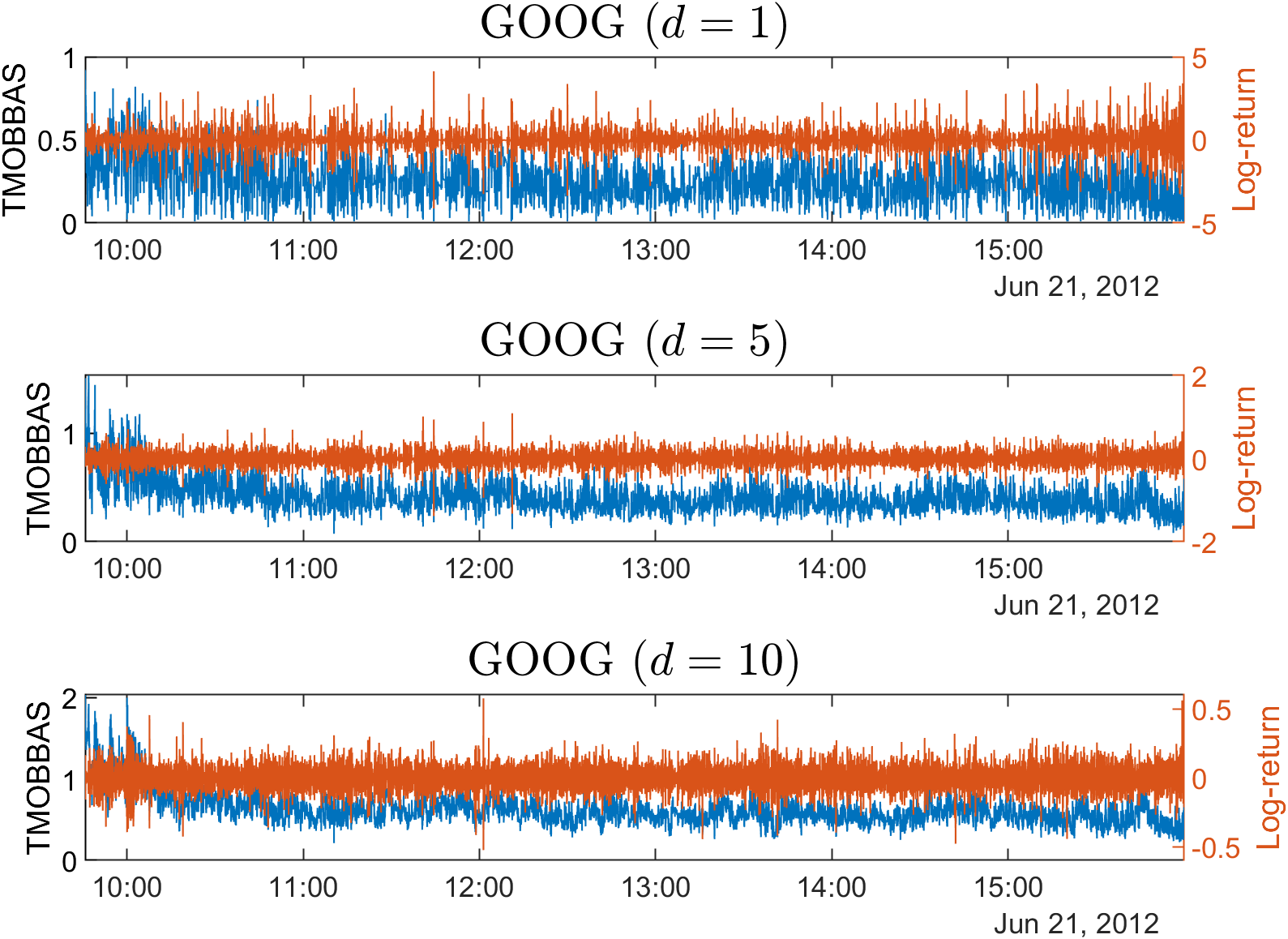}}
\label{figure: TMOBBAS_GOOG}
\subfloat[]{\includegraphics[width=0.49\textwidth]{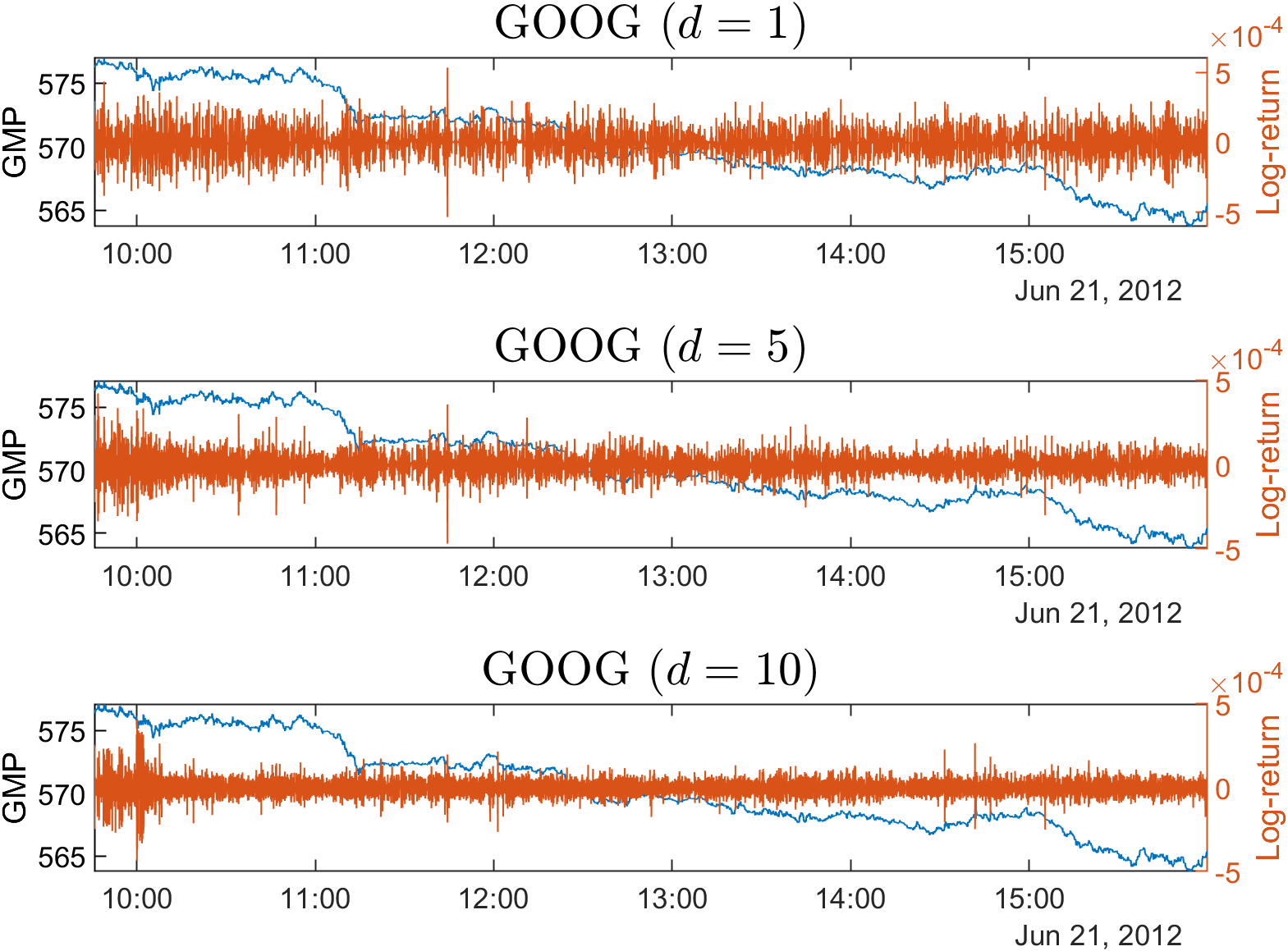}}
\label{figure: GMP_GOOG}
\caption{The evolution of (a) TMOBBAS, (b) GMP and their log-returns over time at depths of 1, 5, and 10 for GOOG (each graph contains two vertical axes. The left vertical axis is for TMOBBAS (or GMP), and the right vertical axis is for the log-return. The evolution of TMOBBAS (or GMP) is represented by the blue line while the orange line the evolution of the log returns)}
\label{figure: TMOBBAS and GMP_GOOG}
\end{figure}
\clearpage

\subsubsection{Comparison of the Kernel Density of Log-Returns of TMOBBAS and GMP with the Corresponding Gaussian Distribution}
\begin{figure}[htbp]
\centering
\subfloat[]{\includegraphics[width=0.8\textwidth, height=0.3\textheight, keepaspectratio]{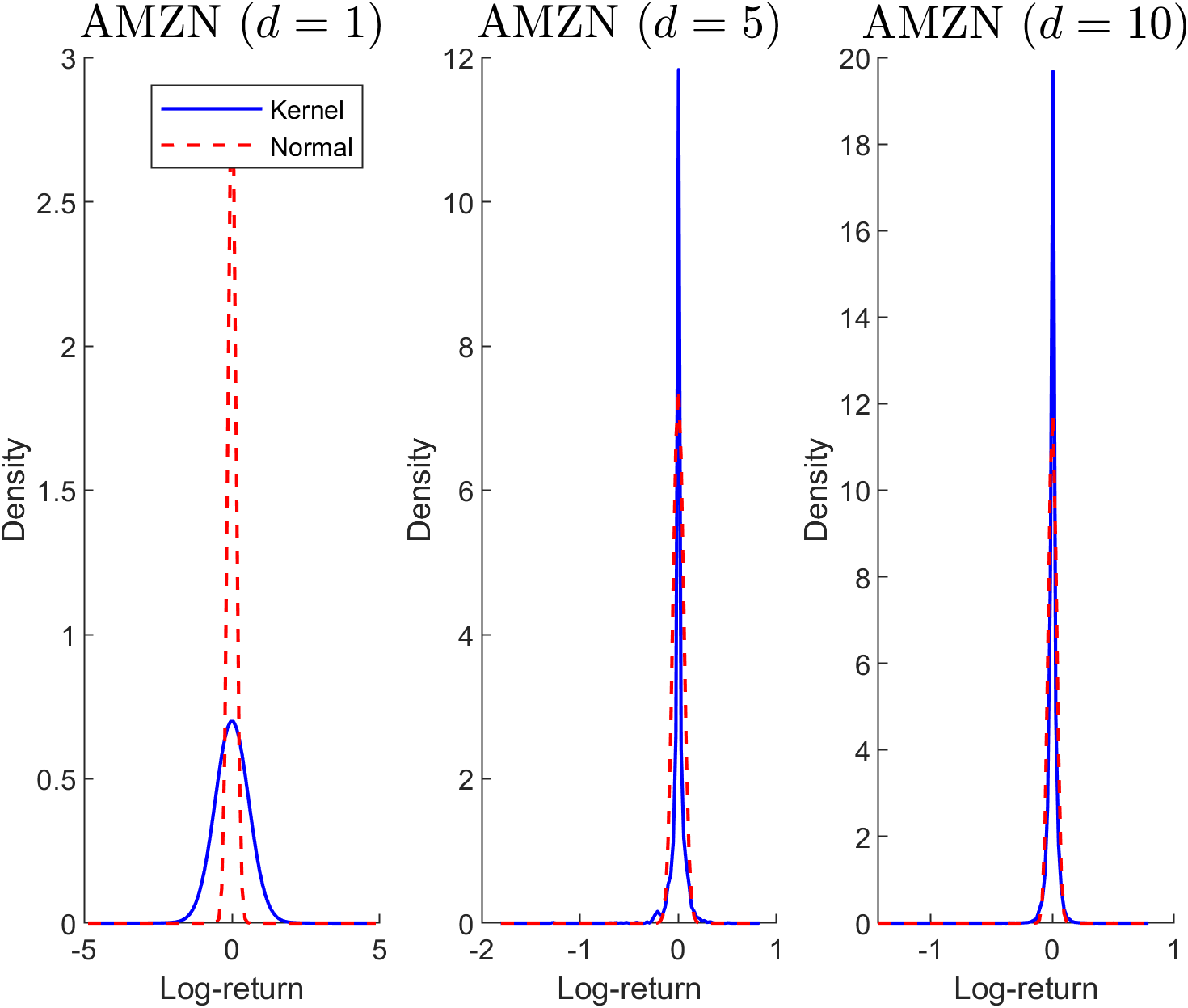}}
\label{figure: KD_TMOBBAS_amzn}
\subfloat[]{\includegraphics[width=0.8\textwidth, height=0.3\textheight, keepaspectratio]{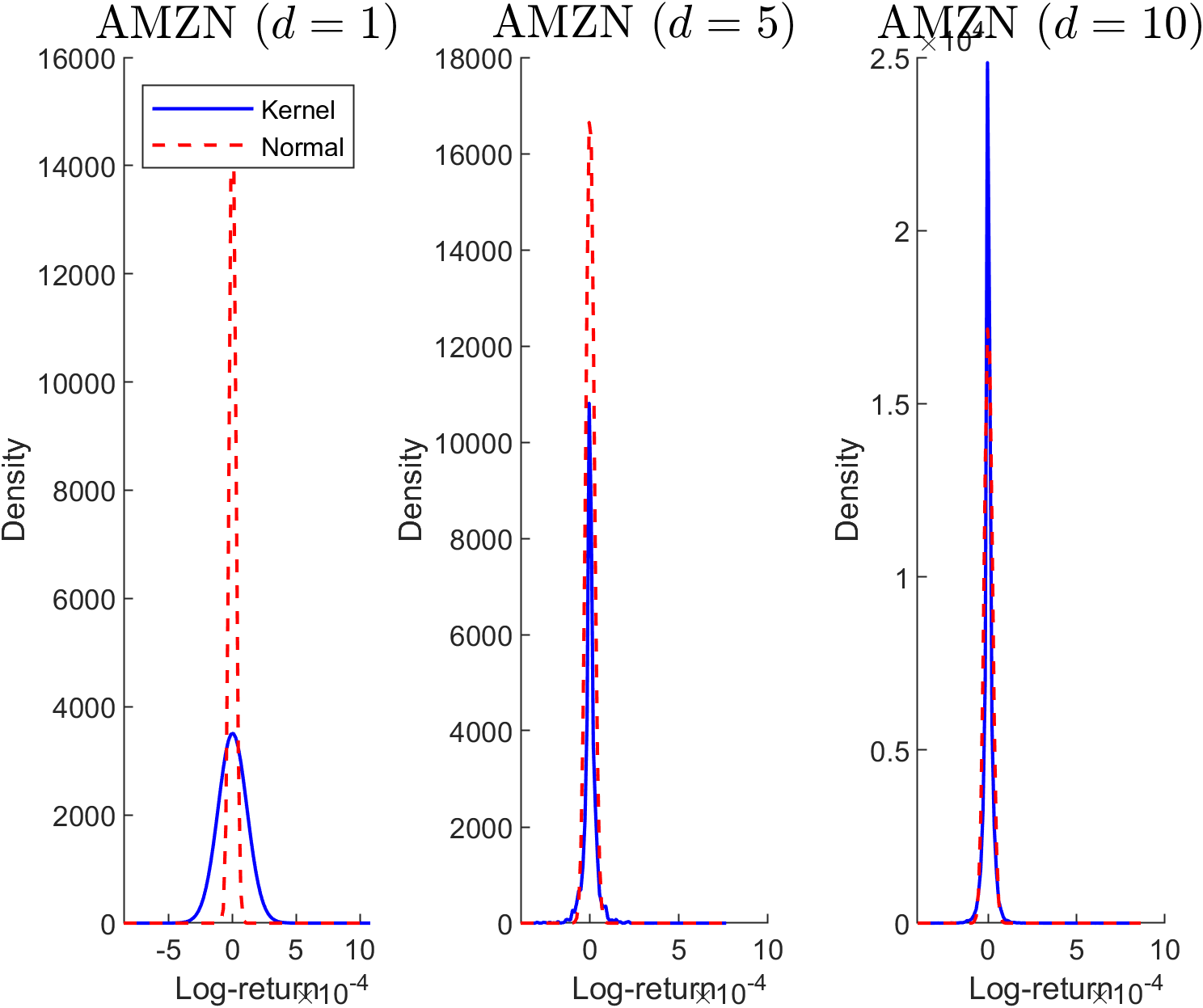}}
\label{figure: KD_GMP_amzn}
\caption{Comparison of the kernel density (represented by black solid lines) of log returns of (a) TMOBBAS, (b) GMP with the corresponding Gaussian distribution with the same sample mean and standard deviation (represented by the red dashed lines) at depths of 1, 5, and 10 for AMZN}
\label{figure: KD_amzn}
\end{figure}

\begin{figure}[htbp]
\centering
\subfloat[]{\includegraphics[width=0.8\textwidth, height = 0.3\textheight, keepaspectratio]{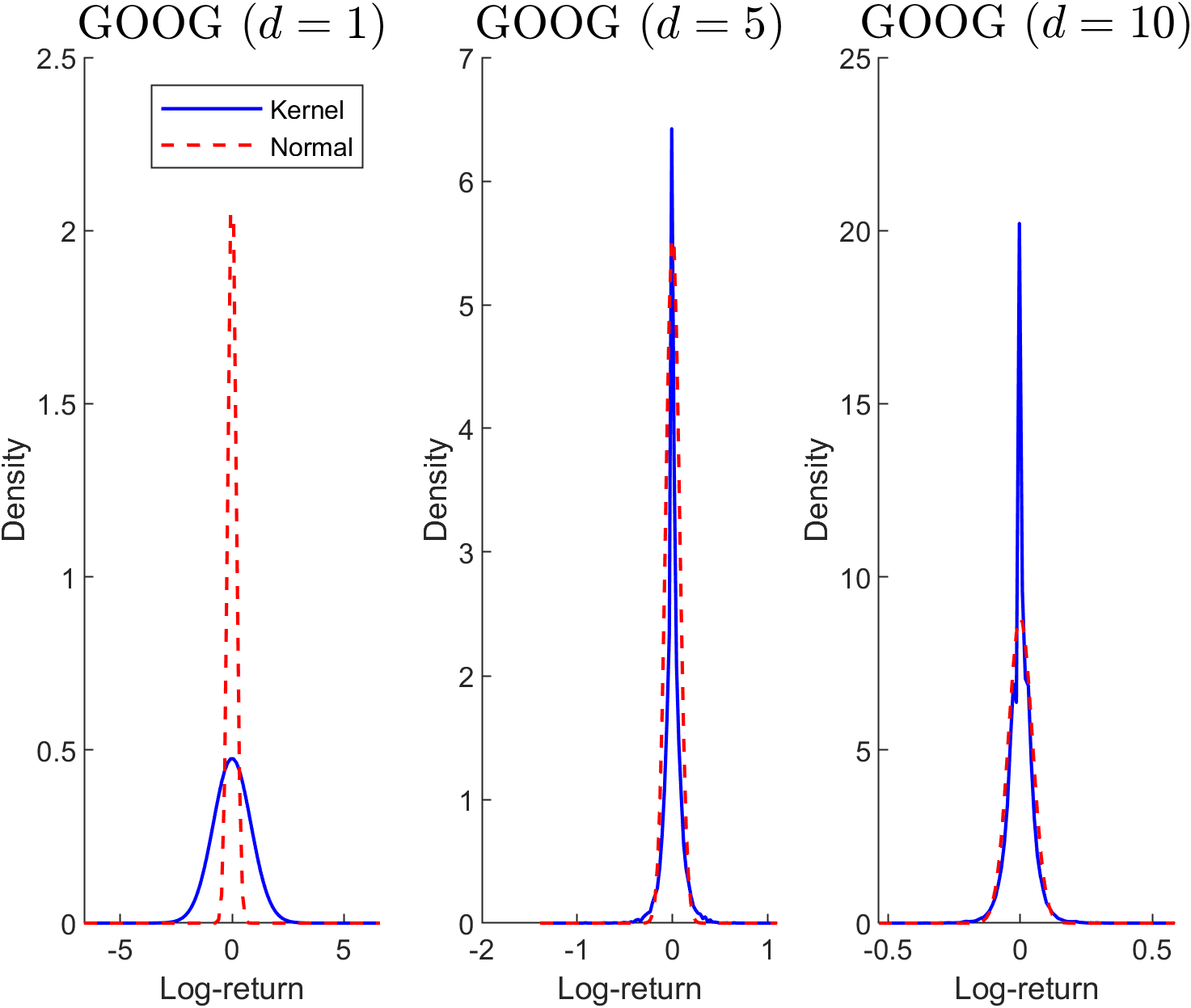}}
\label{figure: KD_TMOBBAS_GOOG}
\subfloat[]{\includegraphics[width=0.8\textwidth, height = 0.3\textheight, keepaspectratio]{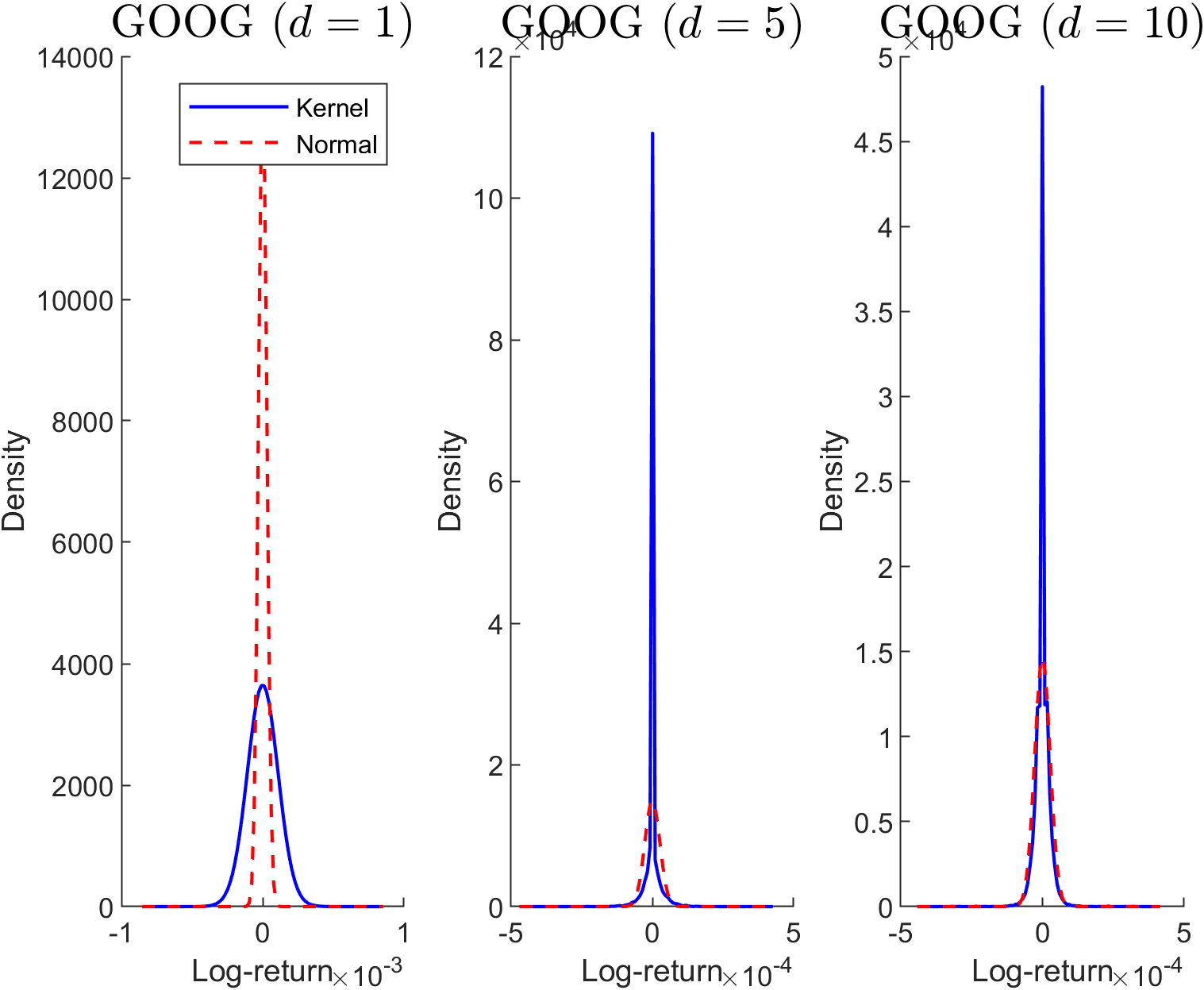}}
\label{figure: KD_GMP_GOOG}
\caption{Comparison of the kernel density (represented by black solid lines) of log returns of (a) TMOBBAS, (b) GMP with the corresponding Gaussian distribution with the same sample mean and standard deviation (represented by the red dashed lines) at depths of 1, 5, and 10 for GOOG}
\label{figure: KD_GOOG}
\end{figure}
\clearpage

\subsubsection{Normal QQ-Plots of the Return Distributions for TMOBBAS and GMP}
\begin{figure}[htbp]
\centering
\subfloat[]{\includegraphics[width=0.8\textwidth, height=0.3\textheight, keepaspectratio]{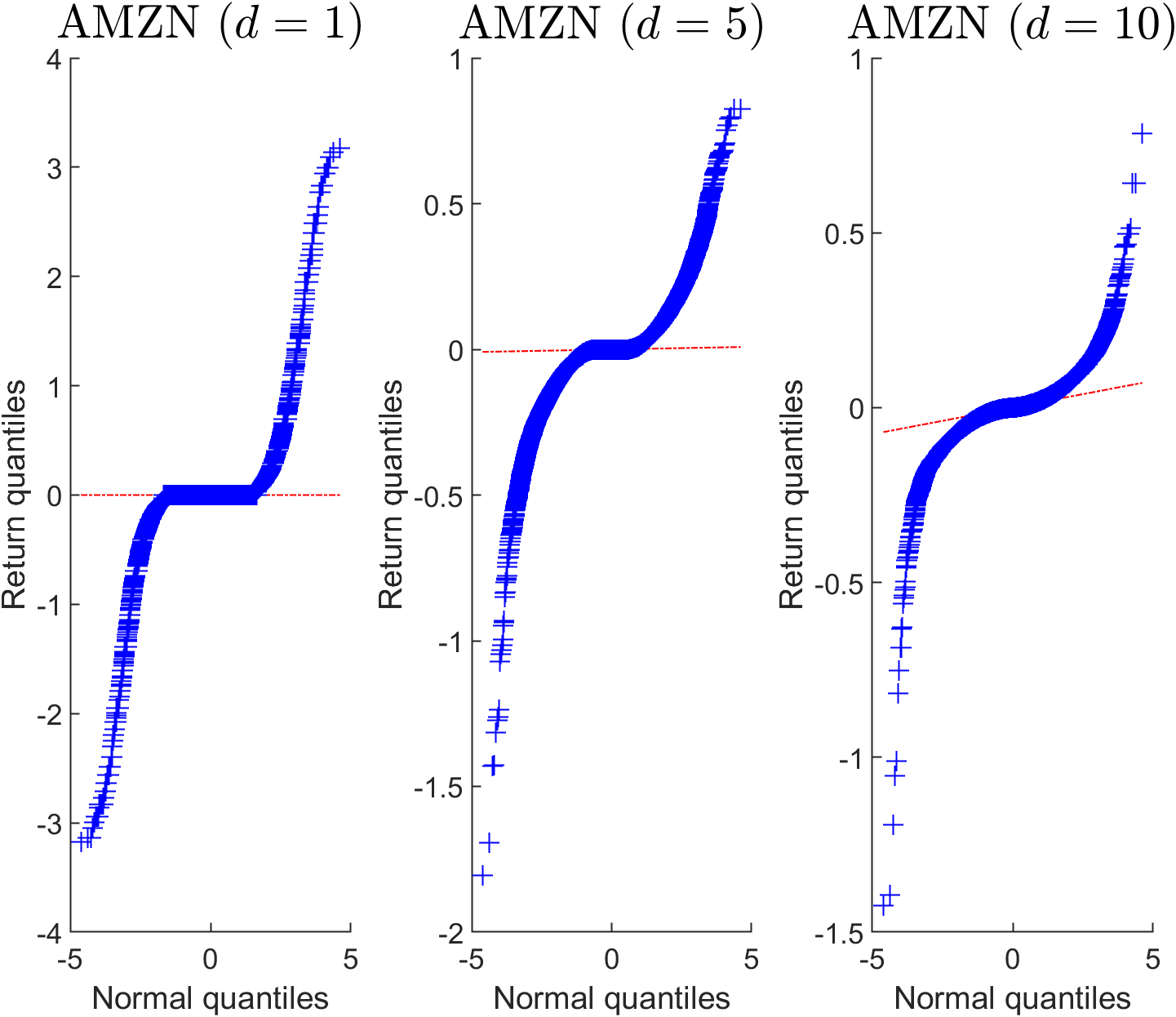}}
\label{figure: tmobbas_qq_plot_amzn}
\subfloat[]{\includegraphics[width=0.8\textwidth, height=0.3\textheight, keepaspectratio]{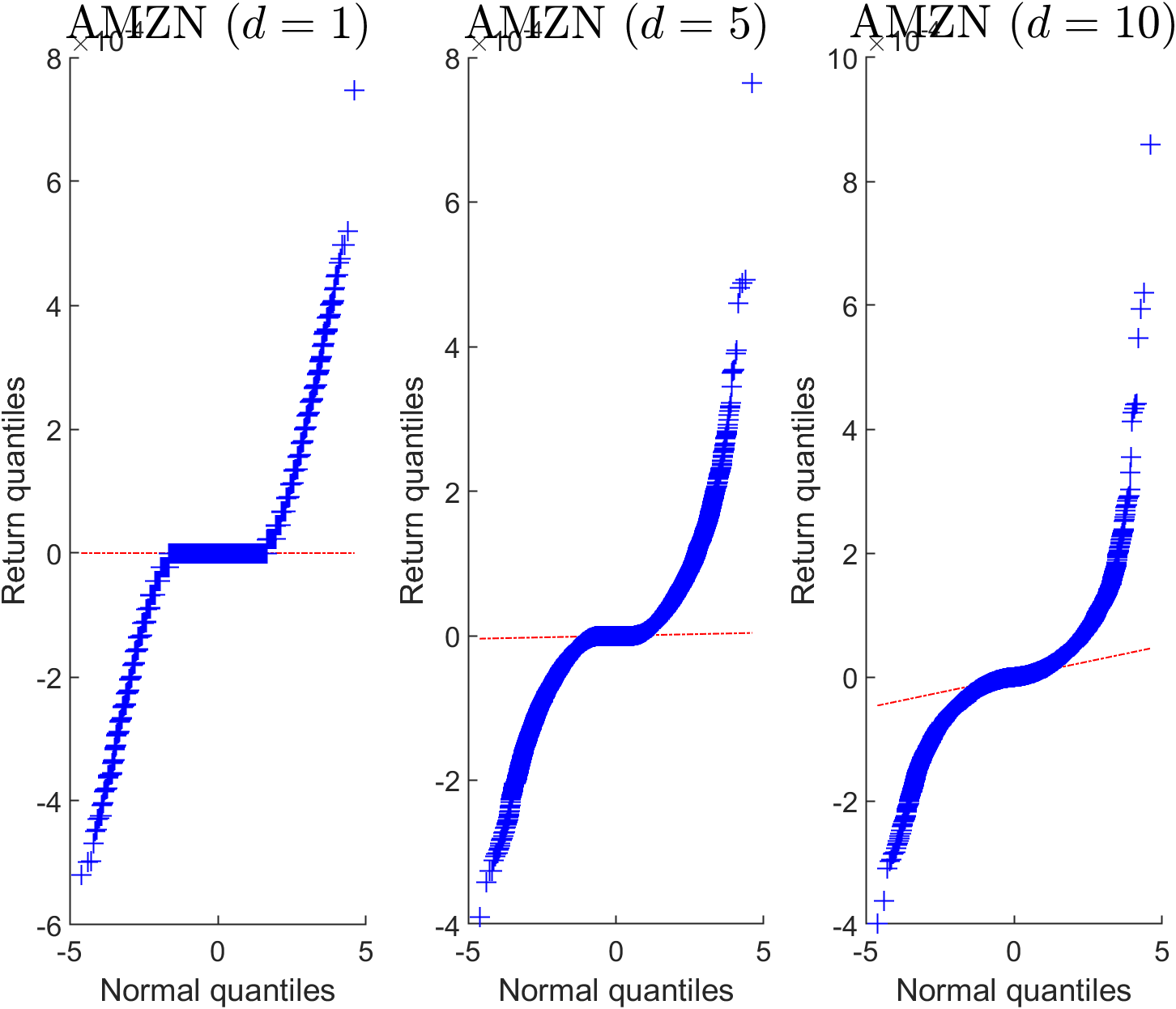}}
\label{figure: gmp_qq_plot_amzn}
\caption{Normal QQ plots of the return distributions for the (a) TMOBBAS and (b) GMP for AMZN at depths of 1, 5, and 10}
\label{figure: qq_plot_amzn}
\end{figure}

\begin{figure}[htbp]
\centering
\subfloat[]{\includegraphics[width=0.8\textwidth, height = 0.3\textheight, keepaspectratio]{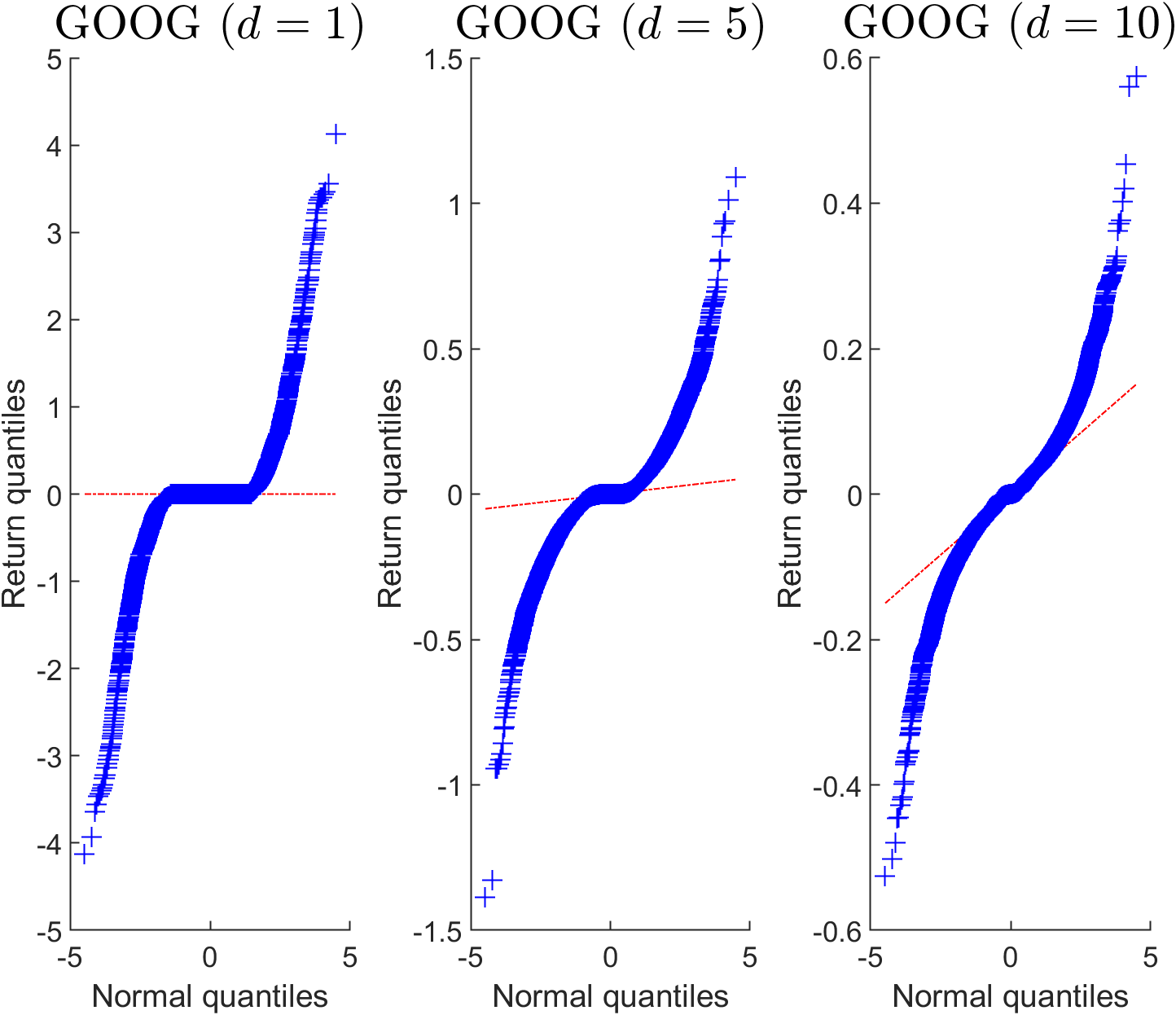}}
\label{figure: tmobbas_qq_plot_goog}
\subfloat[]{\includegraphics[width=0.8\textwidth, height = 0.3\textheight, keepaspectratio]{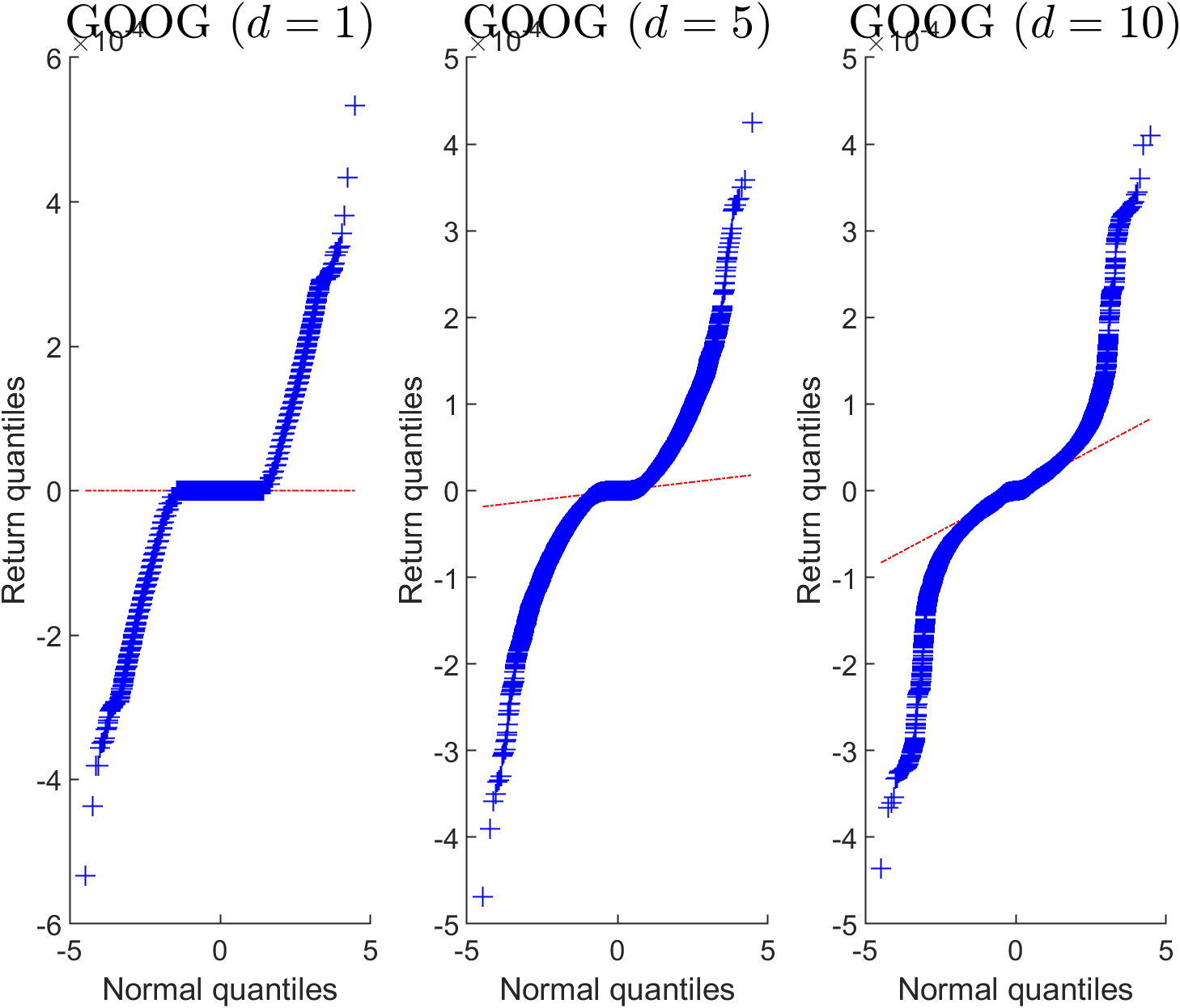}}
\label{figure: gmp_qq_plot_goog}
\caption{Normal QQ plots of the return distributions for the (a) TMOBBAS and (b) GMP for GOOG at depths of 1, 5, and 10}
\label{figure: qq_plot_goog}
\end{figure}
\clearpage

\subsubsection{Comparison of Empirical CDF of Tail and Its Fitted CDF of GPD of TMOBBAS and GMP}
\begin{figure}[htbp]
\centering
\subfloat[]{\includegraphics[width=0.8\textwidth, height=0.3\textheight, keepaspectratio]{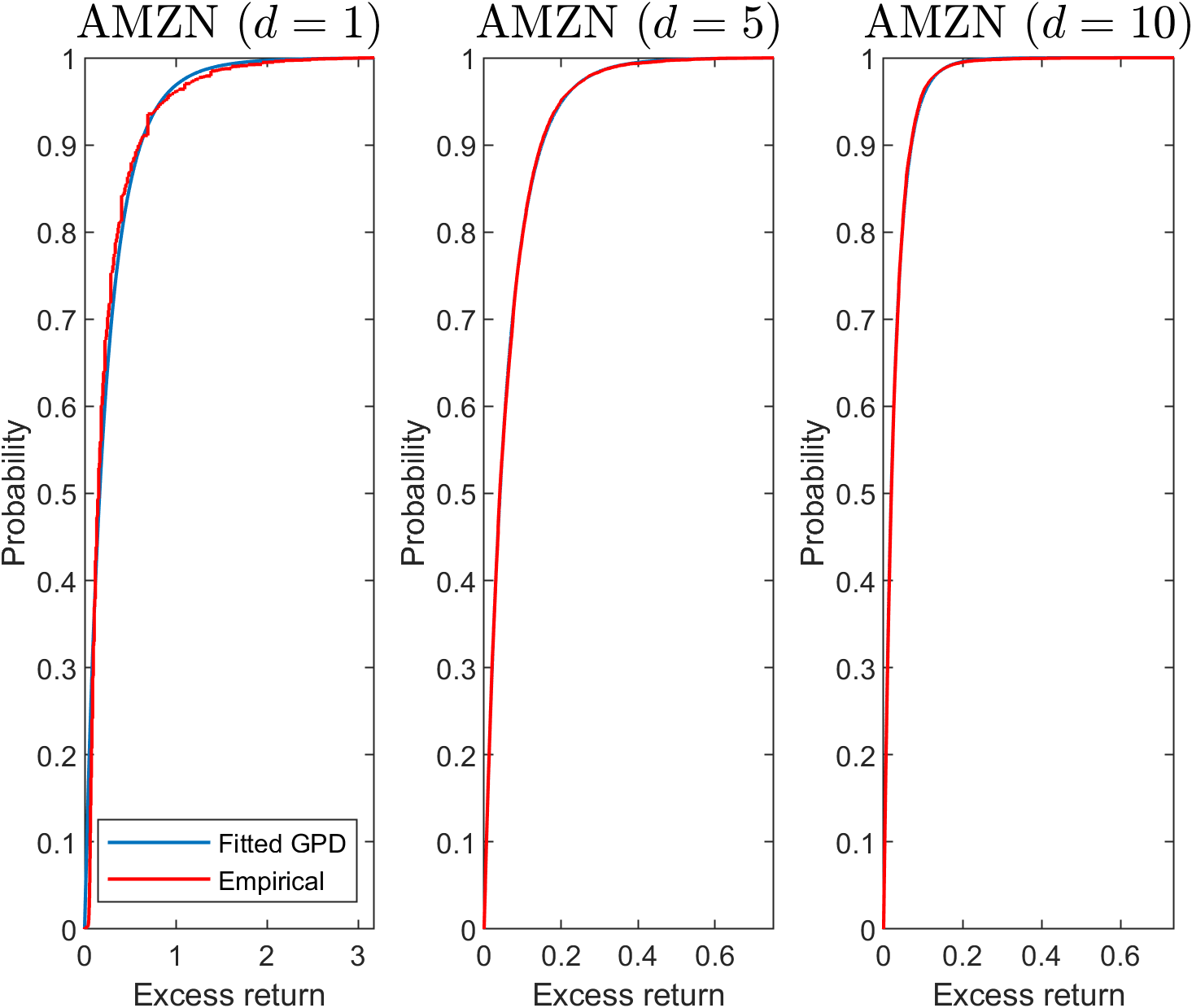}}
\label{figure: TB_TMOBBAS_amzn}
\subfloat[]{\includegraphics[width=0.8\textwidth, height=0.3\textheight, keepaspectratio]{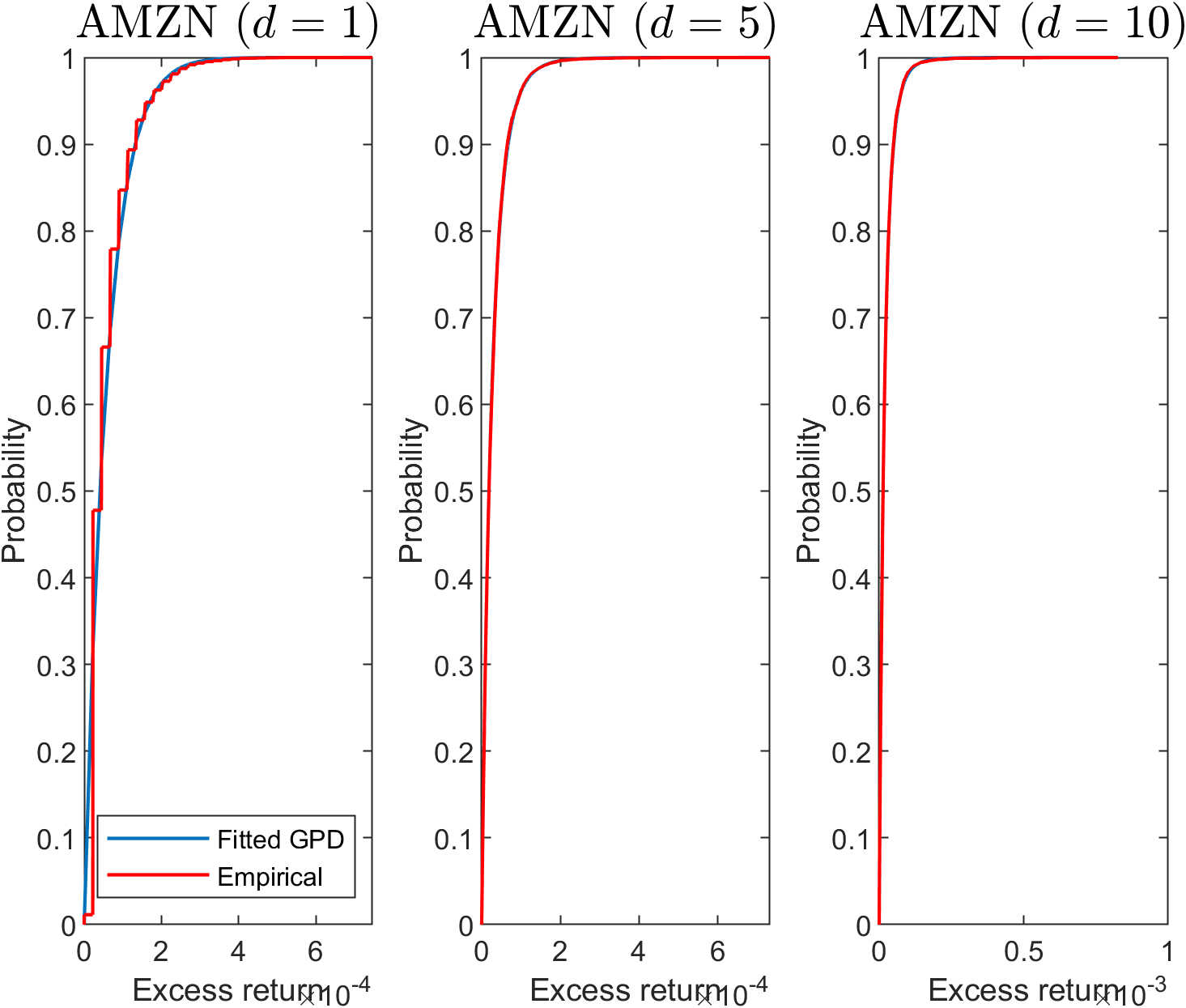}}
\label{figure: TB_GMP_amzn}
\caption{Comparison of empirical CDF (represented by black solid lines) of tail and its fitted CDF of GPD of (a) TMOBBAS and (b) GMP at depths of 1, 5, and 10 for AMZN}
\label{figure: TB_amzn}
\end{figure}

\begin{figure}[htbp]
\centering
\subfloat[]{\includegraphics[width=0.8\textwidth, height = 0.3\textheight, keepaspectratio]{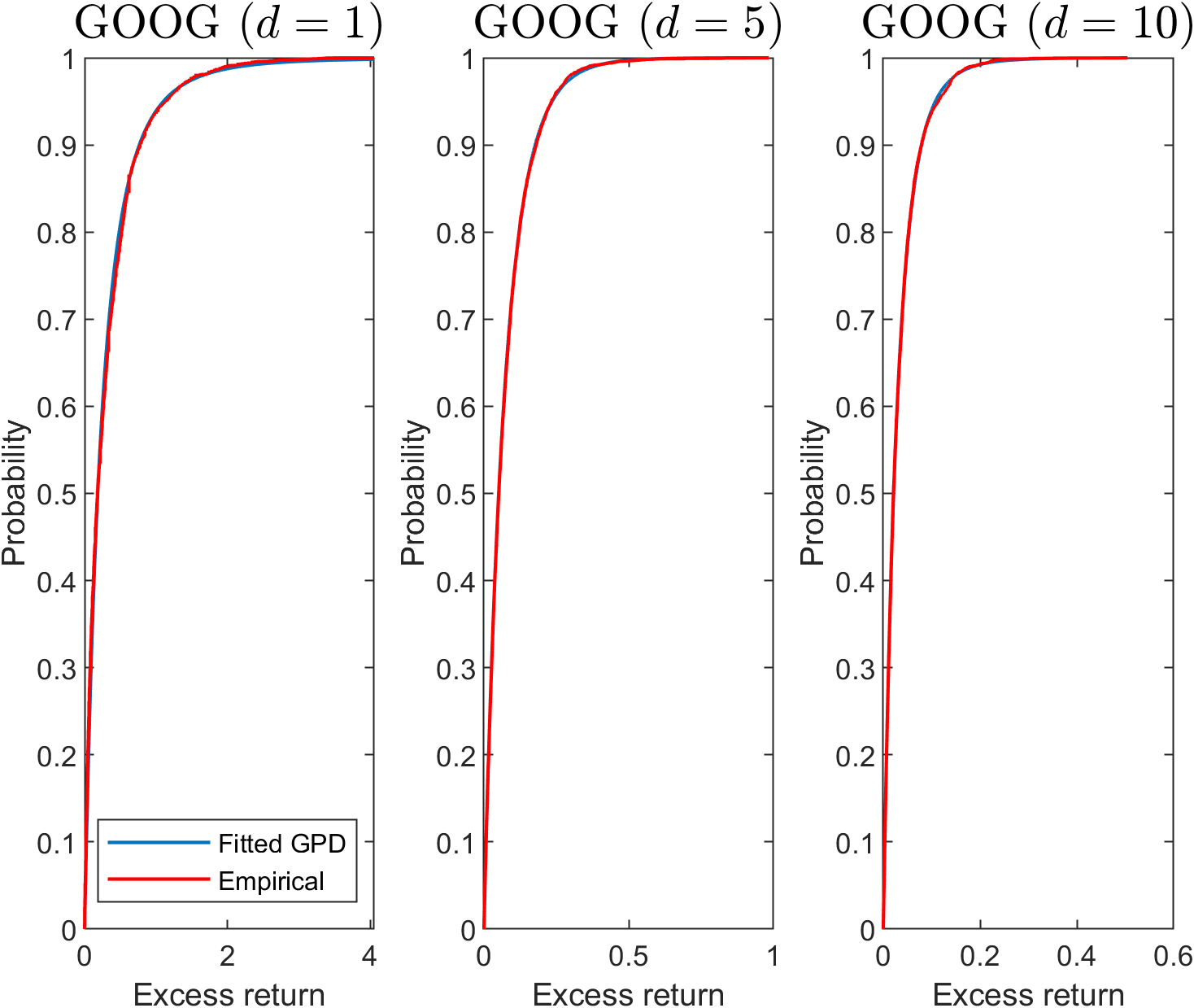}}
\label{figure: TB_TMOBBAS_GOOG}
\subfloat[]{\includegraphics[width=0.8\textwidth, height = 0.3\textheight, keepaspectratio]{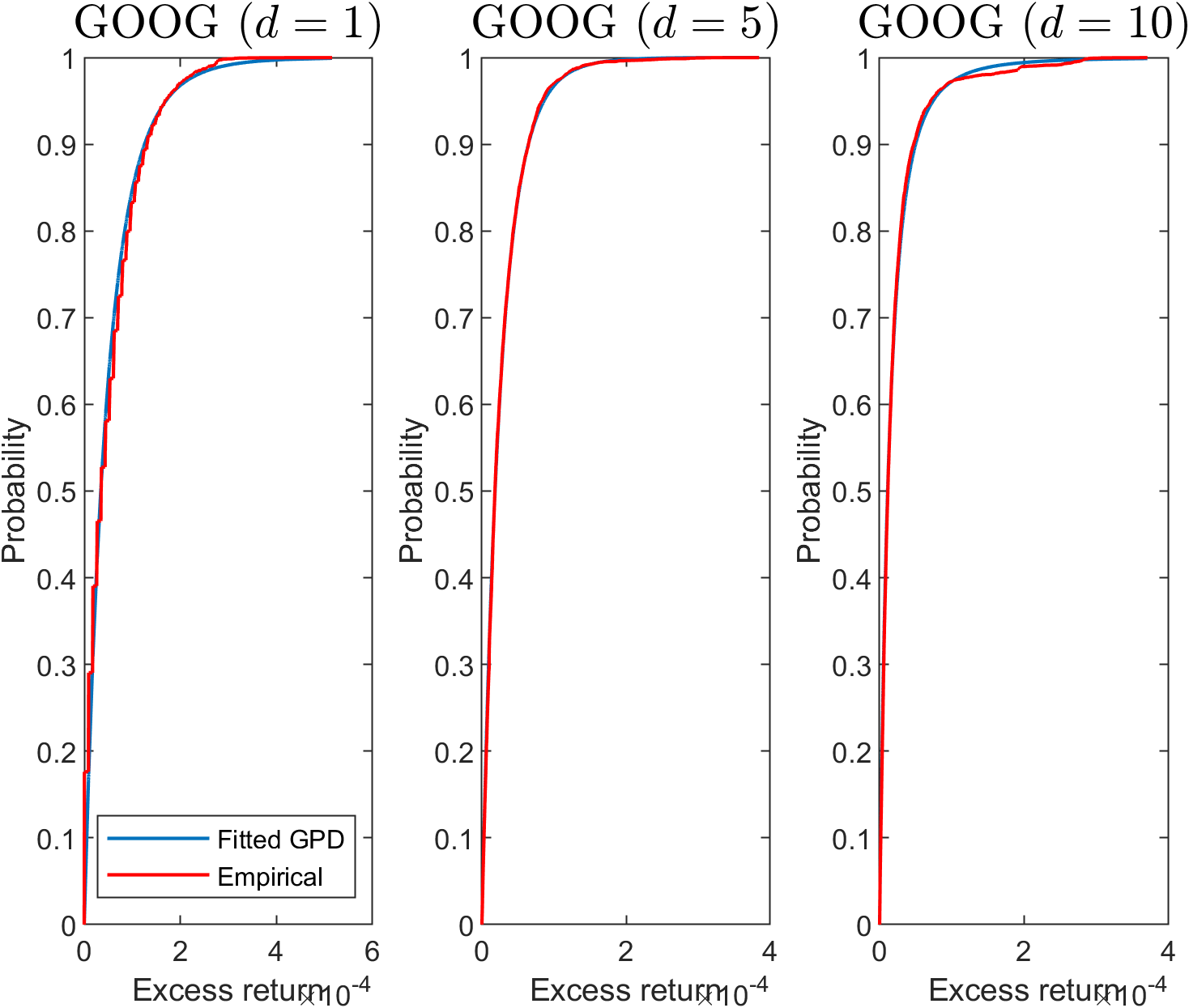}}
\label{figure: TB_GMP_GOOG}
\caption{Comparison of empirical CDF (represented by black solid lines) of tail and its fitted CDF of GPD of (a) TMOBBAS and (b) GMP at depths of 1, 5, and 10 for GOOG}
\label{figure: TB_GOOG}
\end{figure}
\clearpage

\subsubsection{The Dependence of Estimated Tail Index and Its Corresponding 95\% Confidence Interval for TMOBBAS and GMP}
\begin{figure}[htbp]
\centering
\subfloat[]{\includegraphics[width=0.8\textwidth, height=0.3\textheight, keepaspectratio]{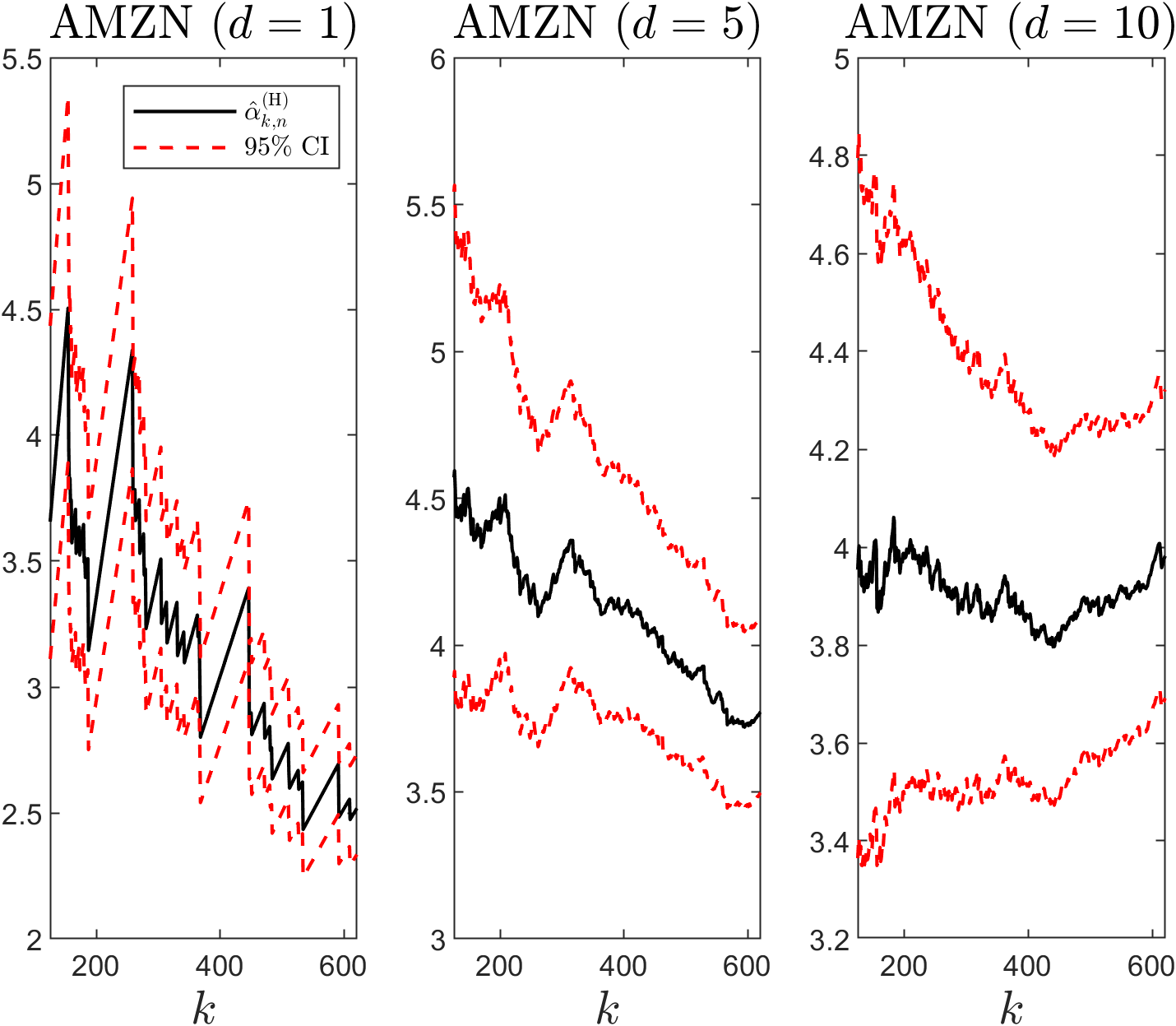}}
\label{figure: Hill_TMOBBAS_amzn}
\subfloat[]{\includegraphics[width=0.8\textwidth, height=0.3\textheight, keepaspectratio]{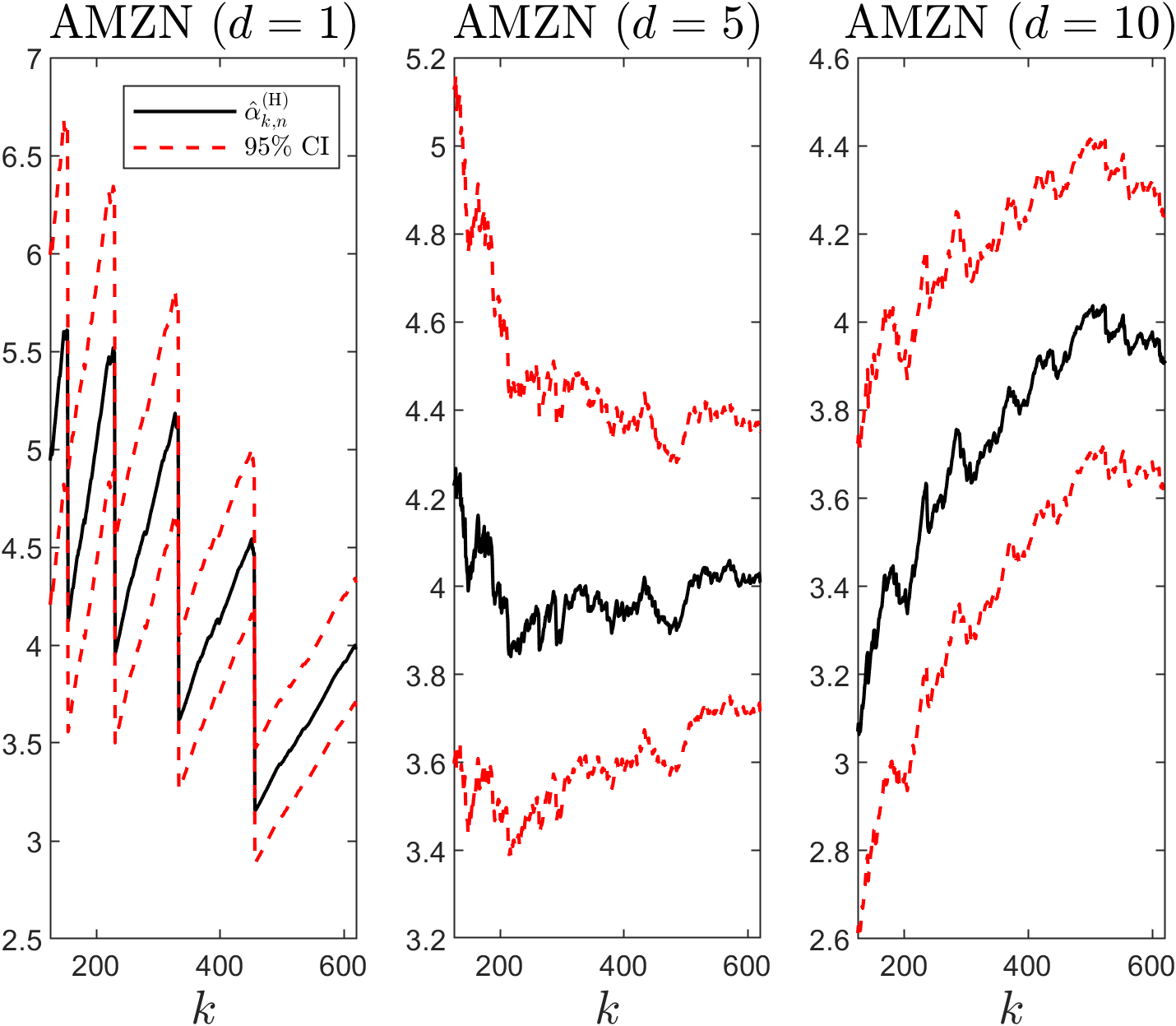}}
\label{figure: Hill_GMP_amzn}
\caption{The dependence of estimated tail index $\alpha$ (represented by blue solid line) and its corresponding 95\% CI (represented by two red dashed lines) for (a) TMOBBAS and (b) GMP on the number of order statistics at depths of 1, 5, and 10 for AMZN}
\label{figure: Hill_amzn}
\end{figure}

\begin{figure}[htbp]
\centering
\subfloat[]{\includegraphics[width=0.8\textwidth, height = 0.3\textheight, keepaspectratio]{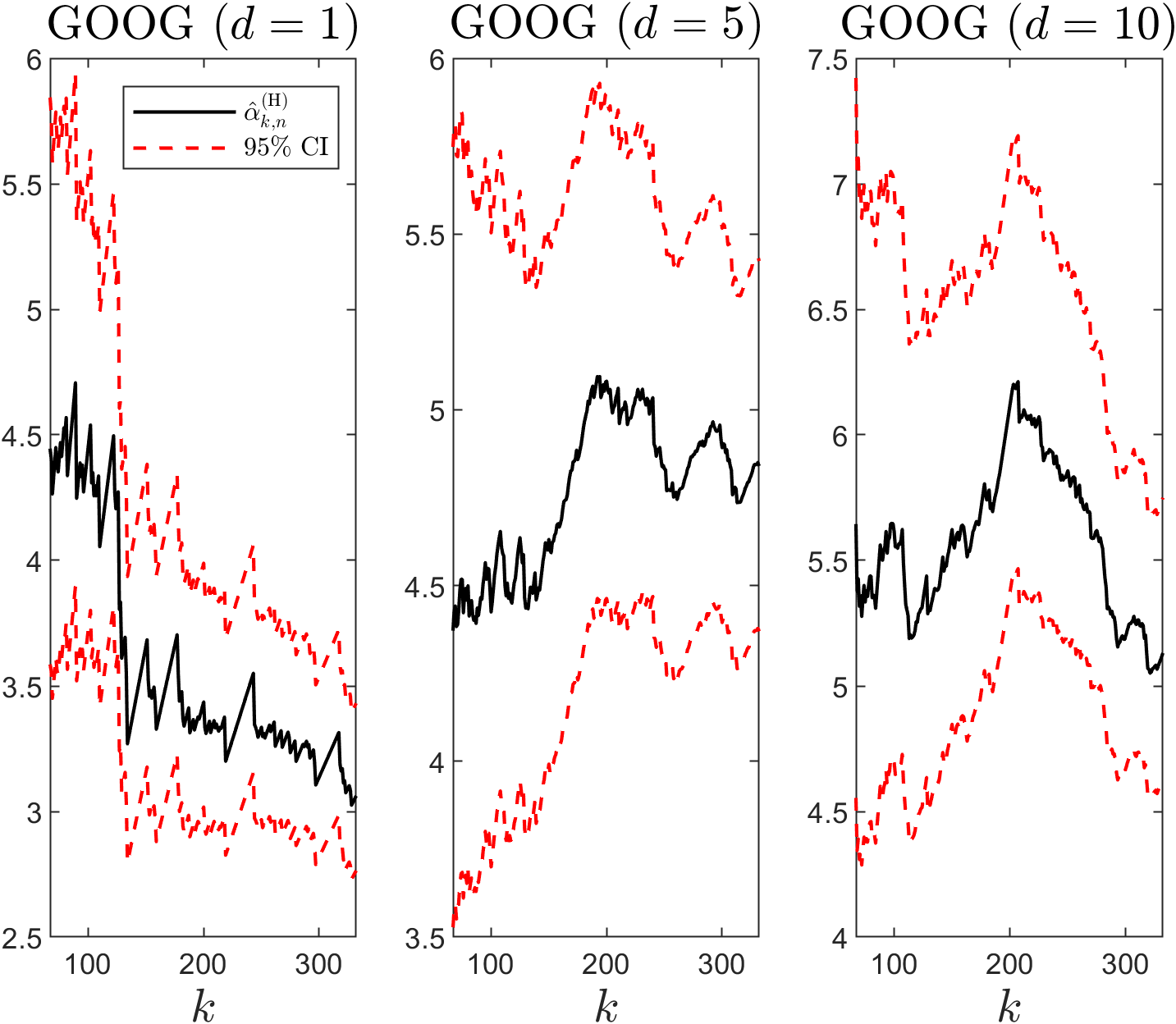}}
\label{figure: Hill_TMOBBAS_GOOG}
\subfloat[]{\includegraphics[width=0.8\textwidth, height = 0.3\textheight, keepaspectratio]{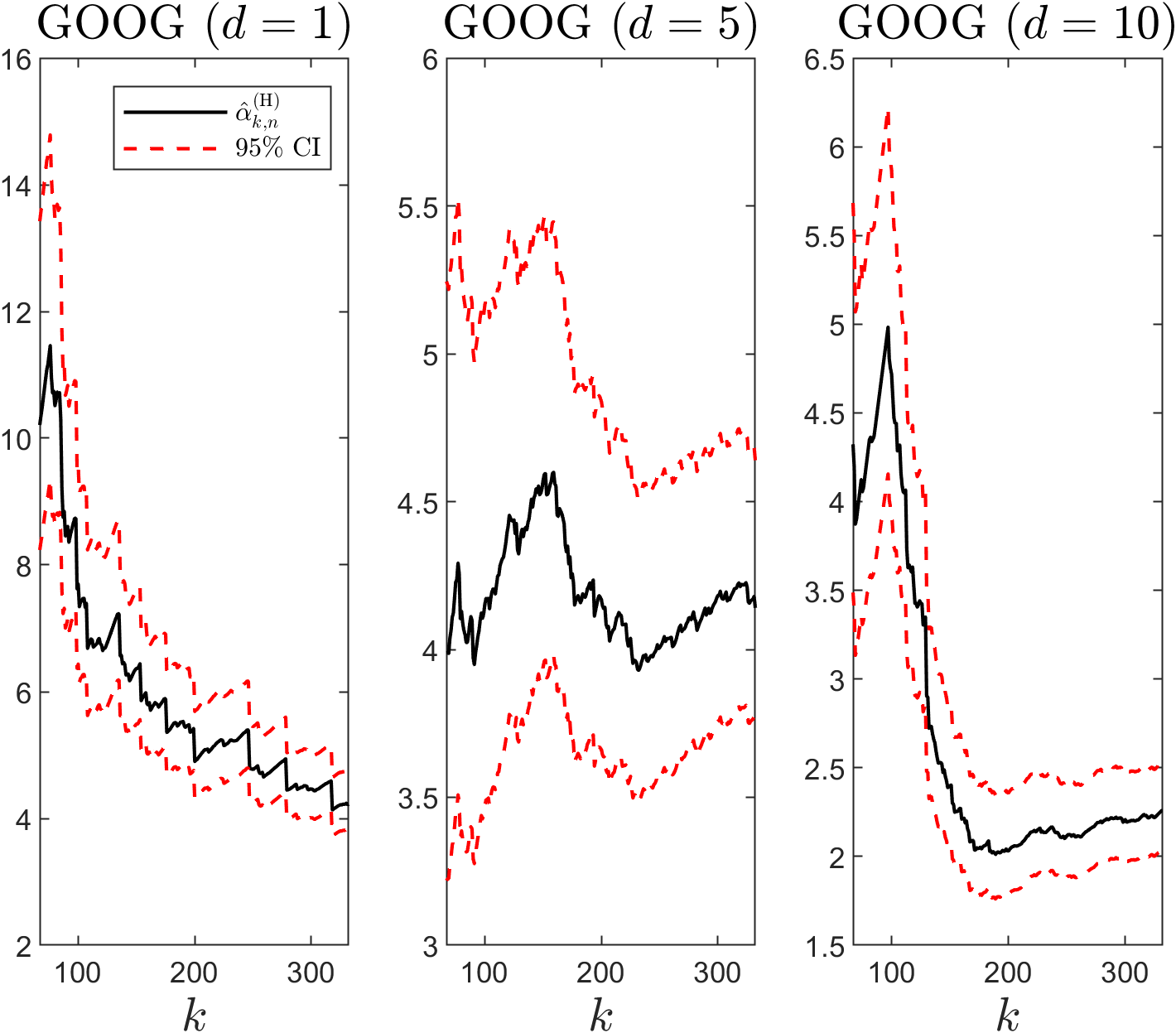}}
\label{figure: Hill_GMP_GOOG}
\caption{The dependence of estimated tail index $\alpha$ (represented by blue solid line) and its corresponding 95\% CI (represented by two red dashed lines) for (a) TMOBBAS and (b) GMP on the number of order statistics at depths of 1, 5, and 10 for GOOG}
\label{figure: Hill_GOOG}
\end{figure}
\clearpage

\subsection{Dynamic Method}
\subsubsection{The Dependence of Estimated Tail Index and Its Corresponding 95\% Confidence Interval for TMOBBAS and GMP}

\begin{figure}[htbp]
\centering
\subfloat[]{\includegraphics[width=0.24\textwidth, height=0.24\textheight, keepaspectratio]{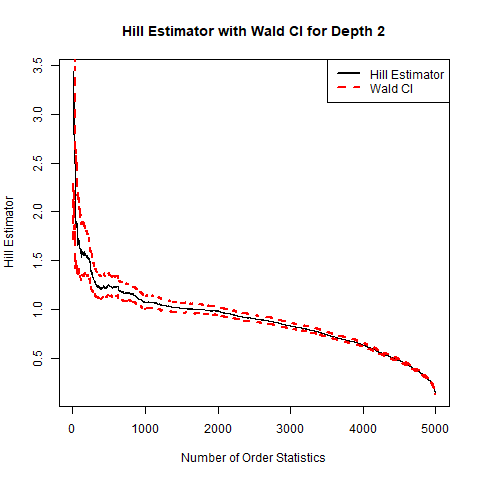}}
\subfloat[]{\includegraphics[width=0.24\textwidth, height=0.24\textheight, keepaspectratio]{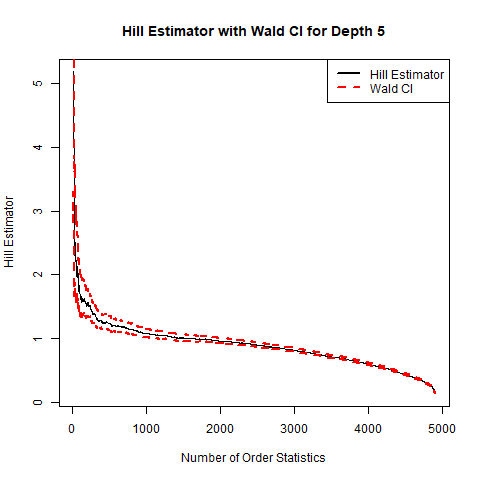}}
\subfloat[]{\includegraphics[width=0.24\textwidth, height=0.24\textheight, keepaspectratio]{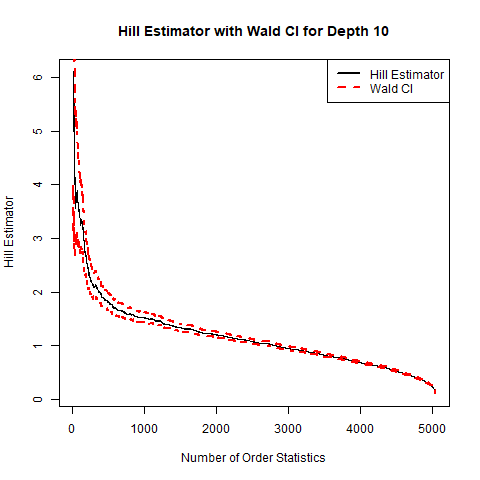}}
\subfloat[]{\includegraphics[width=0.24\textwidth, height=0.24\textheight, keepaspectratio]{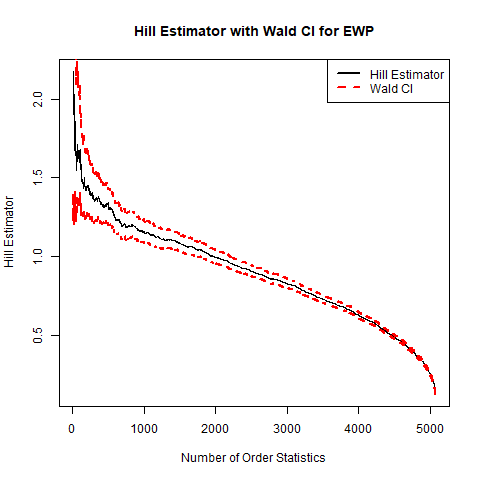}}

\subfloat[]{\includegraphics[width=0.24\textwidth, height=0.24\textheight, keepaspectratio]{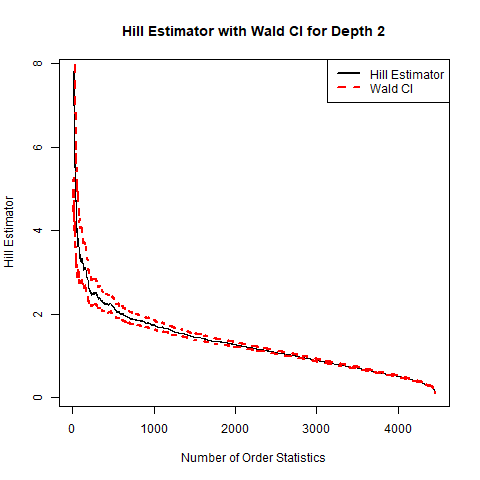}}
\label{GMP_hill_estimator_plot_AMZN_2}
\subfloat[]{\includegraphics[width=0.24\textwidth, height=0.24\textheight, keepaspectratio]{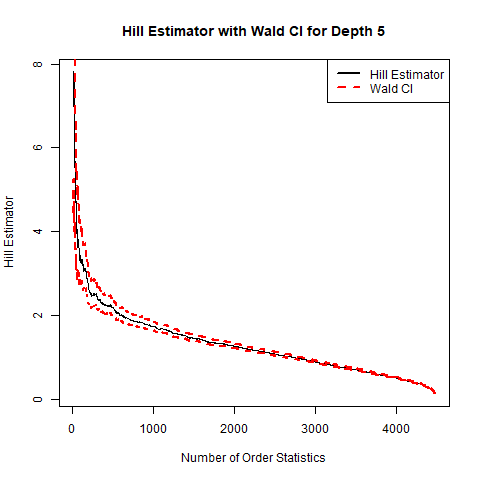}}
\label{GMP_hill_estimator_plot_AMZN_5}
\subfloat[]{\includegraphics[width=0.24\textwidth, height=0.24\textheight, keepaspectratio]{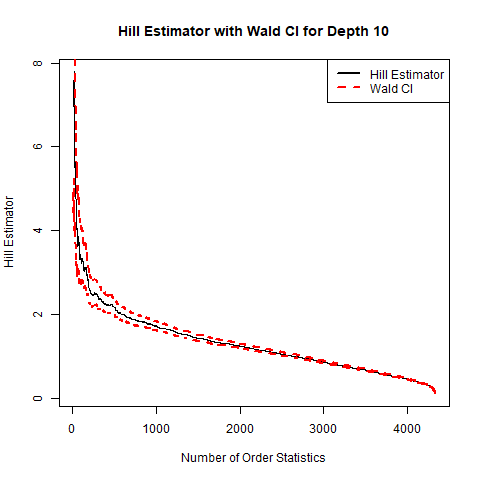}}
\label{GMP_hill_estimator_plot_AMZN_10}
\subfloat[]{\includegraphics[width=0.24\textwidth, height=0.24\textheight, keepaspectratio]{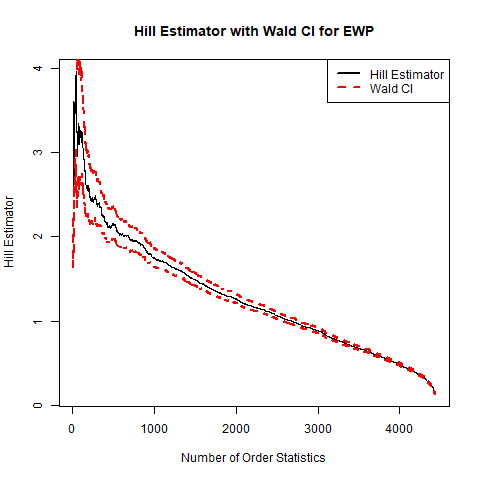}}
\caption{The dependence of estimated tail index $\alpha$ (represented by black solid lines)  and its corresponding 95\% CI (represented by two red dashed lines) on the number of order statistics at depths of 2, 5, 10, and EWP for (a--d) TMOBBAS and (e--h) GMP for AMZN}
\label{Hill_AMZN_dyno}
\end{figure}

\begin{figure}[htbp]
\centering
\subfloat[]{\includegraphics[width=0.24\textwidth, height=0.24\textheight, keepaspectratio]{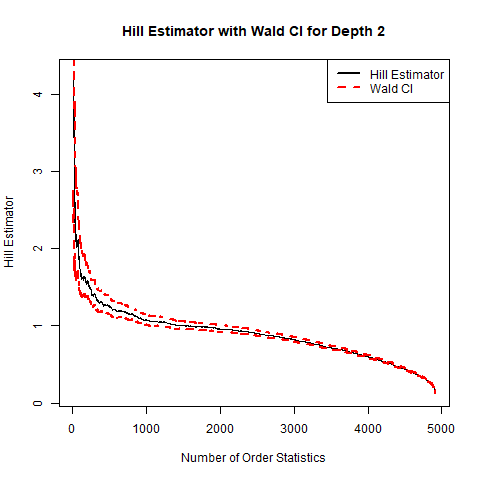}}
\label{hill_estimator_plot_2_GOOG}
\subfloat[]{\includegraphics[width=0.24\textwidth, height=0.24\textheight, keepaspectratio]{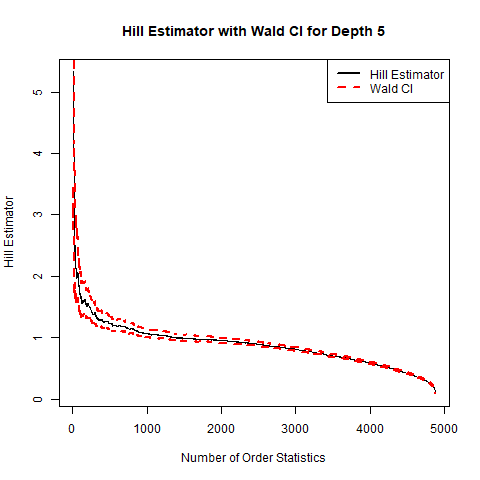}}
\label{hill_estimator_plot_5_GOOG}
\subfloat[]{\includegraphics[width=0.24\textwidth, height=0.24\textheight, keepaspectratio]{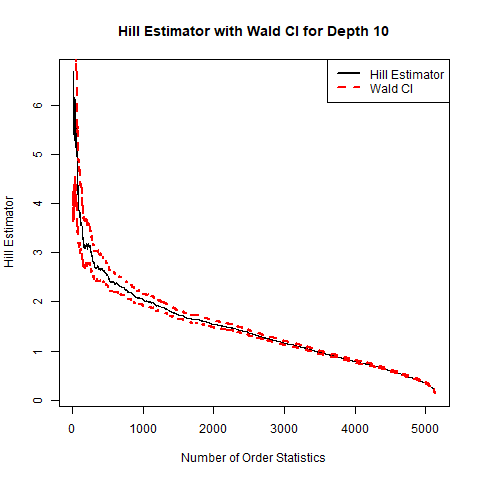}}
\label{hill_estimator_plot_10_GOOG}
\subfloat[]{\includegraphics[width=0.24\textwidth, height=0.24\textheight, keepaspectratio]{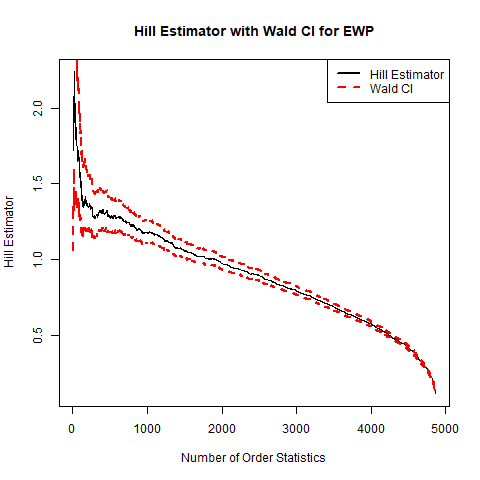}}

\subfloat[]{\includegraphics[width=0.24\textwidth, height=0.24\textheight, keepaspectratio]{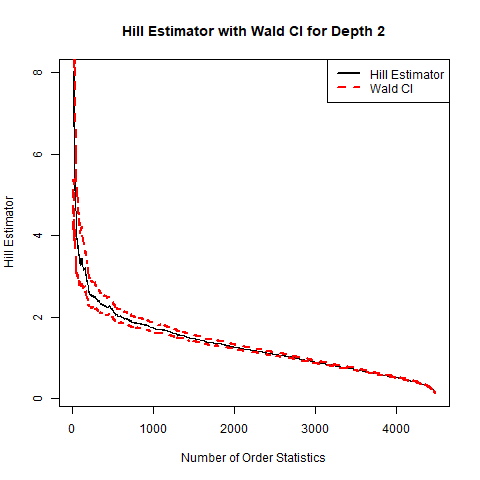}}
\label{GMP_hill_estimator_plot_GOOG_2}
\subfloat[]{\includegraphics[width=0.24\textwidth, height=0.24\textheight, keepaspectratio]{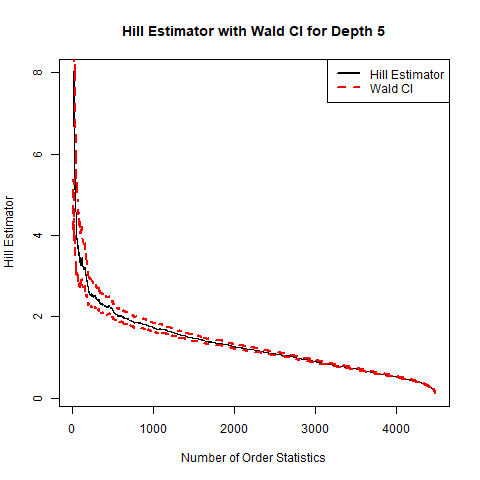}}
\label{GMP_hill_estimator_plot_GOOG_5}
\subfloat[]{\includegraphics[width=0.24\textwidth, height=0.24\textheight, keepaspectratio]{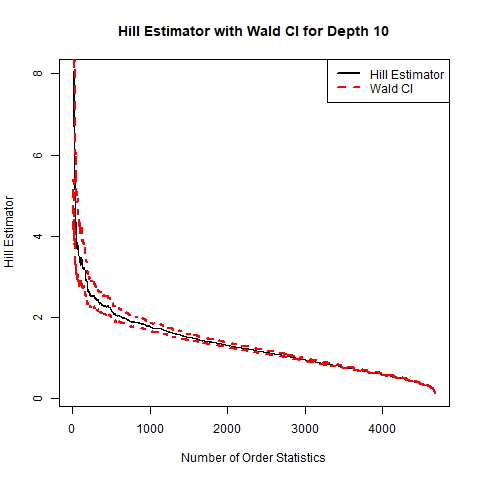}}
\label{GMP_hill_estimator_plot_GOOG_10}
\subfloat[]{\includegraphics[width=0.24\textwidth, height=0.24\textheight, keepaspectratio]{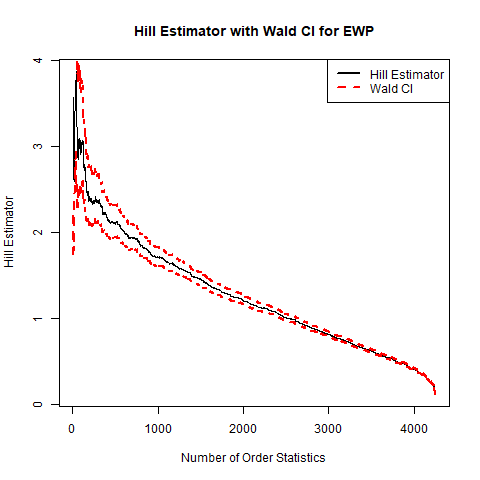}}
\caption{The dependence of estimated tail index $\alpha$ (represented by black solid lines)  and its corresponding 95\% CI (represented by two red dashed lines) on the number of order statistics at depths of 2, 5, 10, and EWP for (a--d) TMOBBAS and (e--h) GMP for GOOG}
\label{Hill_GOOG_dyno}
\end{figure}
\clearpage

\subsubsection{Implied Volatility Surfaces for TMOBBAS and GMP}

\begin{figure}[htbp]
\centering
\subfloat[]{\includegraphics[width=0.24\textwidth]{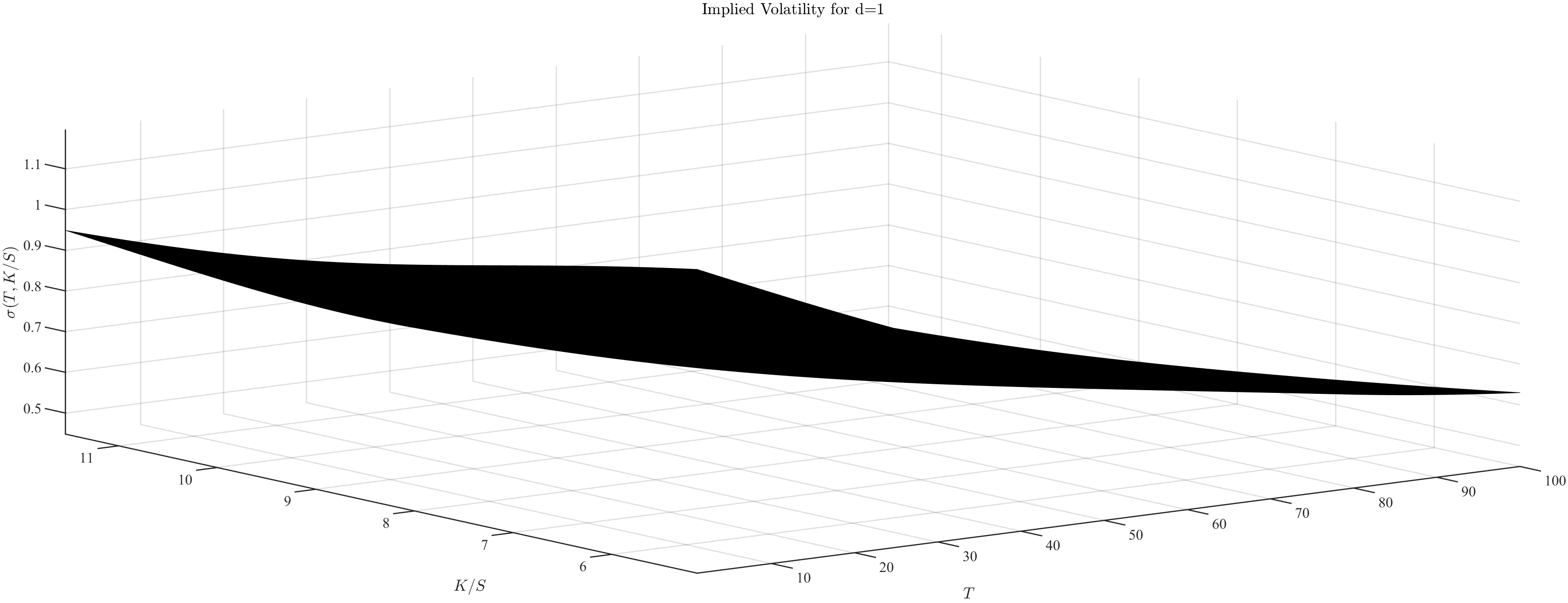}}
\subfloat[]{\includegraphics[width=0.24\textwidth]{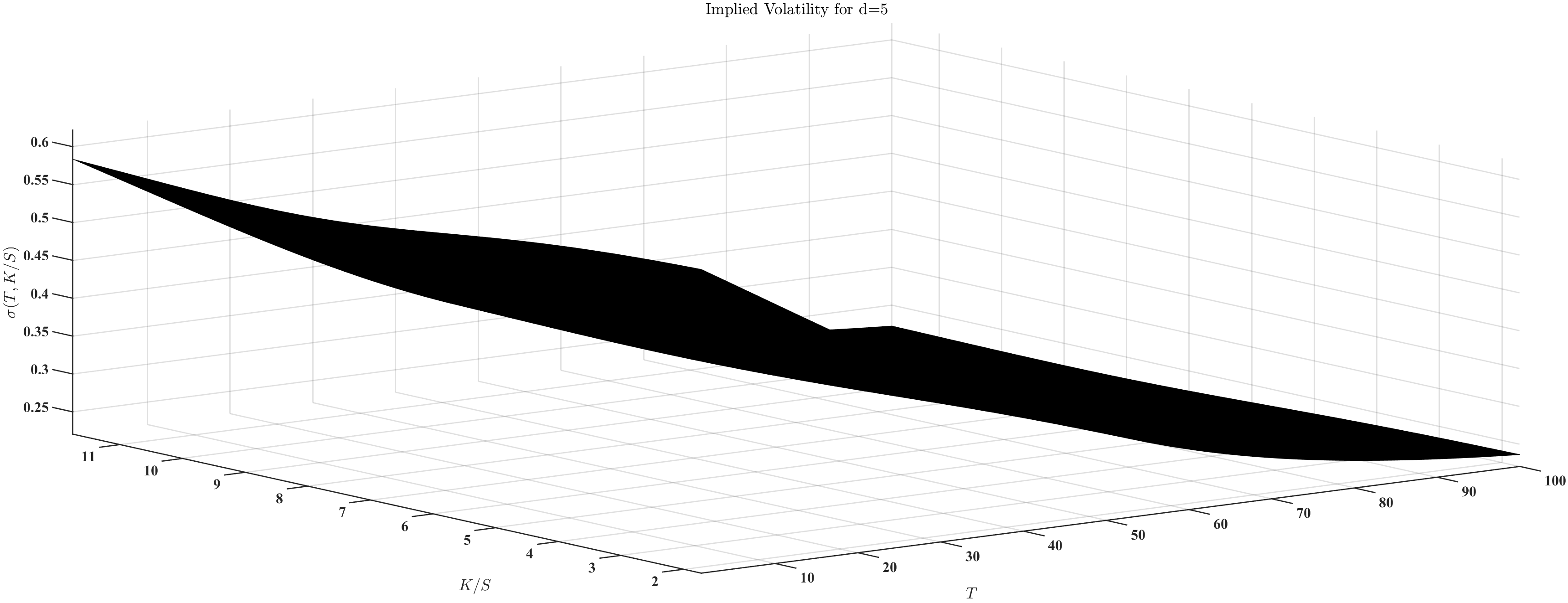}}
\subfloat[]{\includegraphics[width=0.24\textwidth]{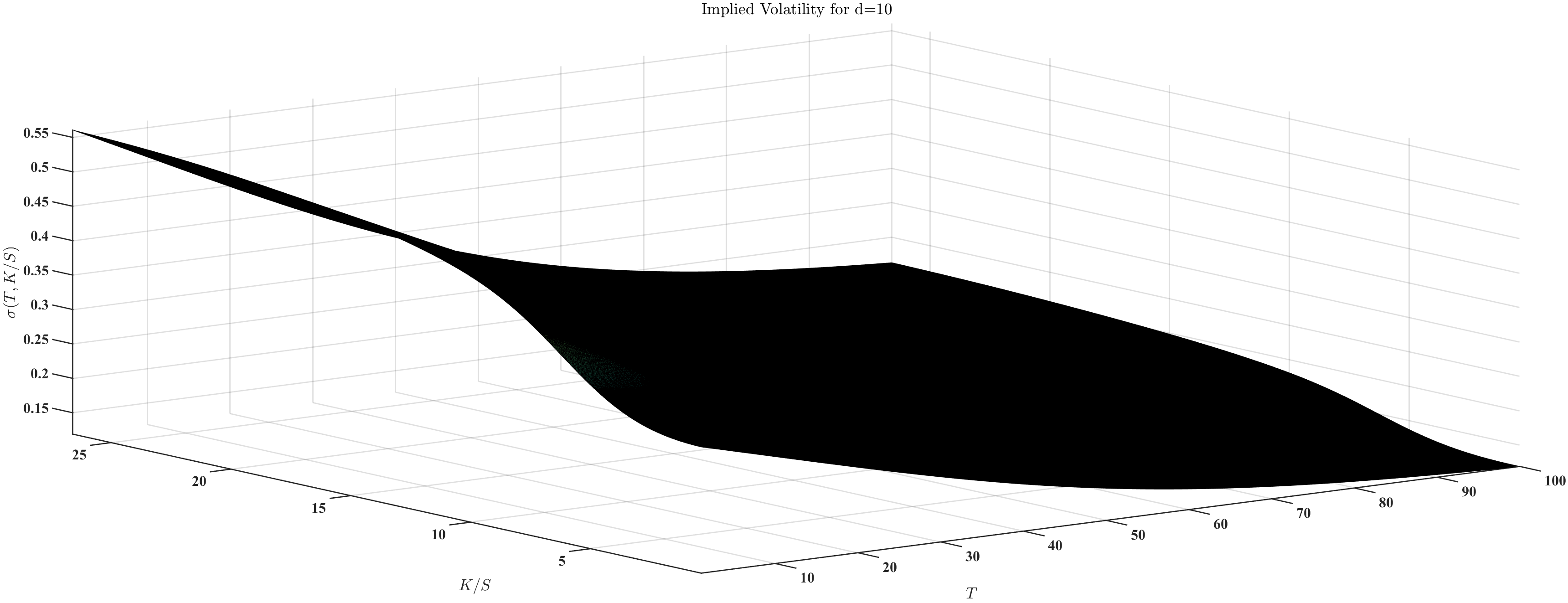}}
\subfloat[]{\includegraphics[width=0.24\textwidth]{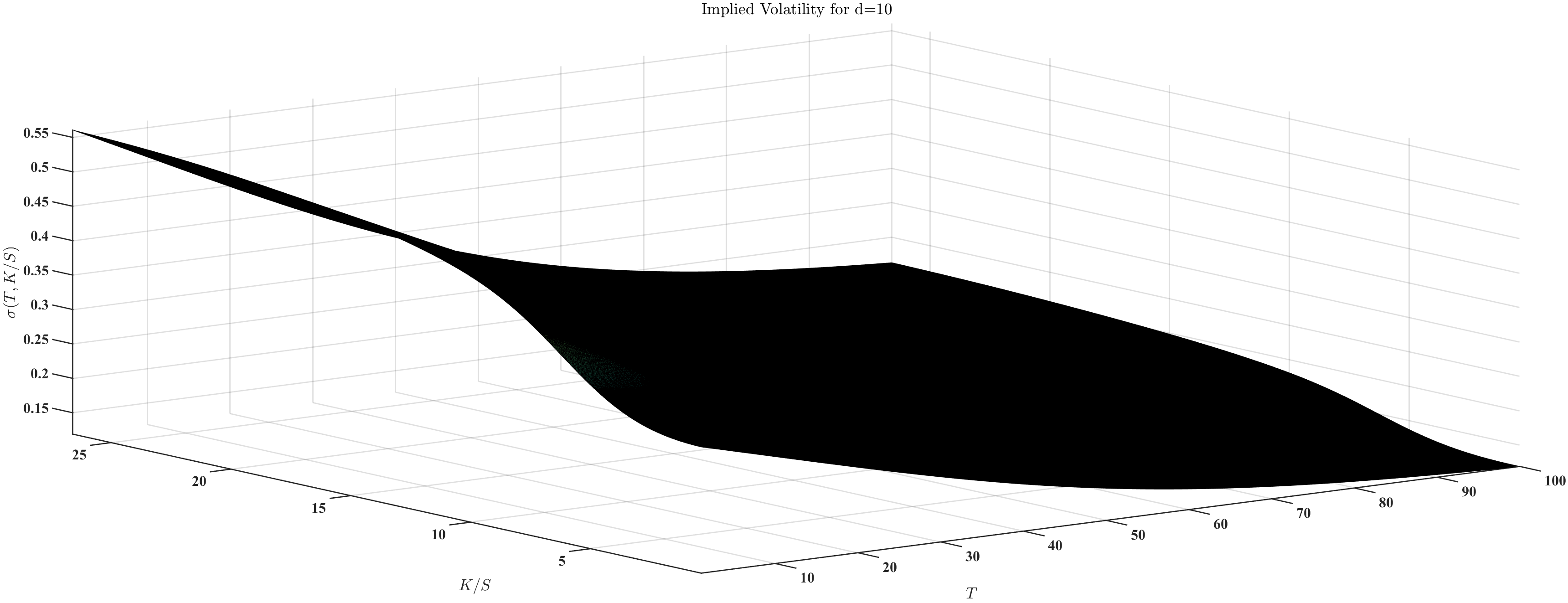}}

\subfloat[]{\includegraphics[width=0.24\textwidth]{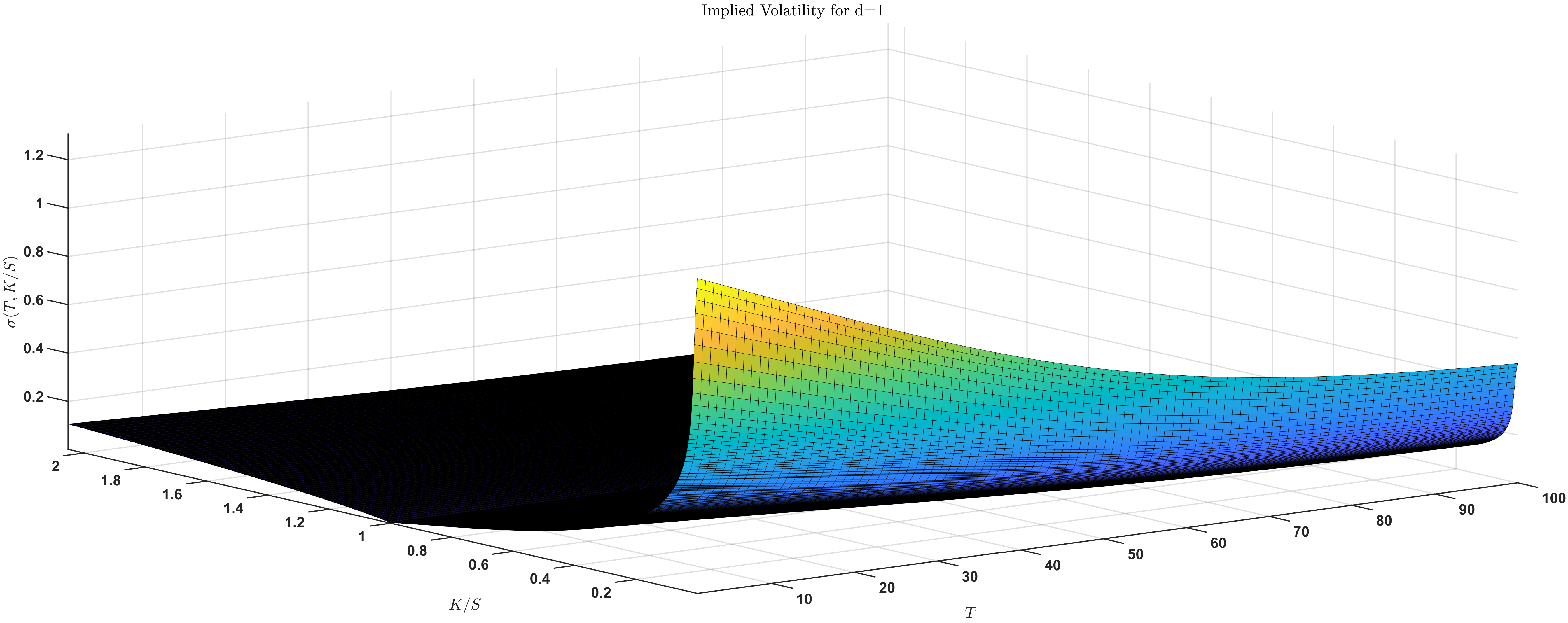}}
\label{Mid_call_impl_vol_AMZN (1)}
\subfloat[]{\includegraphics[width=0.24\textwidth]{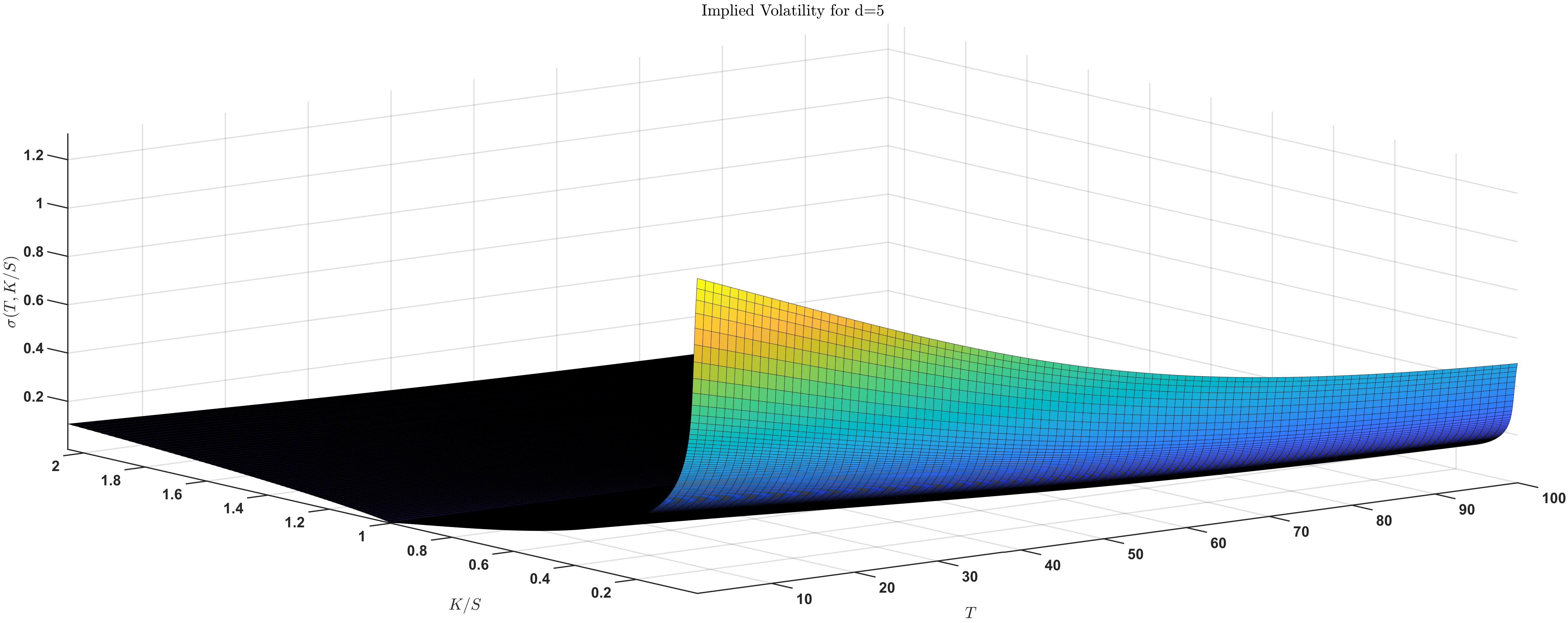}}
\label{Mid_call_impl_vol_AMZN (5)}
\subfloat[]{\includegraphics[width=0.24\textwidth]{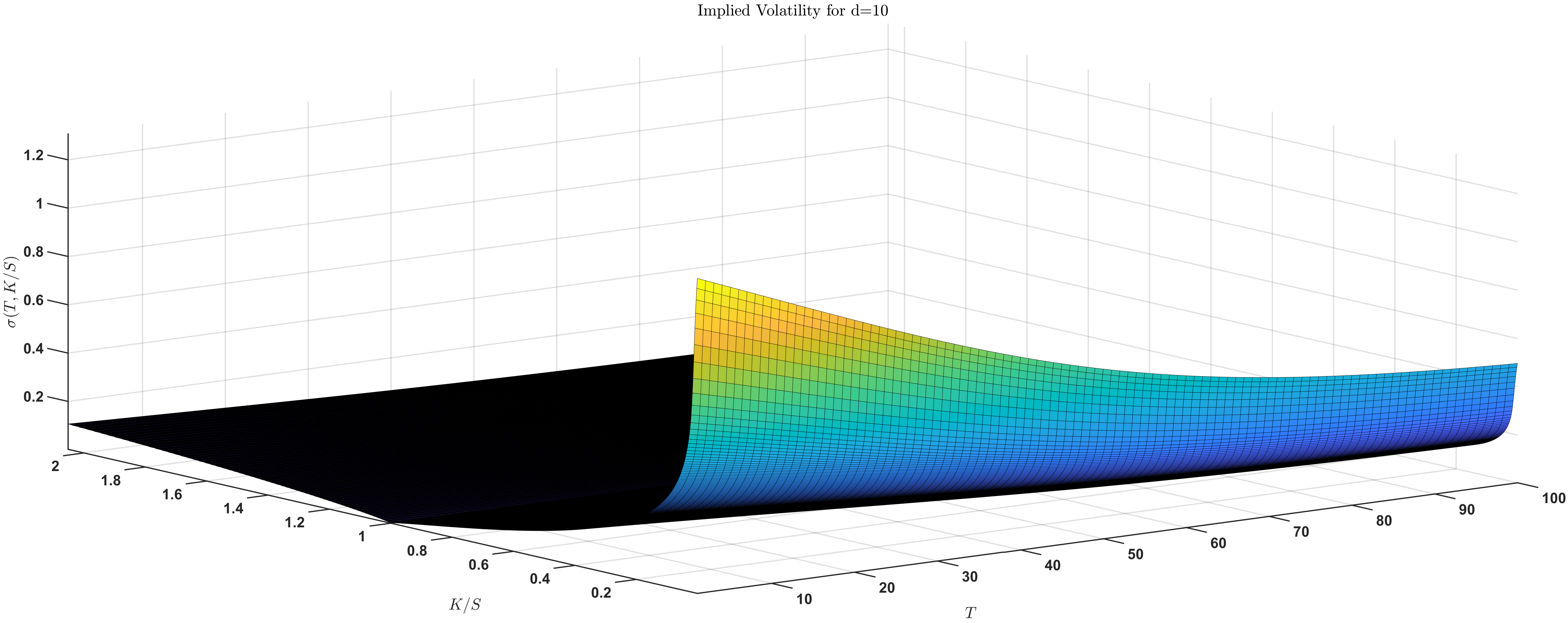}}
\label{Mid_call_impl_vol_AMZN (10)}
\subfloat[]{\includegraphics[width=0.24\textwidth]{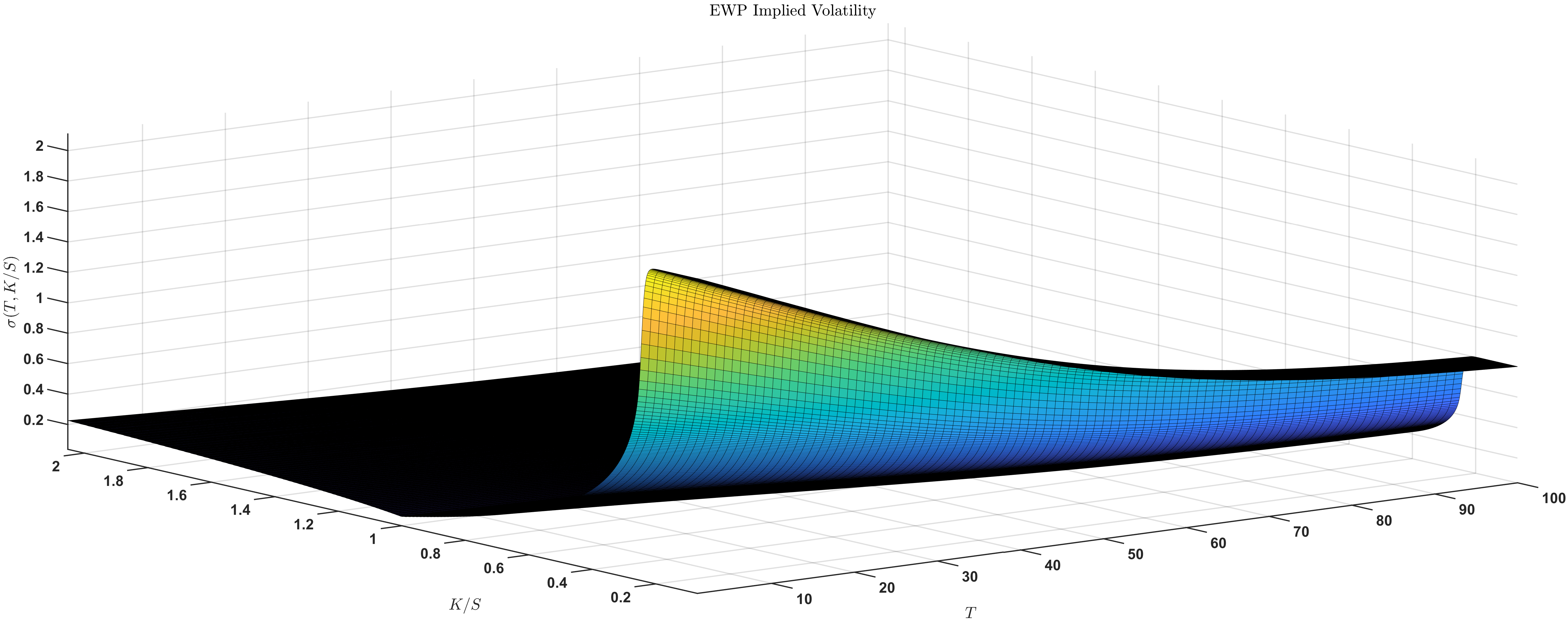}}
\caption{Implied volatility surfaces for (a--d) TMOBBAS and (e--h) GMP at depths of 1, 5, 10, and EWP for AMZN}
\label{IV_dyn_AMZN}
\end{figure}

\begin{figure}[h]
\centering
\subfloat[]{\includegraphics[width=0.24\textwidth]{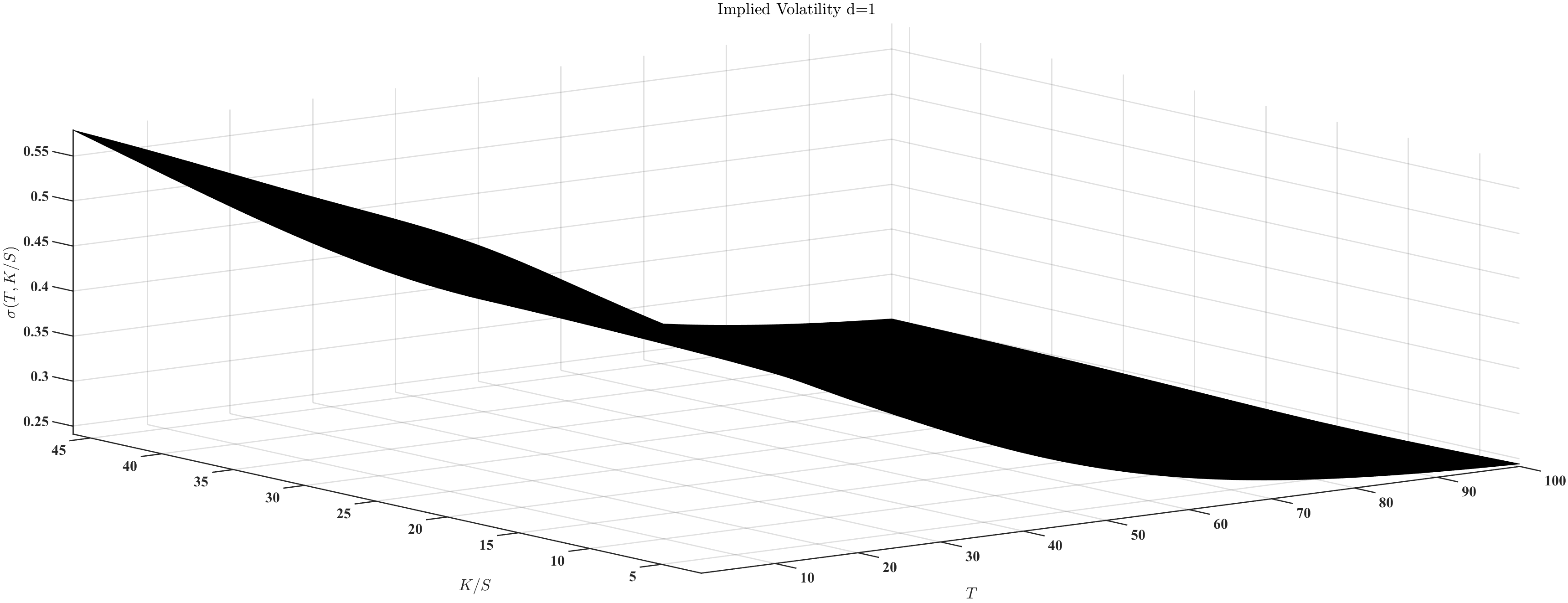}}
\label{call_impl_vol_1}
\subfloat[]{\includegraphics[width=0.24\textwidth]{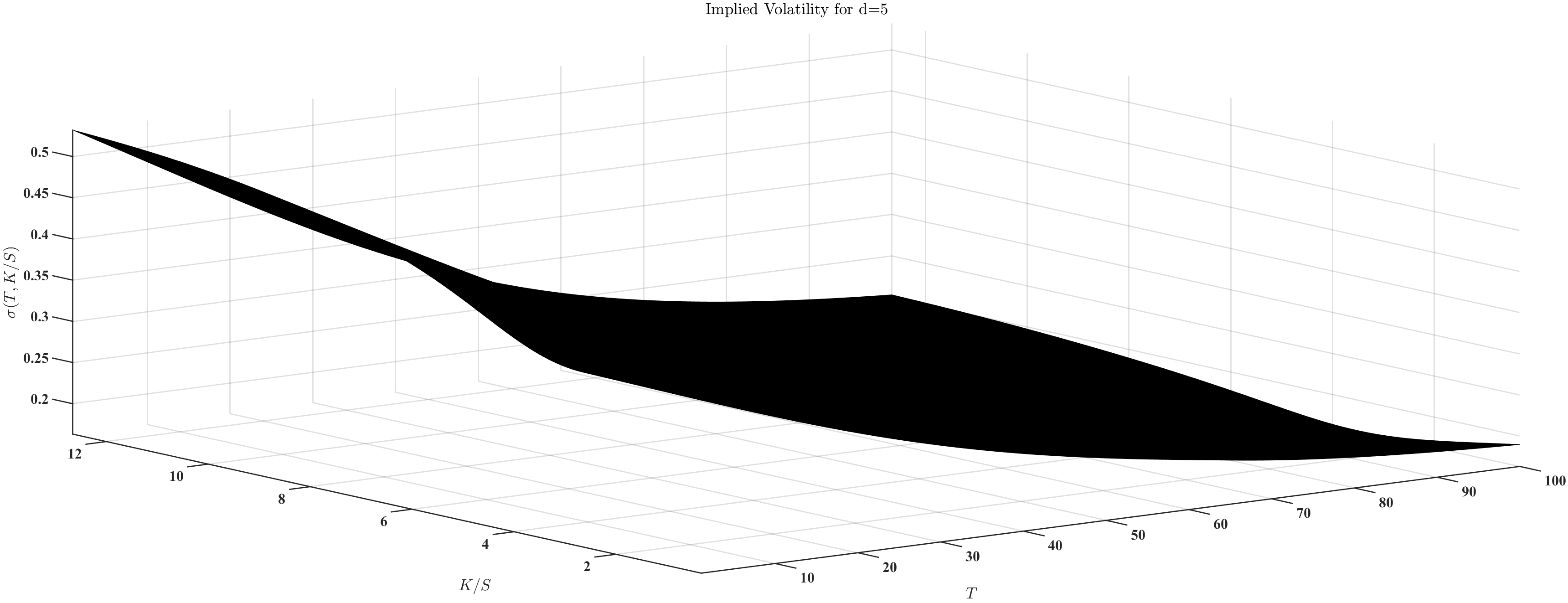}}
\label{call_impl_vol_5}
\subfloat[]{\includegraphics[width=0.24\textwidth]{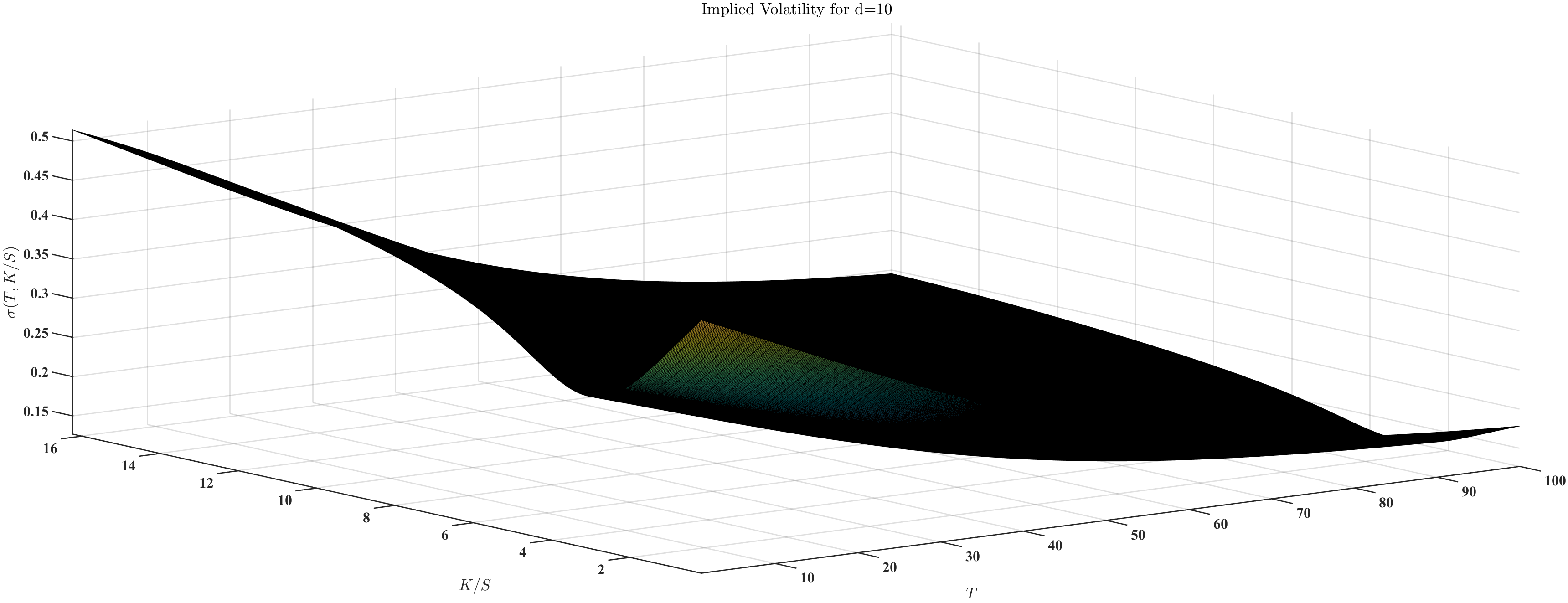}}
\label{call_impl_vol_10}
\subfloat[]{\includegraphics[width=0.24\textwidth]{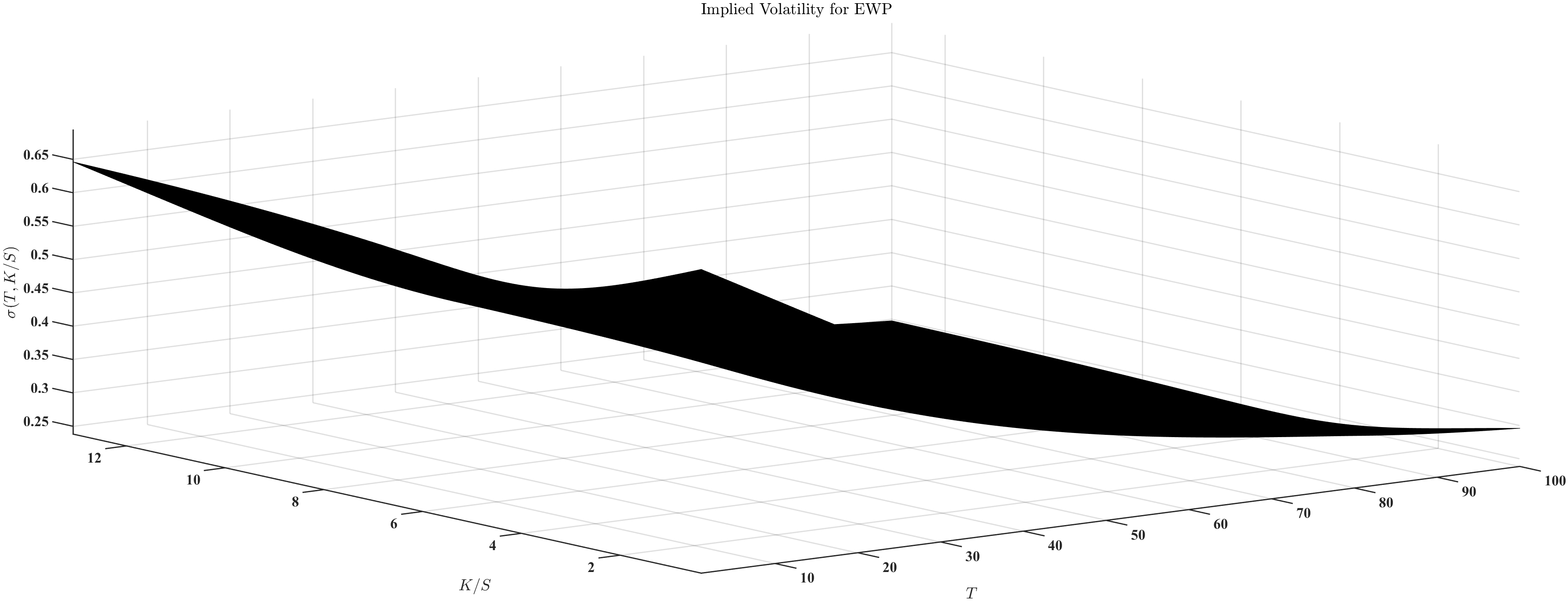}}

\subfloat[]{\includegraphics[width=0.24\textwidth]{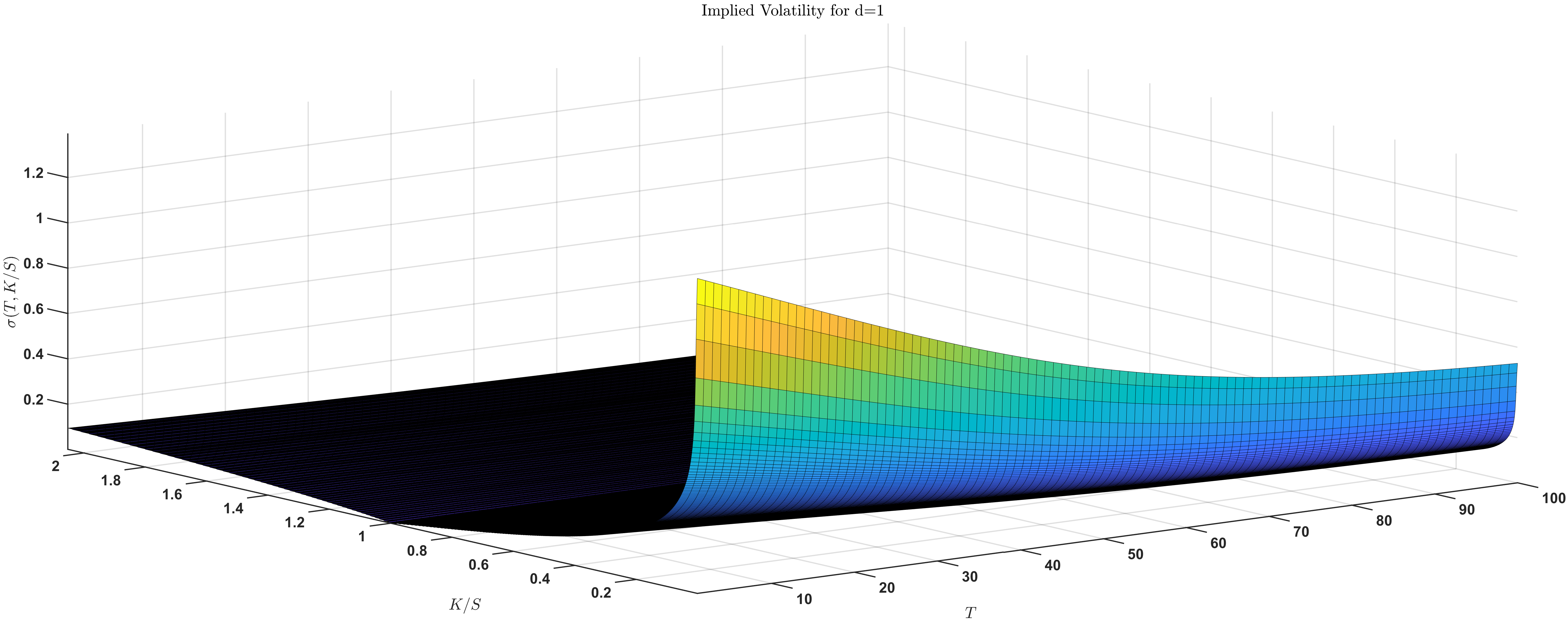}}
\label{Mid_call_impl_vol_GOOG (1)}
\subfloat[]{\includegraphics[width=0.24\textwidth]{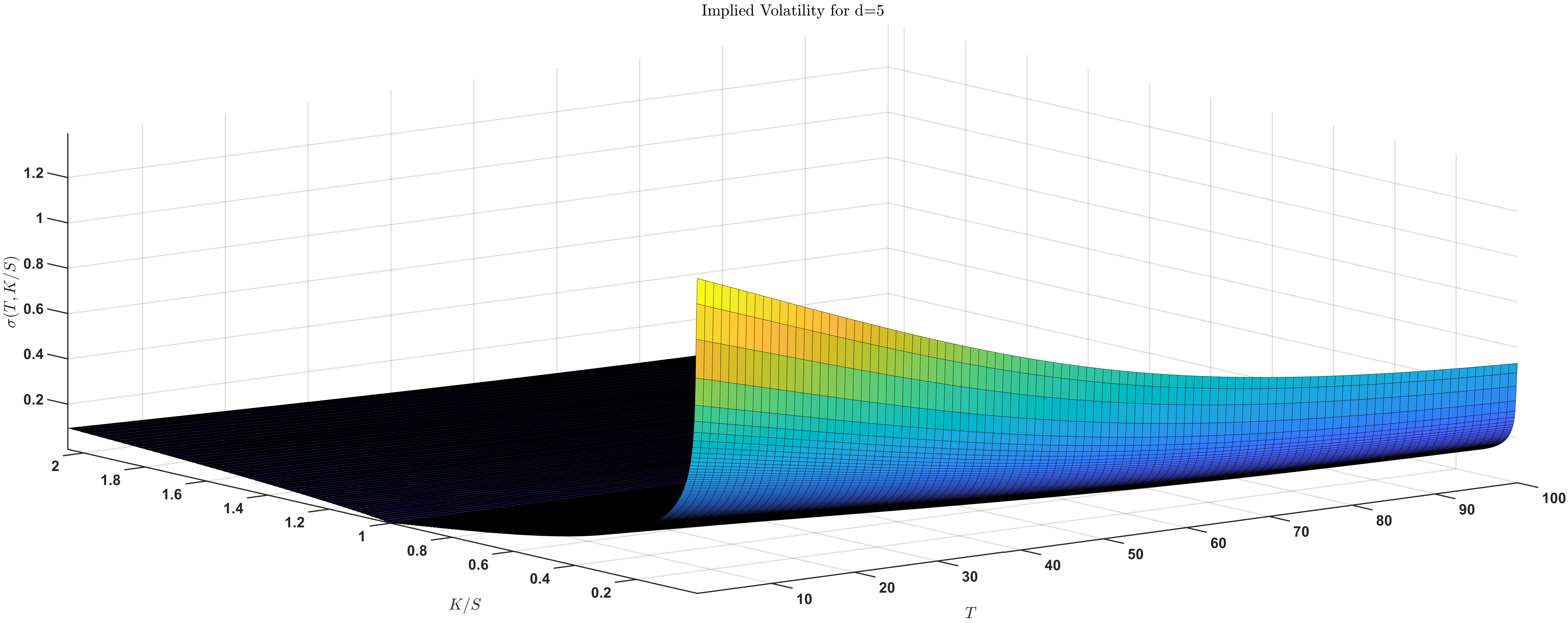}}
\label{Mid_call_impl_vol_GOOG (5)}
\subfloat[]{\includegraphics[width=0.24\textwidth]{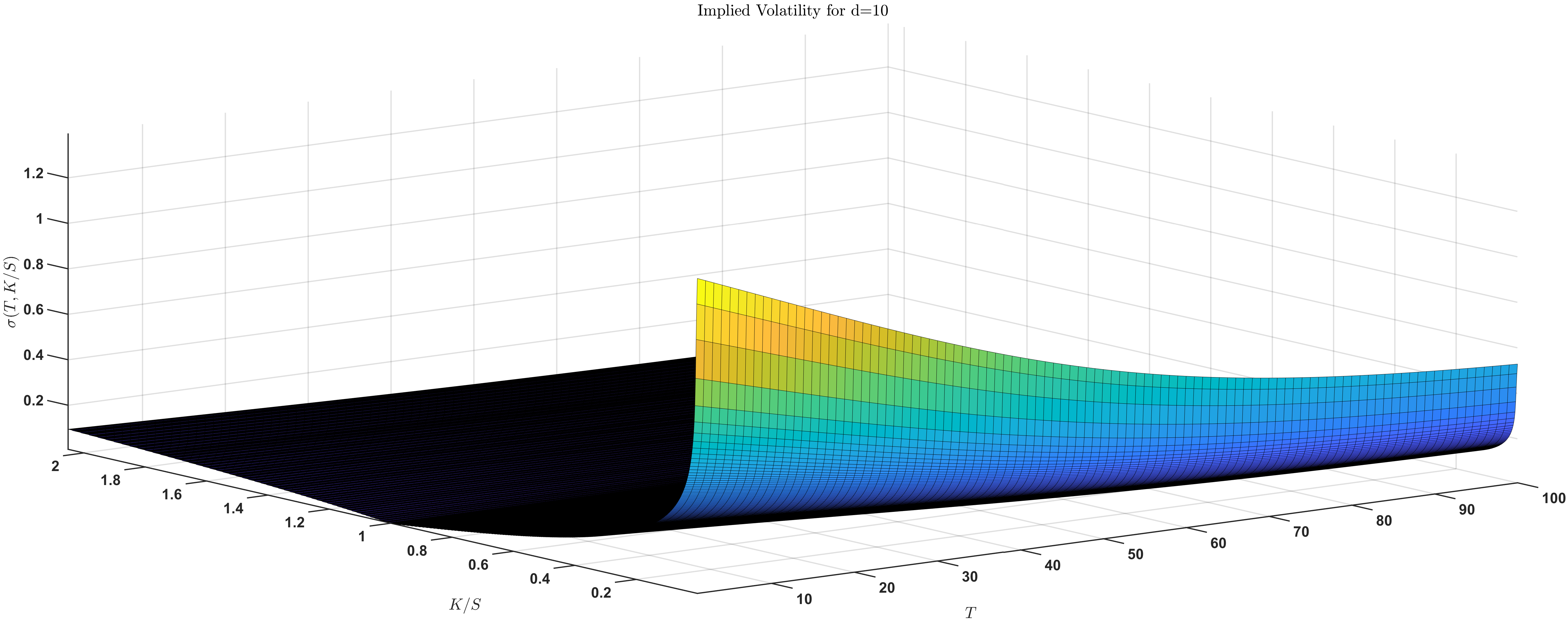}}
\label{Mid_call_impl_vol_GOOG (10)}
\subfloat[]{\includegraphics[width=0.24\textwidth]{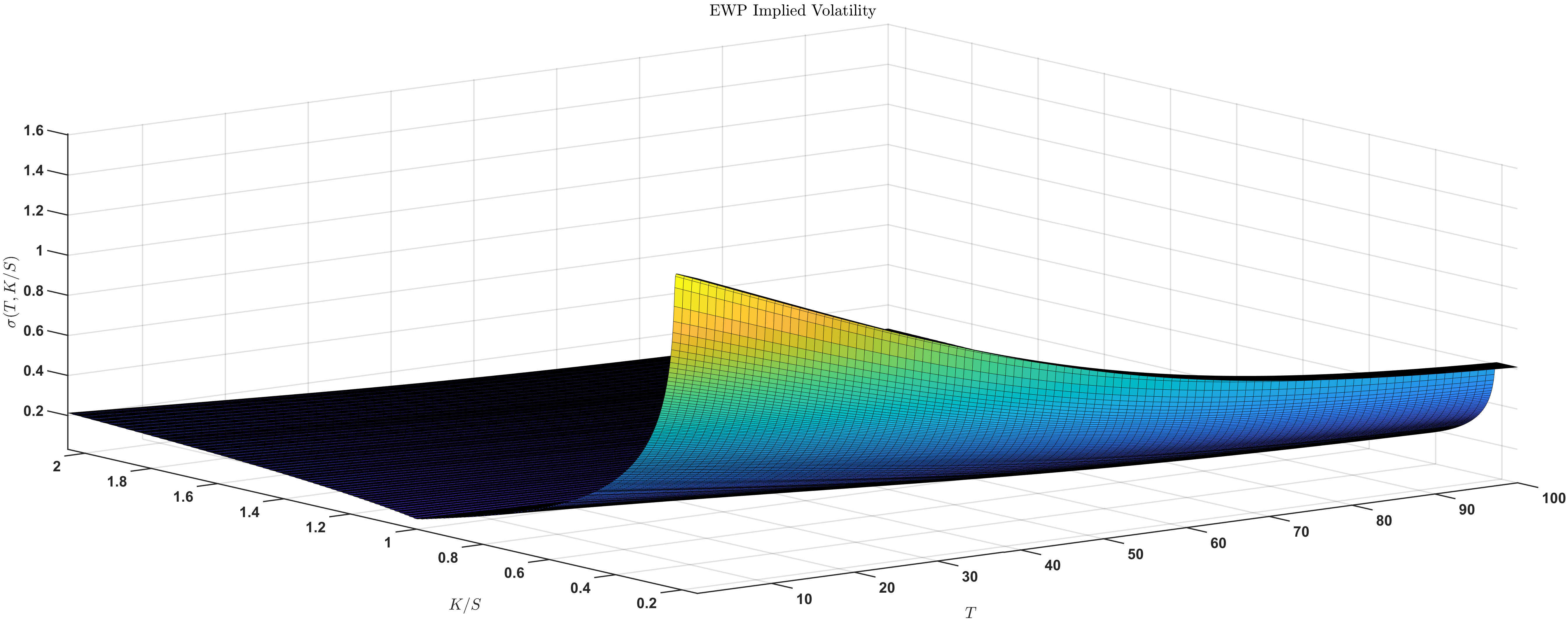}}
\caption{Implied volatility surfaces for (a--d) TMOBBAS and (e--h) GMP at depths of 1, 5, 10, and EWP for GOOG}
\label{IV_dyn_GOOG}
\end{figure}
\clearpage

\section{Table}\label{appendix: table}
These tables present the outcomes of the analyses conducted for AMZN and GOOG. As these analyses parallel those performed for AAPL, we  only furnish  the results, omitting the explanatory narratives for brevity and consistency.

\subsection{Estimating the Shape Parameter by Fitting a Generalized Pareto Distribution}
\begin{table}[htbp]
\tbl{The value of $\hat{\xi}$ and its 95\% CIs of TMOBBAS for AMZN at all 10 depths}
{\begin{tabular}{c c c | c c c}
\hline
Depth $d$ & $\hat{\xi}$ & 95\% CI & Depth $d$ & $\hat{\xi}$ & 95\% CI\\ \hline
1 & 0.135 & (0.118, 0.152) & 2 & 0.203 & (0.182, 0.225)\\
3 & 0.141 & (0.120, 0.162) & 4 & 0.108 & (0.088, 0.127)\\
5 & 0.105 & (0.086, 0.125) & 6 & 0.081 & (0.062, 0.099)\\
7 & 0.106 & (0.087, 0.125) & 8 & 0.101 & (0.083, 0.120)\\
9 & 0.127 & (0.108, 0.146) & 10 & 0.114 & (0.096, 0.133)\\ \hline
\end{tabular}}
\label{table: tmobbas_gpd_fit_est_amzn}
\end{table}

\begin{table}[htbp]
\tbl{The value of $\hat{\xi}$ and its 95\% CIs of TMOBBAS for GOOG at all 10 depths}
{\begin{tabular}{c c c | c c c}
\hline
Depth $d$ & $\hat{\xi}$ & 95\% CI & Depth $d$ & $\hat{\xi}$ & 95\% CI\\ \hline
1 & 0.265 & (0.231, 0.299) & 2 & 0.100 & (0.073, 0.127)\\
3 & 0.110 & (0.084, 0.137) & 4 & 0.100 & (0.073, 0.127)\\
5 & 0.056 & (0.031, 0.081) & 6 & 0.055 & (0.029, 0.081)\\
7 & 0.063 & (0.037, 0.090) & 8 & 0.103 & (0.075, 0.131)\\
9 & 0.084 & (0.057, 0.110) & 10 & 0.129 & (0.101, 0.156)\\ \hline
\end{tabular}}
\label{table: tmobbas_gpd_fit_est_goog}
\end{table}

\begin{table}[htbp]
\tbl{The value of $\hat{\xi}$ and its 95\% CIs of GMP for AMZN at all 10 depths}
{\begin{tabular}{c c c | c c c}
\hline
Depth $d$ & $\hat{\xi}$ & 95\% CI & Depth $d$ & $\hat{\xi}$ & 95\% CI\\ \hline
1 & 0.025 & (0.012, 0.039) & 2 & 0.148 & (0.128, 0.169)\\
3 & 0.109 & (0.090, 0.129) & 4 & 0.106 & (0.087, 0.124)\\
5 & 0.114 & (0.095, 0.133) & 6 & 0.105 & (0.086, 0.123)\\
7 & 0.121 & (0.103, 0.140) & 8 & 0.134 & (0.115, 0.153)\\
9 & 0.147 & (0.128, 0.166) & 10 & 0.159 & (0.140, 0.178)\\ \hline
\end{tabular}}
\label{table: gmp_gpd_fit_est_amzn}
\end{table}

\begin{table}[htbp]
\tbl{The value of $\hat{\xi}$ and its 95\% CIs of GMP for GOOG at all 10 depths}
{\begin{tabular}{c c c | c c c}
\hline
Depth $d$ & $\hat{\xi}$ & 95\% CI & Depth $d$ & $\hat{\xi}$ & 95\% CI\\ \hline
1 & 0.114 & (0.078, 0.149) & 2 & 0.026 & (0.002, 0.051)\\
3 & 0.074 & (0.048, 0.100) & 4 & 0.069 & (0.045, 0.093)\\
5 & 0.077 & (0.053, 0.102) & 6 & 0.094 & (0.068, 0.119)\\
7 & 0.116 & (0.090, 0.141) & 8 & 0.218 & (0.191, 0.245)\\
9 & 0.221 & (0.193, 0.248) & 10 & 0.341 & (0.310, 0.371)\\ \hline
\end{tabular}}
\label{table: gmp_gpd_fit_est_goog}
\end{table}
\clearpage

\subsection{Estimating the Tail Index by Employing the Rank Minus 1/2 Method}
\begin{table}[htbp]
\tbl{The value of $\hat{b}$ and its 95\% CI of TMOBBAS for AMZN}
{\begin{tabular}{c c c | c c c}
\hline
Depth $d$ & $\hat{b}$ & 95\% CI & Depth $d$ & $\hat{b}$ & 95\% CI\\ \hline
1 & 1.035 & (1.028, 1.043) & 2 & 0.412 & (0.409, 0.415)\\
3 & 0.341 & (0.338, 0.344) & 4 & 0.447 & (0.444, 0.450)\\
5 & 0.466 & (0.464, 0.469) & 6 & 0.489 & (0.486, 0.491)\\
7 & 0.498 & (0.495, 0.500) & 8 & 0.505 & (0.502, 0.507)\\
9 & 0.512 & (0.510, 0.515) & 10 & 0.515 & (0.513, 0.517)\\ \hline
\end{tabular}}
\label{table: tmobbas_rank_estimator_amzn}
\end{table}

\begin{table}[htbp]
\tbl{The value of $\hat{b}$ and its 95\% CI of TMOBBAS for GOOG}
{\begin{tabular}{c c c | c c c}
\hline
Depth $d$ & $\hat{b}$ & 95\% CI & Depth $d$ & $\hat{b}$ & 95\% CI\\ \hline
1 & 0.801 & (0.793, 0.808) & 2 & 0.358 & (0.355, 0.361)\\
3 & 0.390 & (0.387, 0.394) & 4 & 0.416 & (0.413, 0.419)\\
5 & 0.444 & (0.441, 0.448) & 6 & 0.449 & (0.446, 0.452)\\
7 & 0.452 & (0.448, 0.455) & 8 & 0.458 & (0.455, 0.461)\\
9 & 0.462 & (0.459, 0.465) & 10 & 0.464 & (0.461, 0.467)\\ \hline
\end{tabular}}
\label{table: tmobbas_rank_estimator_goog}
\end{table}

\begin{table}[htbp]
\tbl{The value of $\hat{b}$ and its 95\% CI of GMP for AMZN}
{\begin{tabular}{c c c | c c c}
\hline
Depth $d$ & $\hat{b}$ & 95\% CI & Depth $d$ & $\hat{b}$ & 95\% CI\\ \hline
1 & 1.293 & (1.284, 1.301) & 2 & 0.413 & (0.409, 0.416)\\
3 & 0.450 & (0.447, 0.453) & 4 & 0.466 & (0.463, 0.468)\\
5 & 0.471 & (0.468, 0.474) & 6 & 0.482 & (0.480, 0.485)\\
7 & 0.492 & (0.489, 0.494) & 8 & 0.501 & (0.499, 0.504)\\
9 & 0.511 & (0.508, 0.513) & 10 & 0.516 & (0.514, 0.518)\\ \hline
\end{tabular}}
\label{table: gmp_rank_estimator_amzn}
\end{table}

\begin{table}[htbp]
\tbl{The value of $\hat{b}$ and its 95\% CI of GMP for GOOG}
{\begin{tabular}{c c c | c c c}
\hline
Depth $d$ & $\hat{b}$ & 95\% CI & Depth $d$ & $\hat{b}$ & 95\% CI\\ \hline
1 & 0.852 & (0.844, 0.861) & 2 & 0.352 & (0.349, 0.355)\\
3 & 0.406 & (0.403, 0.409) & 4 & 0.427 & (0.424, 0.430)\\
5 & 0.436 & (0.433, 0.439) & 6 & 0.439 & (0.436, 0.442)\\
7 & 0.441 & (0.438, 0.444) & 8 & 0.447 & (0.444, 0.450)\\
9 & 0.447 & (0.444, 0.450) & 10 & 0.449 & (0.446, 0.452)\\ \hline
\end{tabular}}
\label{table: gmp_rank_estimator_goog}
\end{table}
\clearpage

\subsection{Assessing Systemic Risk Contagion from Log-Return of TMOBBAS to Individual Spread Levels}
\begin{table}[htbp]
\tbl{CoES and CoETL at different depths for two levels of $\alpha$ for AMZN}
{\begin{tabular}{c c c c c}
\hline
Depth $d$ & CoES(95\%) & CoETL(95\%) & CoES(99\%) & CoETL(99\%) \\ \hline
1     & -0.002    & -0.120     & -0.008    & -0.122     \\ 
2     & -0.003    & -0.180     & -0.015    & -0.197     \\ 
3     & -0.003    & -0.118    & -0.015    & -0.120     \\ 
4     & -0.003    & -0.152     & -0.014    & -0.162     \\ 
5     & -0.008    & -0.282     & -0.032    & -0.318     \\ 
6     & -0.022    & -0.491     & -0.086    & -0.522     \\ 
7     & -0.027    & -0.268     & -0.097    & -0.293     \\ 
8     & -0.044    & -0.373     & -0.143    & -0.412     \\ 
9     & -0.043    & -0.548     & -0.110   & -0.714     \\ 
10    & -0.053    & -0.293     & -0.191    & -0.374     \\ \hline
\end{tabular}}
\label{AMZN_CoES_CoETL_alpha}
\end{table}

\begin{table}[htbp]
\tbl{CoES and CoETL at different depths for two levels of $\alpha$ for GOOG}
{\begin{tabular}{c c c c c}
\hline
Depth $d$ & CoES(95\%) & CoETL(95\%) & CoES(99\%) & CoETL(99\%) \\ \hline
1     & -0.002    & -0.283     & -0.009    & -0.291     \\ 
2     & -0.007    & -0.607     & -0.032    & -0.617     \\ 
3     & -0.012    & -0.479     & -0.052    & -0.496     \\ 
4     & -0.025    & -0.480     & -0.088    & -0.522     \\ 
5     & -0.050    & -1.266     & -0.166    & -1.852     \\ 
6     & -0.062    & -0.687     & -0.175    & -0.958     \\ 
7     & -0.074    & -1.059     & -0.250    & -2.139     \\ 
8     & -0.113    & -0.673    & -0.384    & -0.997     \\ 
9     & -0.103    & -0.915     & -0.238    & -1.282     \\ 
10    & -0.092    & -0.782     & -0.303    & -0.995     \\ \hline
\end{tabular}}
\label{GOOG_CoES_CoETL_alpha_2}
\end{table}
\end{document}